\documentclass[journal]{IEEEtran}
\usepackage[numbers,square,sort&compress]{natbib}
\usepackage{float}
\usepackage{amsmath,amssymb}
\usepackage[utf8]{inputenc}
\usepackage{graphicx}    
\usepackage{url}     
\usepackage{array}
\usepackage{multirow}
\usepackage{tabularx}
\usepackage{placeins}
\usepackage{tikz}
\usetikzlibrary{positioning,arrows.meta,shapes.geometric}
\usepackage{xurl}
\usepackage[hidelinks]{hyperref}
\usepackage{pifont}
\usepackage[table]{xcolor}
\usepackage{titlesec}
\titlespacing*{\subsubsection}{0pt}{*1}{*1}
\usepackage[dvipsnames]{xcolor}
\usepackage{xcolor}
\usepackage{hyperref}
\hypersetup{colorlinks=true,citecolor=blue,linkcolor=blue,urlcolor=blue}
\usepackage{acronym}
\usepackage{hypernat}

\makeatletter
\renewcommand{\NAT@open}{\textcolor{blue}{[}}
\renewcommand{\NAT@close}{\textcolor{blue}{]}}

\makeatother

\newcommand{\cellD}{\cellcolor{BlueGreen!90}\D}
\newcommand{\cellM}{\cellcolor{BlueGreen!50}\M}
\newcommand{\cellB}{\cellcolor{BlueGreen!15}\B}
\newcommand{\cellNA}{\cellcolor{BlueGreen!5}\NA}

\newcommand{\NA}{--}
\newcolumntype{C}[1]{>{\centering\arraybackslash}m{#1}}
\newcommand{\B}{\textbf{B}}
\newcommand{\M}{\textbf{M}}
\newcommand{\D}{\textbf{D}}
\title{Internet of Everything in the 6G Era: Paradigms, Enablers, Potentials and Future Directions}

\author{
\IEEEauthorblockN{****}
\IEEEauthorblockA{
Affiliation\\
Email: driss.choukri.doc@uhp.ac.ma
}
}
\author{Driss~Choukri,~\IEEEmembership{Student Member,~IEEE,}
        Essaid Sabir,~\IEEEmembership{Senior Member,~IEEE,}\\
        Elmahdi Driouch,~\IEEEmembership{Senior Member,~IEEE,}
        and~Abdelkrim Haqiq,~\IEEEmembership{Senior Member,~IEEE}

\thanks{D. Choukri is with Computer Networks, Mobility
and Modeling Laboratory (IR2M), FST, Hassan I University of Settat, Morocco, and the Department of Science and Technology, TÉLUQ, University of Quebec, Montreal, H2S 3L4, Canada, e-mail: driss.choukri.doc@uhp.ac.ma.}
\thanks{E. Sabir is with the Department of Science and Technology, TÉLUQ, University of Quebec, Montreal, H2S 3L4, Canada, e-mail: essaid.sabir@teluq.ca.}
\thanks{E. Driouch is with the Department of Computer Science, University of Quebec at Montreal (UQAM), Montreal, H2L 2C4, Canada, e-mail: driouch.elmahdi@uqam.ca.}
\thanks{A. Haqiq is with Computer Networks, Mobility
and Modeling Laboratory (IR2M), FST, Hassan I University of Settat, Morocco, e-mail: abdelkrim.haqiq@uhp.ac.ma.}
\thanks{Manuscript received April 19, 2026; revised July 26, 2026.}}

\begin{document}

\maketitle

\begin{abstract}
The Internet of Everything (IoE) represents an evolution of the Internet of Things (IoT) by integrating people, data, processes, and things into a unified intelligent ecosystem. IoE aims to enhance automation, decision-making, and service efficiency across multiple application domains such as smart cities, healthcare, industry, and next-generation wireless networks. This paper provides a structured overview of the IoE concept, its core components, architectural foundations, enabling technologies, and major research challenges. Finally, open research directions toward 6G-enabled intelligent IoE systems are discussed, with emphasis on scalability, security, privacy, and energy efficiency.
\end{abstract}


\section{Introduction}
\label{sec:intro}

\begin{quote}
\emph{“The most profound technologies are those that disappear. They weave themselves into the fabric of everyday life until they are indistinguishable from it.”} 
\end{quote}
\vspace{-0.3em}
\begin{flushright}
--- Mark Weiser (1991) \cite{weiser1991computer}
\end{flushright}

\subsection{From Internet of Things to Internet of Everything}

\begin{figure}[!t]
  \centering
  \includegraphics[width=0.5\textwidth]{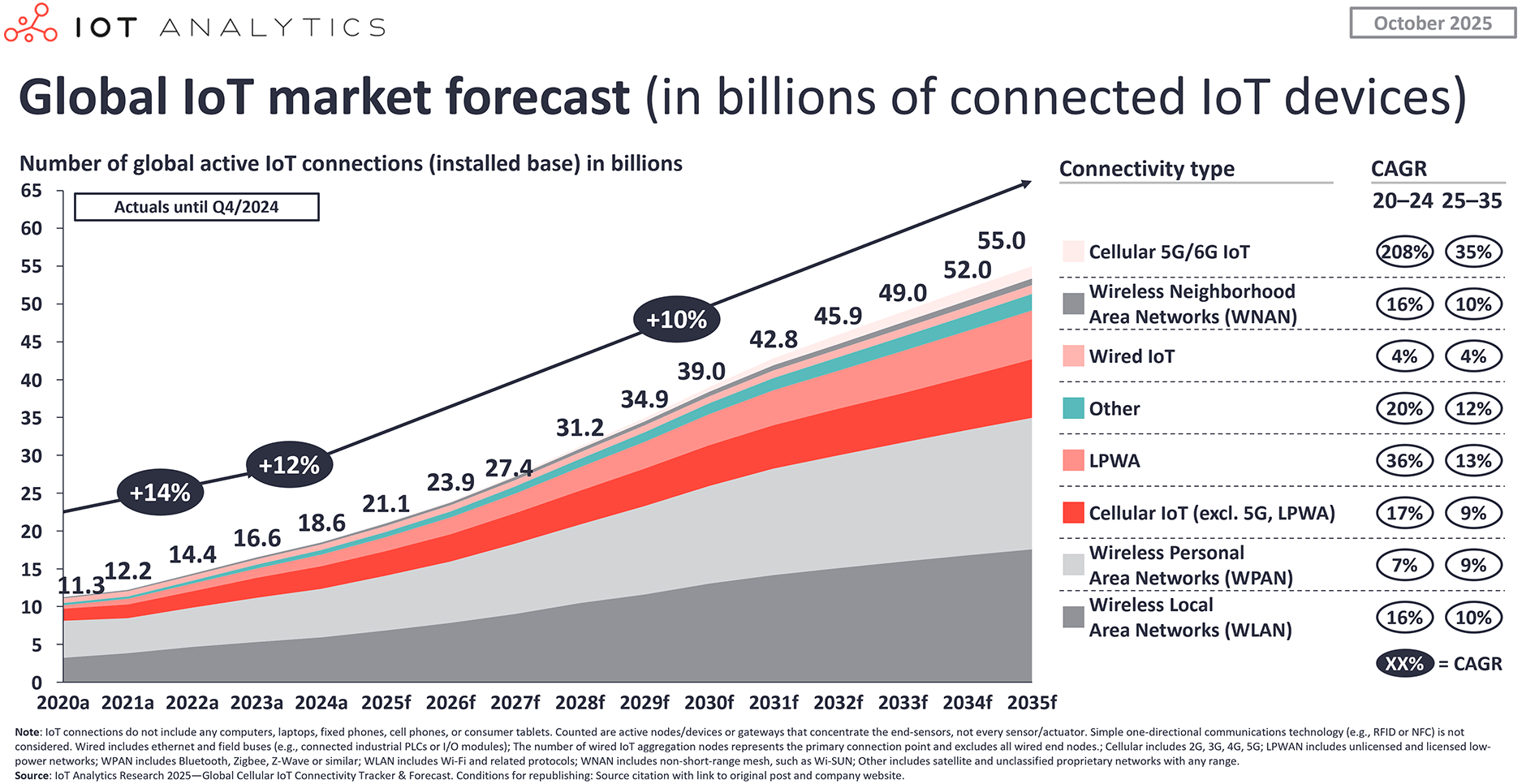}\vspace{-0.3cm}
  \caption{Projected growth of global connected IoT devices, reaching approximately 39 billion by 2030 and 55 billion by 2035 (\textit{IoT Analytics}, \url{https://iot-analytics.com/number-connected-iot-devices/}).}
  \label{fig:evolution}\vspace{-0.2cm}
\end{figure}

\IEEEPARstart{T}{he} rapid proliferation of connected devices, embedded sensors, and intelligent services has accelerated the evolution of communication networks from traditional Internet connectivity toward large-scale cyber--physical ecosystems. The IoT has emerged as a key paradigm for interconnecting a vast array of heterogeneous physical objects to the Internet, leveraging sensors and actuators to enable seamless sensing, monitoring, and actuation across diverse application domains \cite{chang2019_iot_new_paradigms,borgia2014_iot_vision,yun2019_onem2m_sensing_actuation}. Recent industry estimates reported that the global installed base of connected IoT devices had already reached approximately 21.1 billion by the end of 2025 \cite{iotanalytics_connected_iot_devices_2025}, reflecting the strong momentum of IoT adoption. Building on this trajectory, long-term projections illustrated in 
Fig.~\ref{fig:evolution} anticipate continued expansion to nearly 39.0 billion devices by 2030 and up to 55.0 billion by 2035, driven in particular by advances in cellulars such as the fifth-generation (5G) and sixth-generation (6G) mobile systems and low-power wide-area connectivity. However, as the ecosystem scales toward these milestones, deployments continue to face fundamental challenges related to architectural heterogeneity, network scalability, and the need for efficient data-driven operation and management \cite{atzori2010iot,alfuqaha2015iot}. 

The term \emph{Internet of Everything} was first formally introduced by Cisco in 2012\footnote{Dave Evans, "The Internet of Everything: How More Relevant and Valuable Connections Will Change the World", Cisco, \url{https://www.cisco.com/c/dam/global/en_my/assets/ciscoinnovate/pdfs/IoE.pdf}}, where it was defined as the intelligent interconnection of \emph{people, process, data, and things} to create more relevant and valuable networked interactions \cite{cisco2012ioe,cisco_ioe_faq}. Unlike the earlier IoT, which primarily emphasized connectivity among physical devices, IoE broadened the scope to include human interaction, contextual data processing, and service orchestration as integral components of value creation.
Since its inception, the concept of IoE has evolved significantly within the
academic community. Recent research frames IoE not merely as an extension of IoT,
but as a large-scale cyber–physical–social ecosystem characterized by distributed
intelligence, edge–cloud integration, trust-aware coordination, and cross-domain
service automation \cite{saad2020vision6g,sharma2025_6g_overview,ding2025_economic_ioe}.
In the emerging 6G context, IoE is increasingly envisioned as an AI-native,
energy-aware, and security-by-design architecture, where intelligence is
deeply embedded across communication and computing layers to enable
massive connectivity, immersive services, and autonomous decision-making
throughout the device–edge–cloud continuum \cite{cui2025ai6g}.

\subsection{Trending IoE Topics}

The IoE is rapidly evolving toward a tightly integrated cyber–physical–social fabric where intelligence, connectivity, trust, and sustainability must co-exist by design. Beyond incremental IoT enhancements, IoE broadens the scope by integrating \emph{people}, \emph{data}, and \emph{processes} in addition to \emph{things}, and emphasizes value creation from relevant, contextual, and actionable connections at scale \cite{cisco_ioe_faq,cisco2012ioe}. From this perspective, the 6G-enabled IoE vision demands a fundamental rethinking of system architecture across cognition, networking, computing, sustainability, security, service design, and economic models.. The following emerging directions capture the dominant research momentum shaping next-generation IoE ecosystems.\vspace{.2cm}

\textbf{Agentic Artificial Intelligence for Autonomous IoE: } 
IoE generates massive, real-time, multimodal data streams from interconnected people, devices, infrastructures, and processes. 
In this context, \emph{Agentic Artificial Intelligence} (AAI) can be viewed as an emerging paradigm aligned with Artificial Intelligence (AI)-native networking and distributed intelligence visions, where intelligent agents enable autonomous perception, reasoning, and decision-making across distributed environments. 
While IoE provides the sensing and actuation substrate of the physical–digital world, agentic AI facilitates adaptive coordination, distributed optimization, and self-organizing control mechanisms. 
The 6G vision increasingly integrates AI-native networking and distributed intelligence as core architectural principles rather than optional overlays \cite{saad2020vision6g,sharma2025_6g_overview}. 
The convergence of AI and blockchain further strengthens trusted autonomous ecosystems, where AI-driven mechanisms enhance distributed coordination while blockchain ensures data provenance, security, and accountability \cite{10251433,10769427}. 
Together, these advances pave the way toward self-adaptive, self-optimizing, and self-organizing IoE systems.\vspace{.2cm}

\textbf{Global Connectivity as the IoE Backbone: }
Global connectivity is foundational to IoE because it enables seamless interaction among billions of heterogeneous entities across geographic boundaries. Unlike traditional IoT deployments, IoE must operate at planetary scale, supporting massive device density, ultra-low latency, high reliability, and integrated terrestrial–non-terrestrial networking. Emerging 6G frameworks target peak data rates up to 1~Tb/s, sub-millisecond latency, enhanced spectral efficiency, and expanded operation across sub-6~GHz, millimeter Wave (mmWave), centimeter wave (cmWave), and Terahertz (THz) bands \cite{li2025_6g_isci_review,shamsabadi2024_ihrllc_6g,sharma2025_6g_overview}. These quantitative Key Performance Indicators (KPIs) directly map to large-scale IoE demands, including immersive services, mission-critical automation, and distributed intelligence. A wide range of enabling technologies have been proposed to support large-scale IoE deployments. Among these, cell-free massive Multiple-Input Multiple-Output (MIMO) and wireless-powered IoE stand out as promising solutions to ensure uniform Quality of Service (QoS) and scalable massive access over wide geographical areas \cite{chen2020wirelesspowered}.\vspace{.2cm}

\textbf{Edge Computing and Decentralized Intelligence: }
As IoE scales, centralized cloud-centric pipelines become insufficient due to latency, bandwidth, privacy, and resilience constraints. Edge computing and fog architectures distribute intelligence closer to data sources, reducing round-trip delays and enhancing operational robustness. The Fog-of-Everything paradigm argues that IoE and fog computing should be treated as complementary constructs, balancing performance and energy efficiency across the device–edge–cloud continuum \cite{foe2017_fog_of_everything}. Blockchain has also been positioned as a decentralization enabler for Edge-of-Things coordination, improving transparency, auditability, and trust across heterogeneous systems \cite{9569742}. Moreover, blockchain-based incentive mechanisms have been proposed to stimulate participation and fair resource sharing in open IoE ecosystems, introducing economic and governance dimensions into decentralized architectures \cite{10788349}. However, performance evaluations reveal non-trivial latency, storage growth, and consensus overheads that challenge real-time IoE scalability \cite{9424688}. Consequently, decentralization must be carefully co-designed with system-level constraints. Federated Learning (FL) further strengthens decentralized intelligence, particularly in industrial IoE where privacy regulations prevent data centralization \cite{gulhane2025intro_fl_industry,mohana2025fl_ioe}. FL enables collaborative model training while preserving data locality, making it a key enabler for distributed, regulation-aware IoE ecosystems.\vspace{.2cm}

\textbf{IoE Architectures  for economic and environmental sustainability: }
The massive scale of IoE risks unsustainable energy consumption if not architected responsibly. Sustainability therefore emerges as a structural requirement rather than a secondary optimization objective. Green 6G enablers include zero-carbon network slicing and energy-aware vertical edge services \cite{10086577}, energy-efficient autonomous wireless networks \cite{babbar2023_massiveiot_ioe}, and green industrial IoT deployments \cite{app14188558}. FL is also evolving toward green-aware designs, emphasizing computation–communication co-optimization, adaptive participation, and energy-efficient aggregation \cite{9745426,QUY2024101061,10.1145/3718363}. Beyond technical metrics, economic sustainability is equally critical. Incentive mechanisms, pricing models, and governance strategies must align stakeholder objectives with system-wide performance goals to ensure long-term scalability of IoE ecosystems \cite{ding2025_economic_ioe}.\vspace{.2cm}

\textbf{IoE–Metaverse Convergence: The Physical–Digital Continuum: }
The convergence between IoE and the metaverse extends connectivity into immersive, interactive, and experiential domains. While IoE provides the intelligent connective fabric, the metaverse offers a real-time digital interface layer built upon digital twins, Augmented Reality (AR), and virtual reality (VR). Such integration enables synchronized physical–cyber operation, remote actuation, immersive industrial control, and context-aware decision support, reinforcing the role of IoE as the infrastructural substrate of next-generation cyber–physical ecosystems \cite{li2025_6g_isci_review}.\vspace{.2cm}

\textbf{Security and Privacy-by-Design as a Structural Requirement: }
Security and privacy can no longer be retrofitted in IoE deployments. A Security/Privacy-by-Design (SPbD/PSbD) paradigm promotes secure-by-default configurations, lifecycle-wide governance, embedded trust anchors, and cross-layer protection \cite{spbd_ioe}. Recent 6G security roadmaps emphasize identity-centric architectures, privacy-preserving computation, 
\begin{figure*}[!ht]
\centering
\scalebox{0.85}{%
  \begin{tikzpicture}[
    font=\small,
    box/.style={draw, very thick, rounded corners=1pt, align=center, inner sep=3pt},
    org/.style={box, text width=26mm, minimum height=9mm, fill=BlueGreen!30},
    sec/.style={box, text width=44mm, minimum height=9mm, fill=BlueGreen!80},
    sub/.style={box, text width=74mm, minimum height=9mm, align=left, inner sep=4pt, fill=Aquamarine!8},
  ]

  \node[sec] (s1) {I. Introduction};
  \node[sec, below=5mm of s1] (s2) {II. IoE Paradigms \& Architecture};
  \node[sec, below=5mm of s2] (s3) {III. IoE Enabling\\Technologies};
  \node[sec, below=6mm of s3] (s4) {IV. IoE Applications \& Use Cases};
  \node[sec, below=5mm of s4] (s5) {V. Key Trends Shaping Future IoE Architectures};
  \node[sec, below=5mm of s5] (s6) {VI. Key Research Frontiers in 6G-Enabled IoE};
  \node[sec, below=5mm of s6] (s7) {VII. Conclusion \& References};
    \draw[very thick] (s1.south) -- (s2.north);
    \draw[very thick] (s2.south) -- (s3.north);
    \draw[very thick] (s3.south) -- (s4.north);
    \draw[very thick] (s4.south) -- (s5.north);
    \draw[very thick] (s5.south) -- (s6.north);
    \draw[very thick] (s6.south) -- (s7.north);




  \node[sub, right=6mm of s2] (ss2) {%
    \begin{tabular}{@{}l@{}}
      II.1. Core Components of IoE\\
      II.2. IoE as cross-domain synergy among IoXs\\
      II.3. Bio-nano IoE and molecular communication\\
      II.4. IoE System Architecture
    \end{tabular}%
  };
  \draw[very thick] (s2.east) -- (ss2.west);

  \node[sub, left=6mm of s2] (ss3) {%
    \begin{tabular}{@{}l@{}}
      III.1. AI/ML and Edge Intelligence\\
      III.2. Edge/Fog/Cloud Computing\\
      III.3. Energy-as-a-Service for RF-Powered IoE systems\\
      III.4. Big Data Analytics and Data Mining\\
      III.5. Connectivity Layer: 5G/6G and Beyond\\
      III.6. Molecular Communication and Bio-Nano IoE \\ \quad \quad (IoBNT)\\
      III.7. Federated Learning for Distributed Intelligence\\
      III.8. Blockchain / Distributed Ledger Technologies\\
      III.9. Standards and Interoperability Mechanisms
    \end{tabular}%
  };
  \draw[very thick] (s3.181) -- (ss3.339);

  \node[sub, left=6mm of s4] (ss4) {%
    \begin{tabular}{@{}l@{}}
      IV.1. Representative Application Domains\\
      IV.2. Deep-Dive Case Studies
    \end{tabular}%
  };
  \draw[very thick] (s4.west) -- (ss4.east);

  \node[sub, below=4mm of ss4] (ss5) {%
    \begin{tabular}{@{}l@{}}
      V.1. Artificial General Intelligence for Autonomous IoE \\
      V.2. Global Connectivity as the IoE Backbone\\
      V.3. Edge Computing and Decentralized Intelligence\\
      V.4. Multi-Cloud Federation and Resource Virtualization \\\quad \quad in IoE\\
      V.5. Edge-Orchestration and Scalability Challenges\\
      V.6. Sustainability and Green IoE Architectures\\
      V.7. IoE--Metaverse Convergence: The Physical--Digital \\\quad \quad Continuum\\
      V.8. Security and Privacy-by-Design in IoE Systems\\
    \end{tabular}%
  };
  \draw[very thick] (s5.180) -- (ss5.20);

  \node[sub, below=4mm of ss2] (ss6) {%
    \begin{tabular}{@{}l@{}}
      VI.1. AI-Native and Semantic-Aware Orchestration for \\\quad\quad Autonomous IoE\\
      VI.2. Zero-Touch Management and Digital-Twin-Assisted \\\quad\quad Operation\\
      VI.3. Near-Field Communications and RIS-Assisted \\\quad\quad Localization\\
      VI.4. Communication-Efficient Federated Intelligence\\
      VI.5. Trustworthy and Incentivized Participation\\
      VI.6. Security and Privacy for 6G IoE\\
      VI.7. Economic-Aware Mechanism Design for 6G-Scale \\\quad\quad IoE\\
      VI.8. IoE Ecomonics and Business Models\\
      VI.9. Green and Energy-Aware Intelligent IoE
    \end{tabular}%
  };
  \draw[very thick] (s6.0) -- (ss6.209);

  \end{tikzpicture}%
} 
\vspace{-.3cm}
\caption{Paper structure: main sections with their corresponding subsection titles.}
\label{fig:paper_organization_tree}\vspace{-0.5cm}
\end{figure*}
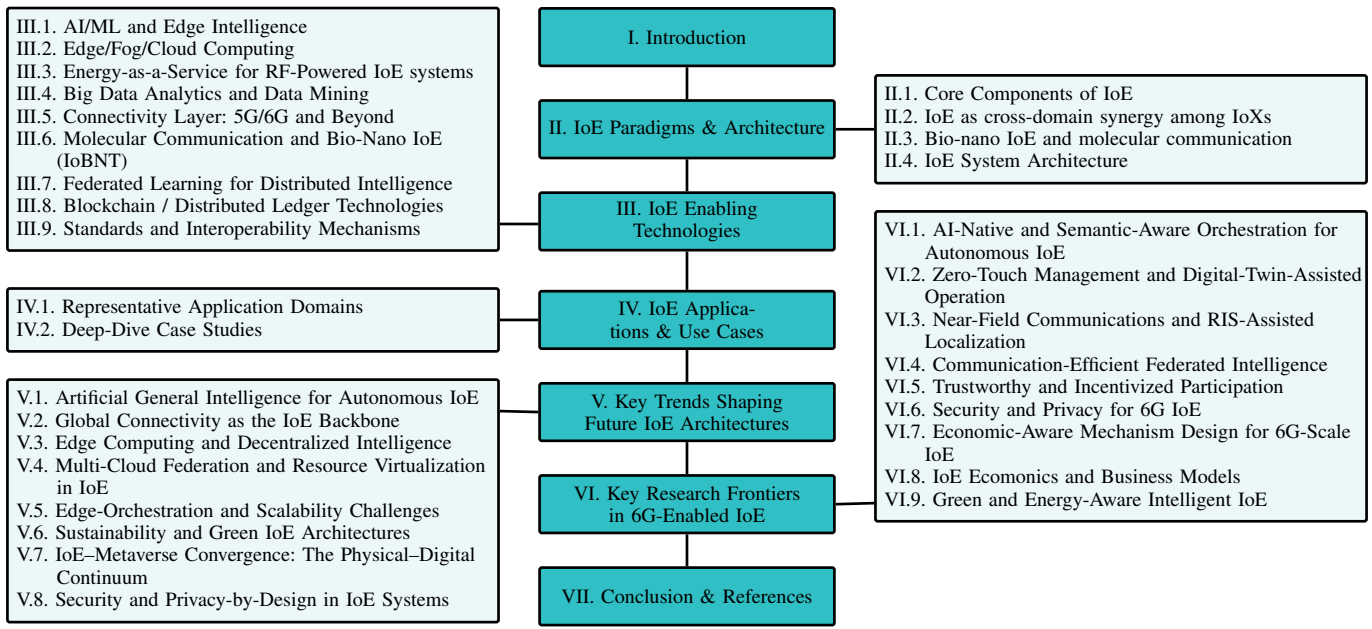
AI-aware defense mechanisms, and zero-trust principles as foundational properties of next-generation networks \cite{10942738,10942542}. Although blockchain can enhance transparency and traceability, it is not a universal security solution and must be embedded within broader cross-layer orchestration frameworks \cite{10251433}. Collectively, these trends illustrate that 6G-enabled IoE systems require unified architectural thinking that integrates cognition, connectivity, decentralization, sustainability, immersive interaction, and native trust as co-equal design pillars.

\subsection{Contributions and scope}

This survey positions IoE as (i) a holistic integration of people--data--processes--things and (ii) a cross-domain convergence of multiple Internet of X (IoX) paradigms, and systematically connects this vision to key 6G requirements, including massive connectivity, Ultra-Reliable Low-Latency Communication (URLLC), and AI-native distributed intelligence. 
Unlike IoT-centric surveys, we further emphasize (a) bio-nano IoE and Molecular Communication (MC), (b) economic sustainability and incentive mechanisms, and (c) SPbD/PSbD as fundamental and cross-layer design principles. 
Our main contributions are summarized as follows:

\begin{itemize}
    \item[--] We provide a comprehensive overview of the evolution from IoT to IoE, highlighting the added value of integrating people, data, processes, and things into a unified and intelligent ecosystem.
    
    \item[--] We present and analyze IoE architectural models, including layered frameworks and the device--edge/fog--cloud continuum, while supporting low-latency, scalable, and AI-driven services.
    
    \item[--] We review and categorize key enabling technologies for IoE, including AI/ML, edge/fog/cloud computing, big-data analytics, 5G/6G connectivity, blockchain, and FL, with a focus on their roles in supporting scalable and autonomous IoE systems.

    \item[--] We provide a comparative analysis of recent IoE surveys and identify key research gaps to position future research directions.
    
    \item[--] We identify and synthesize major challenges and open research directions toward 6G-enabled IoE, with particular emphasis on scalability, energy efficiency, trust, and security/privacy.
\end{itemize}


\begin{figure*}[!t]
  \centering
  \includegraphics[width=\textwidth, height=9cm]{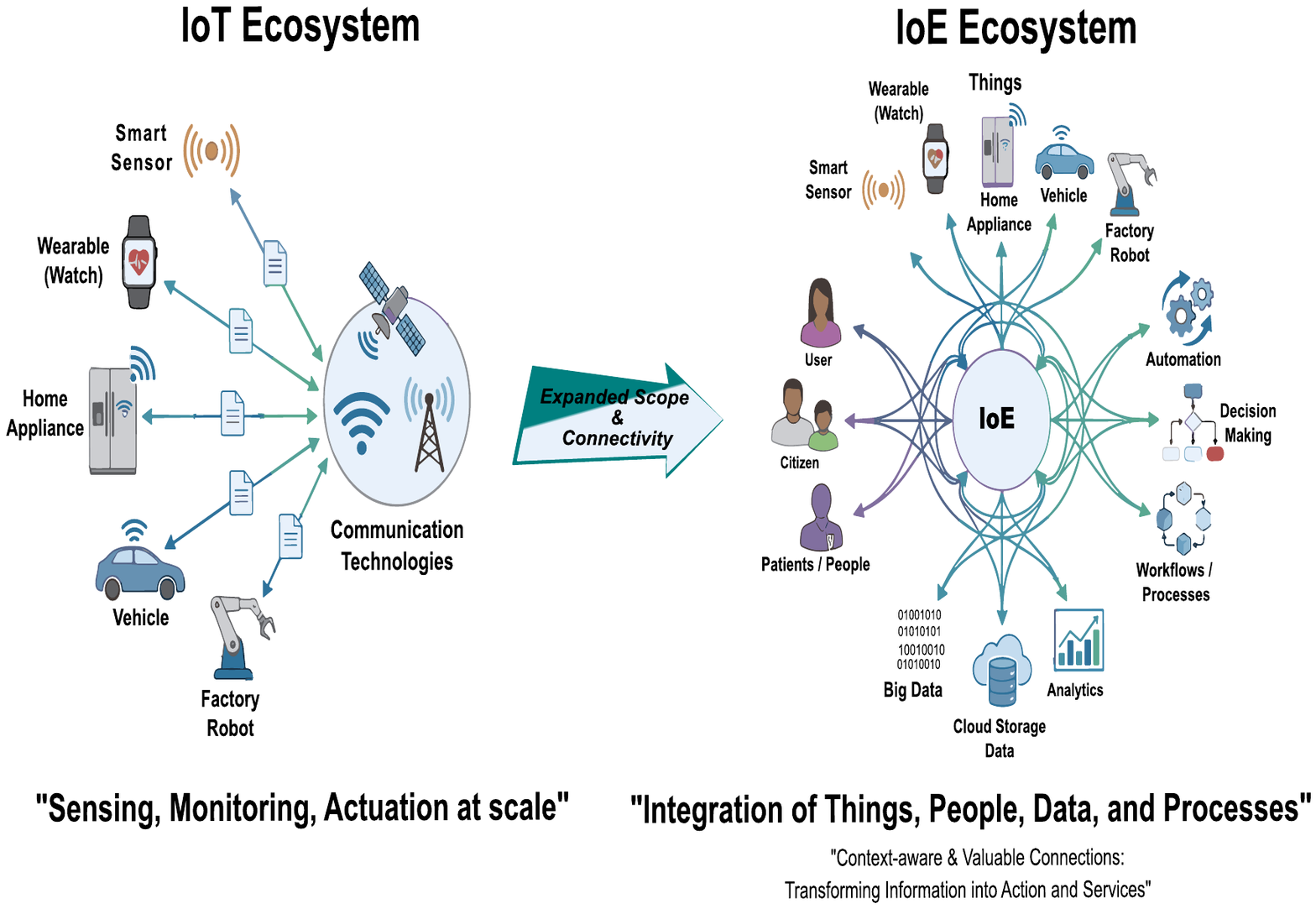}\vspace{-0.5cm}
  \caption{Evolution from the IoT to the IoE.}
  \label{fig:iot_to_ioe_evolution}
  \vspace{-0.5cm}
\end{figure*}

\noindent\textbf{How to read this paper: }
Section~\ref{sec:arch} defines IoE, its core components, and architectural foundations. Section~\ref{sec:enablers} reviews enabling technologies across compute, intelligence, data, connectivity, trust, and sustainability layers.
Section~\ref{sec:apps} summarizes key application domains and representative case studies. Section~\ref{sec:challenges} synthesizes cross-layer challenges (scalability, security/privacy, energy efficiency, and latency/reliability).
Finally, Section~\ref{sec:future} outlines forward-looking research directions toward 6G-enabled IoE. To position our paper relative to existing surveys, Table~\ref{tab:comparison-surveys-depth} summarizes representative high-impact works and compares their coverage depth across main IoE challenges. Fig.~\ref{fig:paper_organization_tree} provides a compact roadmap of the paper, mapping each main section to its key content for quick navigation.

\begin{table*}[!t]
\centering
\caption{Comparative Analysis of Scalable Solutions for IoE from Recent Surveys.}
\label{tab:comparison-surveys-depth}
\small
\setlength{\tabcolsep}{2pt}
\renewcommand{\arraystretch}{1.0}

\begin{tabular}{|C{1.0cm}|C{1.25cm}|C{1.0cm}|C{4cm}|C{4cm}|C{1.5cm}|C{1.5cm}|C{1.4cm}|C{1.4cm}|}
\hline
\rowcolor{Emerald!50}
\textbf{Ref.} & \textbf{Survey} & \textbf{Year} & \textbf{Main Topics} & \textbf{Challenges Addressed} &
\textbf{Network Scalability} & \textbf{Data Management} & \textbf{Energy Efficiency} &
\textbf{Security \& Privacy} \\
\hline

-- & \textbf{This survey} & 2026 &
IoE paradigms; architecture; enabling technologies; sustainability; IoE applications and verticals; 6G alignment &
Cross-layer scalability, data-to-intelligence pipeline, energy--security tradeoffs, 6G readiness, energy efficiency &
\cellD & \cellD & \cellD & \cellD \\ \hline

\cite{11403978} & Lin \emph{et al.} & 2026 &
IoT-enabled hierarchical energy trading; multi-community IoE coordination; dynamic pricing; optimization algorithms &
Inter-community coordination; large-scale energy scheduling; decentralized decision-making; economic scalability &
\cellD & \cellM & \cellD & \cellM \\ \hline

\cite{11381426} & Zhang \emph{et al.} & 2026 &
LLM-driven IoE resource allocation; semantic scheduling; edge intelligence; RL optimization &
Dynamic resource scheduling complexity; heterogeneous task requirements; latency/energy tradeoffs; scalable decision-making &
\cellD & \cellD & \cellD & \cellM \\ \hline

\cite{11037759} & Zheng \emph{et al.} & 2025 &
Embedded cognitive radio (ECR); NOMA-based IoE; spectrum allocation; massive access for 6G &
Spectrum scarcity; QoS guarantees; interference management; massive connectivity scalability &
\cellD & \cellM & \cellM & \cellB \\ \hline

\cite{10736549} & Khoshafa \emph{et al.} & 2025 &
RIS-assisted physical-layer security; RF and optical wireless; ML optimization; 6G integration &
Secure wireless propagation control; scalability with RIS deployment; optimization complexity; future 6G security challenges &
\cellM & \cellB & \cellM & \cellD \\ \hline

\cite{10963909} & Chamola \emph{et al.} & 2025 &
7G smart networks; AI-native connectivity; THz communication; quantum security; IoE integration &
Ultra-massive connectivity; AI-driven network automation; interoperability; energy sustainability; security in future networks &
\cellD & \cellM & \cellD & \cellM \\ \hline

\cite{10654347} & Koca \emph{et al.} & 2025 &
Bacterial communication; molecular communication; bio-nano IoE; integrated sensing--communication--computing &
Biological variability; nano-scale communication reliability; cross-field integration challenges &
\cellB & \cellM & \cellB & \cellB \\ \hline

\cite{ding2025_economic_ioe} & Ding \emph{et al.} & 2025 &
Economic analysis of IoE; modeling methods; business patterns &
Sustainability/scalability via incentives, pricing, learning-based economics &
\cellM & \cellM & \cellM & \cellM \\ \hline

\cite{10609788} & Tang \emph{et al.} & 2024 &
VLC for 6G IoE; channel modeling; RIS \& ISAC integration &
Channel modeling challenges; blockage/sunlight; RIS/ISAC issues &
\cellM & \cellB & \cellB & \cellM \\ \hline

\cite{9762857} & Adhikari \emph{et al.} & 2022 &
Cybertwin-driven 6G edge framework; DRL provisioning &
QoS provisioning; latency/energy tradeoffs; edge intelligence &
\cellD & \cellM & \cellD & \cellB \\ \hline

\cite{9863881} & Kong \emph{et al.} & 2022 &
Edge-driven IoT (end--edge--cloud) &
Offloading, caching, orchestration, scalability, security/privacy &
\cellD & \cellM & \cellM & \cellM \\ \hline

\cite{9509294} & Nguyen \emph{et al.} & 2022 &
6G-enabled IoT survey &
Massive connectivity; latency/reliability; edge intelligence; security &
\cellD & \cellM & \cellB & \cellM \\ \hline

\cite{s21020568} & Costa \emph{et al.} & 2021 &
IoE taxonomies; knowledge-based orchestration &
Knowledge creation; heterogeneous observations &
\cellB & \cellD & \cellNA & \cellB \\ \hline

\cite{9139976} & Qiu \emph{et al.} & 2020 &
Edge computing for IIoT &
Real-time constraints; reliability; scheduling; security &
\cellD & \cellM & \cellM & \cellD \\ \hline

\cite{9044329} & Wang \emph{et al.} & 2020 &
Edge computing + deep learning &
Resource constraints; partitioning; communication overhead; privacy &
\cellD & \cellM & \cellM & \cellM \\ \hline

\cite{9046806} & Habibi \emph{et al.} & 2020 &
Fog computing architectures &
Resource management; interoperability; latency; security aspects &
\cellD & \cellM & \cellM & \cellM \\ \hline

\cite{lim2020fl_men} & Lim \emph{et al.} & 2020 &
Federated learning in mobile edge networks &
Communication cost; scalability; privacy/security; optimization &
\cellD & \cellM & \cellB & \cellD \\ \hline






\end{tabular}

\vspace{2mm}
\footnotesize{\textbf{Legend:} \B = Brief; \M = Moderate; \D = Deep/Structured; -- = Not a primary focus.}

\end{table*}



\section{Paradigms, Architecture and Core Components}
\label{sec:arch}

\subsection{Core Components of IoE}
\label{subsec:Components}
\begin{figure}[!t]
  \centering
  \includegraphics[width=0.9\columnwidth]{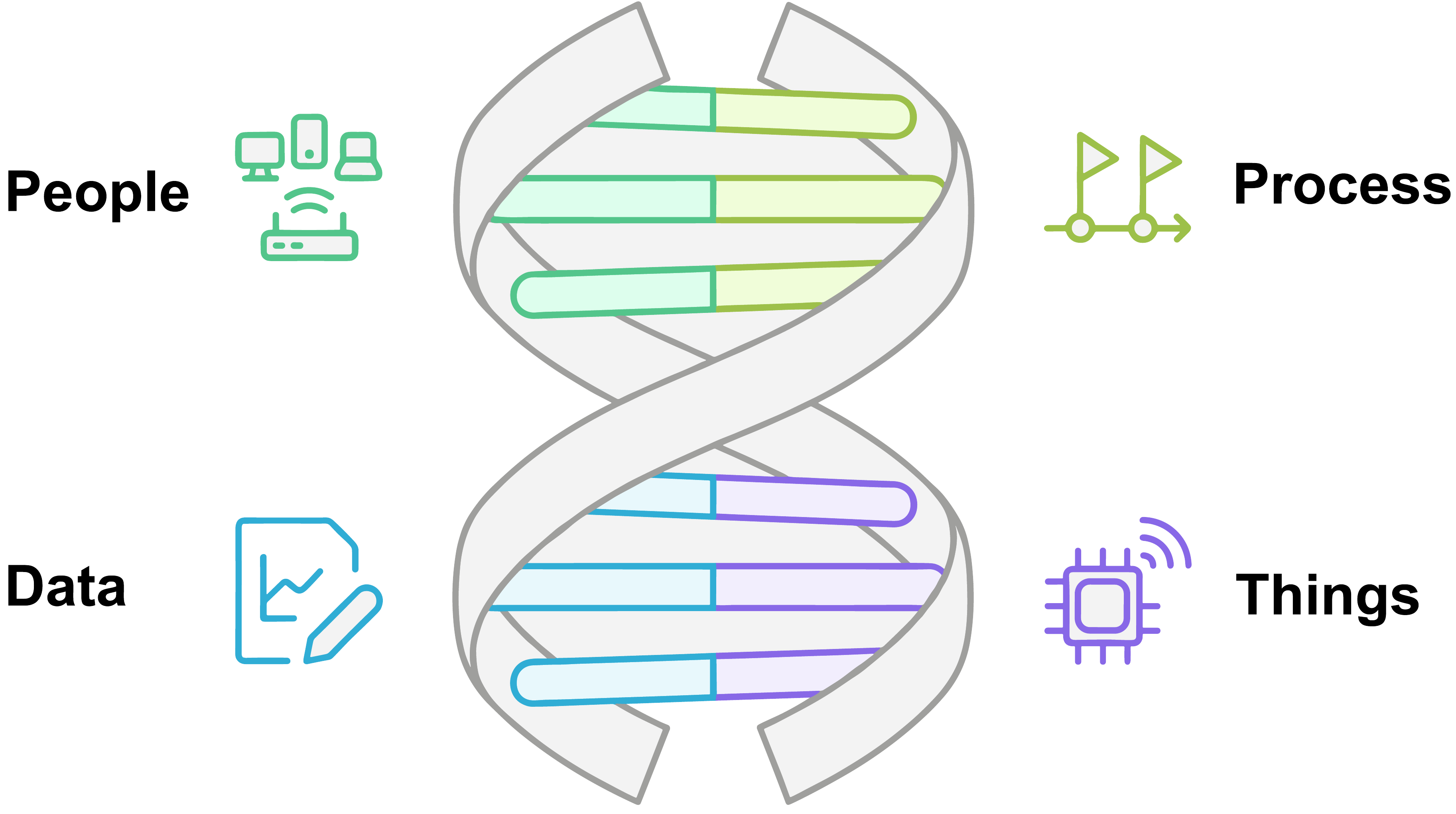}\vspace{-0.1cm}
  \caption{IoE's core components: things, people, data, and processes.}
  \label{fig:ioe_core_components}\vspace{-0.5cm}
\end{figure}
The IoT is commonly described as an ecosystem in which identifiable physical (and often virtual) objects are interconnected through communication technologies to enable large-scale sensing, monitoring, and actuation services \cite{ieee_iot_def2015,atzori2010iot,alfuqaha2015iot}. As shown in Fig.~\ref{fig:iot_to_ioe_evolution}, building on this foundation, Cisco introduced the IoE as a broader paradigm that extends beyond ``things'' by explicitly integrating \emph{people}, \emph{data}, and \emph{processes} to create more context-aware connections and transform information into actions and services \cite{cisco2012ioe}. From a research perspective, IoE is not simply a larger IoT; it requires structured views (e.g., taxonomies) to capture how enabling technologies, observation characteristics, and knowledge processes interact to produce intelligent services across heterogeneous applications \cite{s21020568}. In this work, we describe IoE through four tightly coupled components---\emph{things}, \emph{people}, \emph{data}, and \emph{processes}---where value emerges from their coordinated interaction rather than connectivity alone \cite{cisco2012ioe,s21020568}. To summarize the IoE concept, Fig.~\ref{fig:ioe_core_components} presents the four core components.\vspace{.2cm}

\indent\textbf{Things:}
These connected devices refer to physical and/or virtual objects equipped with sensors, actuators, and/or communication modules that enable data acquisition and interaction with their surrounding environment. In large-scale deployments, such devices are typically heterogeneous and resource-constrained, leading to several practical challenges, including energy limitations, connectivity variability, and increased device management complexity \cite{atzori2010iot,alfuqaha2015iot}.\vspace{.2cm}

\indent\textbf{People:}
They are active participants in IoE systems, contributing data and contextual feedback through smartphones, wearables, and human--machine interfaces. This human-in-the-loop dimension supports user-centric services and richer context awareness beyond conventional machine-to-machine interactions \cite{cisco2012ioe,s21020568}.\vspace{.2cm}

\indent\textbf{Data:}
Information is the central asset of IoE. It encompasses raw measurements collected from things and people as well as derived information obtained through analytics. Managing IoE data involves challenges in volume, heterogeneity, and privacy, and requires robust pipelines to convert raw streams into actionable knowledge for automation and decision support \cite{alfuqaha2015iot,s21020568}.\vspace{.2cm}

\indent\textbf{Processes:}
They define how people, data, and things are orchestrated to deliver end-to-end services (e.g., automation, optimization, and control). In IoE, processes emphasize intelligent coordination across components---including knowledge creation and service composition---to ensure that connectivity translates into measurable value and reliable outcomes \cite{cisco2012ioe,s21020568}.

\subsection{IoE as cross-domain synergy among Internet-of-X}
\label{subsec:IoXs}

Beyond extending IoT by incorporating people, data, and processes, an emerging perspective views future cyber--physical ecosystems as a constellation of
specialized IoX verticals, including the Internet of Vehicles (IoV), Internet of Energy (IoEn), and even nanoscale communication frameworks such as the Internet of Nano Things (IoNT). These domain-specific networks address the requirements of particular application environments such as intelligent transportation systems, smart energy infrastructures, and nano–bio communication systems
\cite{ji2020_iov_survey,rana2025_iont_survey}. Within this context, the IoE can be interpreted as a holistic architecture that enables seamless \emph{interaction, interoperability, and cooperation among heterogeneous IoXs}, thereby enabling cross-domain
services and applications that transcend the capabilities of isolated
vertical ecosystems \cite{akan2023_ioe_molecules_universe,alfuqaha2015iot}.
\begin{figure*}[!t]
\centering
\begin{tikzpicture}[
  font=\small,
  box/.style={draw, rounded corners, align=center, minimum width=4.4cm, minimum height=0.9cm},
  sbox/.style={draw, rounded corners, align=center, minimum width=4.4cm, minimum height=0.75cm},
  nodebox/.style={draw, rounded corners, align=center, minimum width=4cm, minimum height=0.85cm},
  arrow/.style={-Latex, thick}
]

\node[box, fill=Aquamarine!8] (app) {\textbf{Application layer}\\(domain services)};
\node[box, fill=Emerald!50, below=0.5cm of app] (mid) {\textbf{Middleware layer}\\(interoperability, management)};
\node[box, fill=Aquamarine!8, below=0.5cm of mid] (net) {\textbf{Network layer}\\(connectivity)};
\node[box, fill=Emerald!50, below=0.5cm of net] (per) {\textbf{Perception layer}\\(sensing/actuation)};

\draw[arrow] (per.north) -- (net.south);
\draw[arrow] (net.north) -- (mid.south);
\draw[arrow] (mid.north) -- (app.south);

\node[align=center, below=0.3cm of per] {\textbf{(a) Layered Architectural Model}};

\node[nodebox, fill=Aquamarine!8, right=1.5cm of per] (dev) {\textbf{Devices}\\(sensors, actuators)};
\node[nodebox, fill=Emerald!50, above=0.5cm of dev] (edge) {\textbf{Edge}\\(near-source compute)};
\node[nodebox, fill=Aquamarine!8, above=0.5cm of edge] (fog) {\textbf{Fog}\\(multi-tier nodes)};
\node[nodebox, fill=Emerald!50, above=0.5cm of fog] (cloud) {\textbf{Cloud}\\(centralized services)};

\draw[arrow] (dev) -- (edge);
\draw[arrow] (edge) -- (fog);
\draw[arrow] (fog) -- (cloud);

\node[sbox, fill=BlueGreen!20, right=0.25cm of edge, xshift=1.8cm] (lat) {\textbf{Lower latency \& backhaul reduction}\\(process close to data sources)};
\node[circle, fill=black, inner sep=1pt, above=0.25cm of lat] (la) {};
\node[sbox, fill=BlueGreen!20, above=0.25cm of la] (scale) {\textbf{Scalability \& distributed intelligence}\\(support 6G services)};
\draw[arrow] (scale) -- (lat);
\draw[arrow] (lat) -- (scale);
\draw[arrow] (edge.9) -- (la.west);
\draw[arrow] (fog.351) -- (la.west);

\node[align=center, below=0.3cm of dev] {\textbf{(b) Device--Edge/Fog--Cloud Continuum}};

\end{tikzpicture}
\caption{IoE system architecture: (a) layered functional abstraction and (b) device--edge/fog--cloud continuum for low-latency and scalable intelligence.}
\label{fig:ioe_architecture_views_1}\vspace{-0.4cm}
\end{figure*}
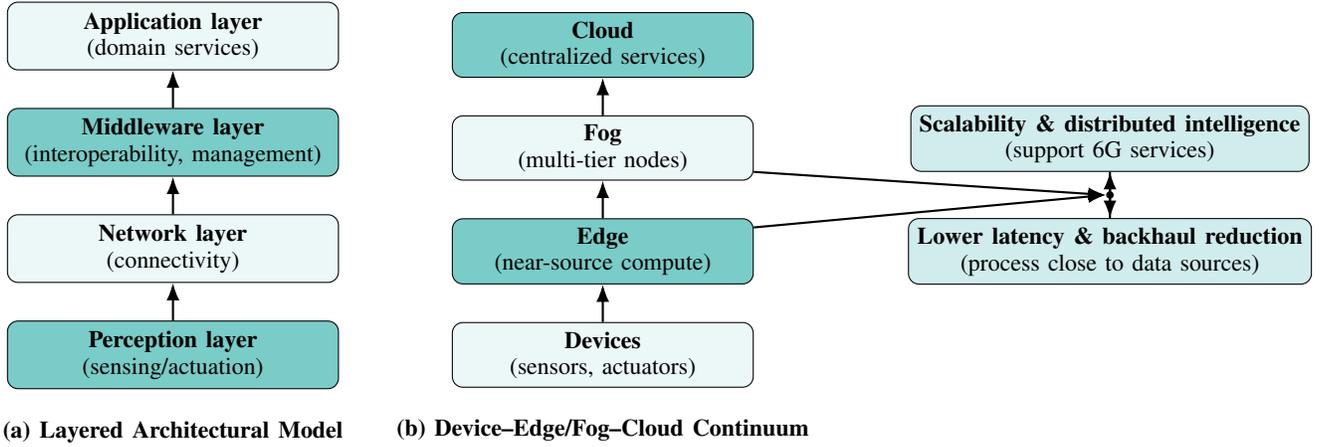

\subsection{Bio-nano IoE and MC}
\label{subsec:Bio-nano}

Beyond conventional cyber--physical connectivity, the IoE paradigm is increasingly extended toward bio--nano networking environments. In this context, MC has emerged as a fundamental nanoscale communication paradigm in which
information is encoded, transmitted, and received through chemical signals exchanged between biological or synthetic nanomachines. MC forms the communication backbone of the Internet of Bio-Nano Things (IoBNT), enabling nanoscale entities to sense, process, and exchange information within biological environments such as the human body
\cite{akyildiz2015iobnt,farsad2016molecular,10654347}. Integrating such bio-nano networks with conventional IoT infrastructures opens new possibilities for healthcare, environmental monitoring, and targeted drug delivery applications, thereby extending IoT ecosystems toward biologically embedded communication systems
\cite{akyildiz2019molecularhealth}. In this perspective, the evolution from IoT to IoBNT further reinforces the broader IoE vision of seamless interconnectedness across heterogeneous digital, physical, and biological domains \cite{10654347}. Recent studies also explore the integration of bio-nano networks with emerging IoE ecosystems and intelligent communication frameworks. For instance, bacterial communication mechanisms and bio-inspired signaling models have been proposed as computational and networking
primitives in IoE-enabled biological systems
\cite{koca2025bacterial_ioe}. Meanwhile, advances in molecular
communication-based sensing and diagnostics demonstrate the potential of IoBNT for healthcare applications such as breath-based disease detection and biomedical monitoring \cite{bhattacharjee2026breath_mc}. AI-driven semantic communication frameworks are also being developed to enhance information efficiency and diagnostic accuracy in IoBNT systems \cite{cai2025semantic_mc}.

\subsection{IoE System Architecture}
\label{subsec:Architecture}

Beyond defining its core components, IoE requires architectural models that support scalability, modularity, and low-latency intelligence across heterogeneous environments. As summarized in Fig.~\ref{fig:ioe_architecture_views_1}, two complementary views are commonly adopted: (i) a layered abstraction that structures system functions, and (ii) a device--edge/fog--cloud continuum that distributes computation and storage close to data sources.

\vspace{0.2cm}

\subsubsection{Architectural Views} To capture the complexity and heterogeneity of IoE systems, two complementary architectural views are commonly adopted. 
The first view relies on a layered abstraction that organizes system functionalities into modular components, while the second view considers a distributed computing continuum spanning devices, edge/fog nodes, and cloud infrastructures. 
Together, these views provide a unified framework for analyzing how sensing, communication, and intelligence are structured and deployed across large-scale IoE environments.

\vspace{0.2cm}

\paragraph{Layered Architectural Model}
IoE systems are often described by building upon the classical layered architectural model originally developed for IoT systems, which separates sensing/actuation, networking, data management, and service logic to improve modularity and scalability \cite{alfuqaha2015iot,atzori2010iot}. 
In this baseline view, the \emph{perception layer} comprises sensors and actuators that generate and consume data, the \emph{network layer} provides connectivity across heterogeneous access technologies, the \emph{middleware layer} supports interoperability, device management, and data handling, and the \emph{application layer} delivers domain-specific services (e.g., smart healthcare, industrial automation, and transportation) \cite{alfuqaha2015iot}. 
However, this layered abstraction primarily reflects an IoT-centric perspective and does not fully capture the broader scope of IoE systems. In particular, IoE extends beyond device-centric interactions to incorporate people, processes, and data-driven intelligence, requiring tighter integration across layers and support for distributed intelligence, real-time analytics, and cross-domain orchestration. 
As a result, practical IoE deployments must address additional challenges such as latency, bandwidth, and reliability constraints under massive-scale data generation, while also enabling intelligent, adaptive, and autonomous system behavior \cite{atzori2010iot}.
\vspace{0.2cm}

\begin{figure*}[t]
  \centering
  \includegraphics[width=\textwidth]{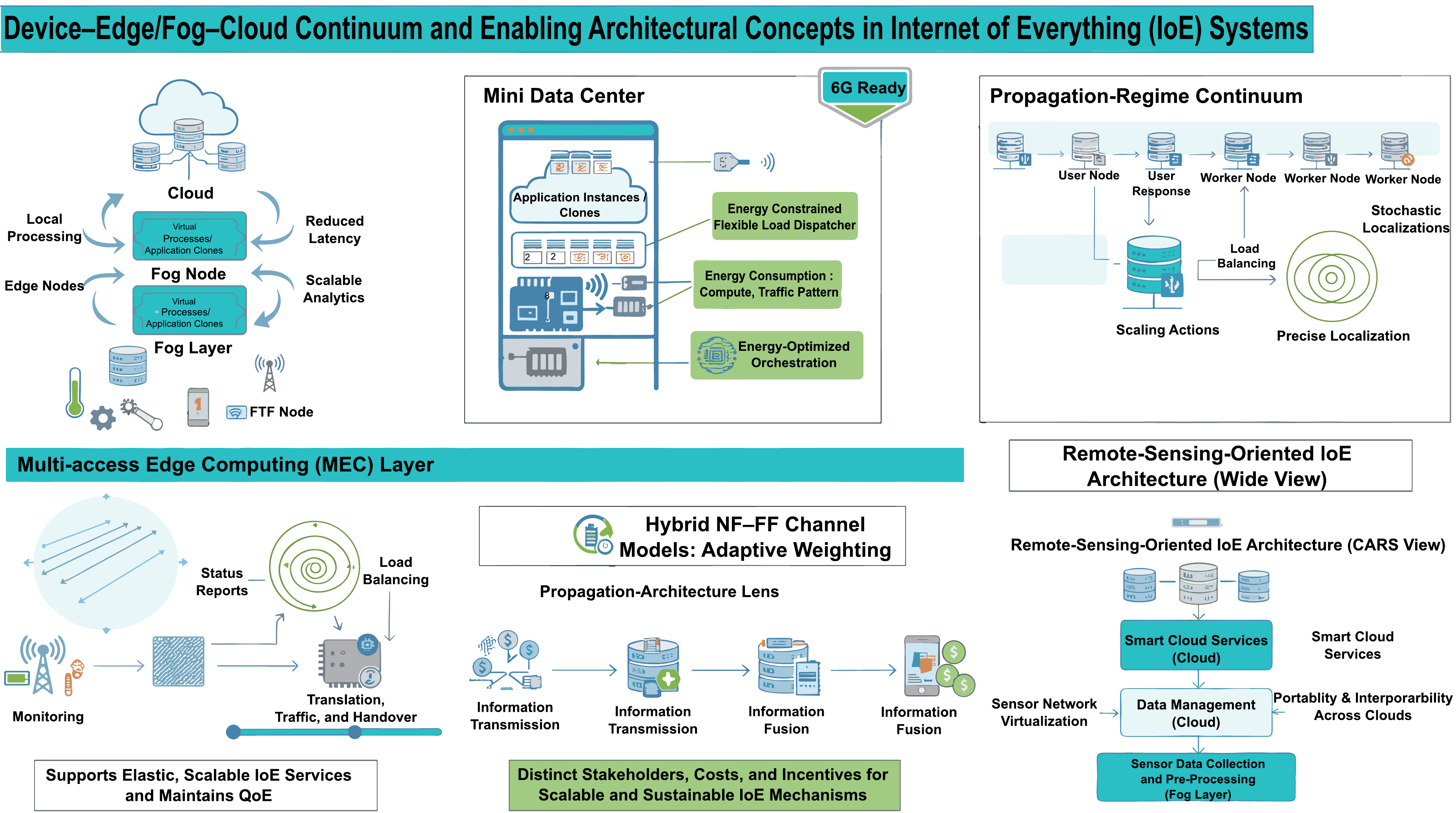}
  \caption{Device--edge/fog--cloud continuum as an enabling architectural concept for scalable and low-latency IoE services.}
  \label{fig:ioe_architecture_views}\vspace{-0.5cm}
\end{figure*}

\paragraph{Device--Edge/Fog--Cloud Continuum}

To satisfy the stringent latency, scalability, and energy requirements
of large-scale IoE deployments, recent architectures increasingly rely
on a \emph{device--edge/fog--cloud continuum} in which sensing,
communication, and computing resources are distributed across end
devices, edge nodes located close to data sources, intermediate fog
layers, and centralized cloud infrastructures (see
Fig.~\ref{fig:ioe_architecture_views}) \cite{yu2017edge_iot,9046806}.
By pushing computation closer to the network edge, such architectures
can significantly reduce backhaul traffic, improve responsiveness, and
enable real-time analytics for mission-critical applications
\cite{yu2017edge_iot,premsankar2018edge_case}.  This distributed continuum aligns with emerging 6G visions that promote native support for distributed intelligence and ultra-low latency services across heterogeneous cyber--physical environments \cite{saad2020vision6g,wang2023road6g}. Therefore, multiple complementary design perspectives emerge, including the virtualized fog computing layer that operates as a distributed mini data center close to data sources, as well as energy-aware wireless frameworks that enable sustainable IoE sensing and data transmission. These perspectives together illustrate how the device--edge/fog--cloud continuum supports scalable, responsive, and energy-efficient operation of future IoE ecosystems.\vspace{0.2cm}

\subsubsection{Key Architectural Enablers and Extensions}
Beyond high-level architectural views, practical IoE systems rely on a set of key enabling mechanisms that support the deployment of scalable, low-latency, and energy-efficient services across heterogeneous environments. 
These enablers complement the layered and continuum-based abstractions by providing concrete implementations for distributed computation, energy management, and service orchestration. 
In particular, technologies such as fog computing, wireless-powered communication, and edge orchestration frameworks play a central role in enabling adaptive and efficient operation of large-scale IoE ecosystems.

\vspace{0.2cm}

\paragraph{Fog computing architecture}
A practical fog layer can be modeled as a set of \emph{virtualized} fog nodes operating as distributed mini--data centers located close to data sources. Each node typically integrates local storage and buffering capabilities, physical compute and networking resources,
and a virtualization layer that hosts multiple application instances (e.g., containers or virtual Machines (VMs)) deployed as lightweight processing "clones" \cite{bonomi2012fog,shi2016edge}. Such architectures enable latency-sensitive services by offloading computation from centralized clouds to nearby edge infrastructures
while maintaining scalable resource management across distributed systems. In practice, fog nodes employ adaptive workload dispatchers that dynamically route incoming tasks toward available clones according to QoS, latency, and energy objectives \cite{mao2017survey_mec,zhou2023edge_survey}. This system-level view is particularly useful in IoE environments because it makes explicit where energy is consumed—across virtualization layers, compute resources, networking interfaces, and traffic handling mechanisms—and clarifies the control points for energy-aware orchestration across the device–edge–cloud continuum \cite{foe2017_fog_of_everything,liu2026cloudedge}.

\vspace{0.2cm}

\paragraph{Wireless-powered IoE}
A practical energy-sustainable sensing IoE transmission architecture can be organized around wireless powered communication network  principles, where wireless devices harvest energy from Radio-Frequency (RF) signals broadcast by a hybrid Access Point (AP) and later use the harvested energy for uplink data transmission
\cite{bi2015wirelesspowered,ju2014throughput}. In such systems, the communication frame is typically structured within Time Division Duplex coherence intervals that allocate (i) uplink pilot signaling for Channel State Information (CSI) estimation, (ii) downlink Wireless Energy Transfer (WET) for energy harvesting, and (iii) uplink payload transmission powered by the harvested energy. Above-mentioned harvest--then--transmit structure introduces a fundamental coupling between CSI acquisition, energy transfer efficiency, and uplink throughput. Accurate CSI enables energy beamforming and efficient resource allocation, while the harvested energy directly determines the achievable transmission power and system lifetime. Consequently, the joint optimization of channel estimation, WET, and data transmission becomes a central challenge for large-scale IoE deployments, particularly under massive sporadic access and energy-constrained device operation
\cite{chen2020wirelesspowered,lu2015rfharvesting,rosabal2023massivewet}.\vspace{0.2cm}

\paragraph{Distributed orchestration and control}

The device--edge/fog--cloud continuum introduces a highly distributed computing environment in which services and workloads must be dynamically coordinated across heterogeneous infrastructures. As computation moves closer to data sources through edge and fog paradigms, the management
of distributed resources becomes increasingly complex, requiring orchestration mechanisms capable of automating service deployment, scaling, and resource allocation while maintaining strictquality-of-service guarantees \cite{yu2017edge_iot,premsankar2018edge_case}.
These orchestration mechanisms are particularly important in large-scale IoE deployments where massive numbers of devices generate highly dynamic workloads across geographically distributed infrastructures. In this context, distributed control frameworks and edge orchestration platforms are emerging as key enablers for scalable IoE operation, supporting automated service placement, elastic resource provisioning, and coordinated management across the device--edge--cloud continuum. Such capabilities align with the broader 6G vision of native support for distributed intelligence and ultra-low-latency services across
large-scale cyber--physical ecosystems \cite{saad2020vision6g}.\vspace{0.2cm}

\paragraph{MEC orchestration plane} At scale, the device--edge/fog--cloud continuum requires a dedicated orchestration plane capable of elastically managing distributed IoE services across heterogeneous infrastructures. In practice,
Multi-access Edge Computing (MEC) nodes operate as containerized execution environments where application components are deployed as lightweight microservices or containers orchestrated across edge clusters and cloud backends \cite{taleb2017mec_orchestration,ullaha2023cloudtothings}. A practical architecture consists of MEC worker nodes hosting containerized services (e.g., pods or microservice instances), which periodically report resource utilization and service status to a central or hierarchical controller responsible for scheduling, load balancing, and service placement decisions. Container orchestration frameworks—such as Kubernetes and its lightweight edge variants—enable automated deployment, monitoring, and dynamic scaling of services across distributed edge infrastructures
\cite{edge_orchestration_survey_2023,liu2024kubernetes_qos}. Such orchestration mechanisms are essential for supporting IoE elasticity under time-varying workloads and massive device access. By dynamically allocating compute, networking, and storage resources across the edge–cloud continuum, orchestration systems help maintain Quality of Experience (QoE) and service reliability while preventing
resource bottlenecks and service downtime in large-scale IoE
deployments \cite{sami2021_ai_resource_provisioning_ioe_6g}.

\vspace{0.2cm}

\subsubsection{System-Level and Cross-Domain Considerations}
Beyond architectural views and enabling technologies, IoE systems must also be analyzed from a broader system-level perspective that captures the impact of wireless environments, economic factors, and application-driven requirements. 
These cross-domain considerations play a critical role in shaping the performance, scalability, and sustainability of IoE deployments, as they influence how resources are allocated, how data is transmitted and processed, and how value is created across the IoE ecosystem.
\vspace{.2cm}

\paragraph{Wireless environment considerations}
Beyond computational and architectural aspects, IoE systems must also account for the characteristics of the underlying wireless propagation environment. As emerging 6G technologies adopt higher carrier frequencies, extremely large antenna arrays, and Reconfigurable Intelligent Surfaces (RIS), the classical assumptions of wireless propagation are evolving \cite{saad2020vision6g,wang2023road6g}. These developments introduce new propagation regimes and signal behaviors that significantly influence connectivity, localization accuracy, and energy focusing capabilities in large-scale IoE
deployments. Understanding such wireless propagation characteristics is therefore essential for designing efficient communication and sensing mechanisms across next-generation IoE infrastructures.\vspace{0.2cm}

\paragraph{Propagation regime continuum}
In addition to distributing computation along the device--edge/fog--cloud
continuum, future IoE deployments must also cope with
\emph{propagation-regime transitions} in the wireless environment.
With the adoption of higher carrier frequencies and the increasing
aperture of extremely large antenna arrays and the deployment of RIS, wireless links may operate beyond the
traditional Far-Field (FF) planar-wave assumption and instead exhibit Near-Field (NF) spherical-wave propagation characteristics \cite{bjornson2020ris_smart_radio_environment,wu2025_ris_nf_ff_ioe_healthcare}. This transition between NF and FF regimes is particularly relevant for emerging 6G and IoE systems employing extremely large aperture arrays and holographic MIMO surfaces, where the classical boundary between
propagation regions becomes increasingly dynamic
\cite{cui2023nearfield_survey}. From a system perspective, NF operation can enable very high-precision localization and sensing due to the spatial focusing properties of spherical wavefronts, while FF operation remains suitable for wide-area coverage and scalable connectivity \cite{bjornson2021holographic_mimo}. In \cite{wu2025_ris_nf_ff_ioe_healthcare}, it was shown that RIS-assisted beam focusing in the radiating NF can dynamically concentrate electromagnetic energy at specific spatial
locations, enabling improved signal quality, localization accuracy, and spatial multiplexing capabilities in next-generation wireless networks.

\vspace{.2cm}
\paragraph{Economic considerations and application perspectives}

Alongside architectural and networking considerations, IoE systems can also be analyzed from economic and application-oriented viewpoints. Such perspectives help explain how data flows across the IoE value chain, how incentives and resource pricing influence system operation, and how specific application domains leverage the device--edge/fog--cloud continuum to deliver large-scale intelligent services \cite{yu2017edge_iot,ullaha2023cloudtothings}. Here, economic models and application-driven architectures provide important insights into how IoE ecosystems create value, coordinate stakeholders, and support domain-specific deployments. These perspectives are particularly relevant for large-scale sensing applications—such as environmental monitoring, smart cities, and healthcare systems—where distributed sensing infrastructures and edge-enabled data processing pipelines must operate efficiently across multiple IoE layers.

\noindent Complementing functional layering and the device--edge/fog--cloud continuum,
an economic perspective can further structure IoE systems into four
interdependent aspects: \emph{information collection},
\emph{information transmission}, \emph{information fusion}, and
\emph{service application} \cite{ding2025_economic_ioe}. This view
reflects the lifecycle of information in large-scale IoE ecosystems,
from sensing and data acquisition to communication, processing,
and value-added service delivery. Such an economic-architecture lens is particularly useful because each stage introduces distinct stakeholders, operational costs, and incentive structures across the IoE value chain. For instance, sensing layers may require mechanisms to incentivize data contributors, communication layers must manage spectrum and network resource pricing, and data fusion stages introduce tradeoffs between computation cost and information quality.
Finally, service layers determine the monetization of processed data and
the distribution of value among ecosystem participants
\cite{ding2025_economic_ioe,zhang2020incentive_iot,li2018economics_edge}. Nguyen \emph{et al.} \cite{nguyen2021edge_pricing} highlight the role of game-theoretic pricing and incentive mechanisms in coordinating resource allocation across
edge-enabled IoT ecosystems. In particular, Stackelberg and auction-based
models have been proposed to regulate interactions among IoT devices,
edge service providers, and network operators, enabling efficient
pricing of communication and computing resources while maintaining
system-wide fairness and efficiency.\vspace{0.2cm}

\paragraph{Remote sensing IoE}
In addition to generic layered abstractions and the device--edge/fog--cloud continuum, remote-sensing-oriented IoE systems can be interpreted through cloud-based sensing service architectures where sensing data are collected and preprocessed close to sensors (e.g., at fog or edge nodes) and then
progressively aggregated, managed, and analyzed across higher cloud layers to deliver intelligent cloud services \cite{abdelwahab2016cars}. This hierarchical architecture is particularly suitable for large-scale remote sensing applications, where massive sensor networks continuously
generate heterogeneous environmental or situational data streams that must be processed across distributed edge and cloud infrastructures \cite{shi2016edge,ullaha2023cloudtothings}. Recent IoT-enabled
remote sensing architectures emphasize collaborative processing across edge, fog, and cloud layers to support real-time analytics, context-aware decision making, and scalable sensing services in domains such as environmental monitoring, smart cities, and healthcare sensing ecosystems \cite{ortiz2024collaborative_iot_architecture}. 

\vspace{.2cm}
\subsubsection{Middleware and Interoperability}
Alongside architectural structures, enabling mechanisms, and system-level considerations, IoE systems require dedicated middleware frameworks to ensure seamless interoperability across heterogeneous domains. 
As IoE integrates multiple IoXs operating at different scales and with diverse data models, interoperability becomes a key challenge that cannot be addressed solely at the device or network level. 
Middleware solutions therefore play a central role in enabling cross-domain integration, service composition, and semantic consistency across large-scale IoE ecosystems.
While IoT middleware typically mediates device-level heterogeneity, IoE compounds heterogeneity by integrating multiple IoXs spanning diverse scales (from biotic/nanoscale entities to satellites). To address this, a layered IoE architecture has been proposed that introduces an additional \emph{IoE middleware layer} above the individual IoX stacks. In this service-oriented design, each IoX is abstracted as a reusable service (\emph{IoX-as-a-Service}), and the IoE middleware performs service discovery, service call/integration, and manages inter-IoX interactions in terms of \emph{data, events, and resources}. Crucially, the IoE middleware targets cross-domain interoperability through alignment/translation of \emph{IoX-specific semantic ontologies}, while leaving syntactic/device-level issues to the corresponding IoX middleware \cite{akan2023_ioe_molecules_universe}.


\section{Enabling Technologies}
\label{sec:enablers}

To support \emph{massive-scale} IoE deployments, enabling technologies must jointly address
(i) \emph{network scalability}, (ii) \emph{data management and analytics}, (iii) \emph{energy efficiency},
and (iv) \emph{security and privacy}. To keep the discussion structured, we group IoE enablers into the
following taxonomy (mapped to the corresponding subsections below):

\begin{itemize}
    \item[--] \textbf{Intelligence layer:} AI/ML for IoE, edge intelligence, and learning-based control (e.g., DRL orchestration), supporting autonomous optimization and adaptive service management in large-scale IoE ecosystems \cite{li2018learning_edge,park2022edge_ai_survey}.
    
    \item[--] \textbf{Compute continuum:} device--edge/fog--cloud computing, MEC orchestration, service placement, and lifecycle management, enabling delay-sensitive and distributed IoE services across heterogeneous infrastructures \cite{shi2016edge,mao2017survey_mec}.

    \item[--] \textbf{Data layer:} big data analytics, storage architectures, streaming pipelines, and semantic data processing for IoE knowledge extraction \cite{gubbi2013iot_vision,rehman2019bigdata_iot}.

    \item[--] \textbf{Connectivity layer:} 5G/6G and beyond, including ultra-dense networks, cell-free massive MIMO, RIS, NTN/LEO, and THz and Integrated Sensing and Communication (ISAC) technologies to meet extreme IoE requirements \cite{akyildiz2020sixg,saad2020vision6g}.

    \item[--] \textbf{Sustainability layer:} energy harvesting, green networking, carbon-aware orchestration, and wireless-powered IoT paradigms for self-sustainable IoE operation \cite{ul2019energyharvesting_iot,bianzino2012green_networking_survey}.
    
  \item[--] \textbf{Trust layer:} SPbD/PSbD, zero-trust architectures, blockchain/DLT, and trusted execution for resilient IoE ecosystems \cite{roman2018iot_security_survey,kim2021zerotrust_iot}.

\end{itemize}

\begin{figure*}[t]
  \centering
  \includegraphics[height=10cm,width=2\columnwidth]{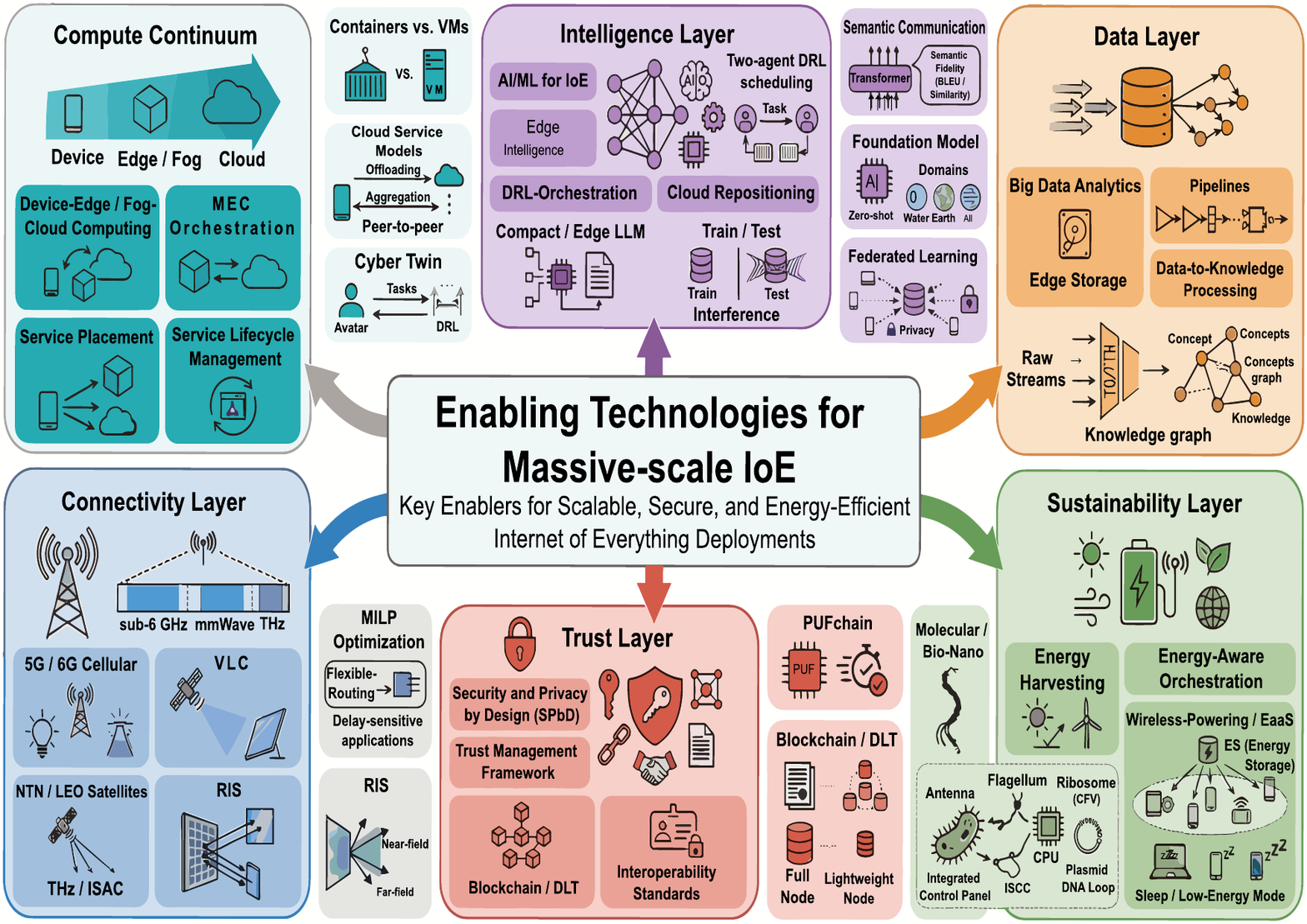}
  \caption{Main IoE enabling technologies.}
  \label{fig:ioe_enablers}\vspace{-0.3cm}
\end{figure*}

We next review these enabler classes and highlight how they jointly support scalable, secure, and energy-efficient IoE.
Fig.~\ref{fig:ioe_enablers} visually summarizes these enabler layers and their coupling: the connectivity substrate enables data exchange,
the compute continuum hosts services and learning, the intelligence layer drives automation, and the trust/sustainability layers constrain design choices through security/privacy and energy limitations.

\subsection{AI/ML and Edge Intelligence}
\label{subsec:AI_ML}

AI and Machine Learning (ML) are key enablers for IoE by transforming raw sensing streams into actionable intelligence
(e.g., prediction, anomaly detection, and decision-making) \cite{alfuqaha2015iot}. Because many IoE applications
require low latency and operate under dynamic wireless conditions, recent research emphasizes \emph{edge intelligence},
where learning and inference are pushed toward the network edge to support real-time adaptation and privacy-aware
analytics \cite{zhu2020intelligent_edge,saad2020vision6g}. In particular, learning-driven communication and edge learning
principles highlight how joint design of wireless resources and ML pipelines can reduce latency and improve scalability
for distributed inference and training at the edge \cite{zhu2020intelligent_edge}. This trend is particularly relevant in
6G contexts, where distributed intelligence is expected to be tightly integrated with the communication fabric \cite{wang2023road6g}.

\vspace{.2cm}
\subsubsection{Edge intelligence for IoE communications and security analytics}
Edge intelligence is emerging as a key enabler for scalable IoE
services, allowing learning-driven communication, security analytics,
and decision making to be executed close to data sources. By combining
edge AI, compact deep-learning models, and communication-efficient
learning frameworks, future IoE systems can support real-time analytics,
robust communication, and low-latency security monitoring in dense,
resource-constrained environments.
In particular, recent advances in compact and distilled models enable
the deployment of efficient learning architectures at edge nodes for
tasks such as anomaly detection, intrusion identification, and automated
threat analysis. This capability allows IoE systems to detect and respond
to security threats in near real time, while leveraging more powerful
cloud-based intelligence for global reasoning and coordinated response.\vspace{.2cm}

\noindent\textbf{Distilled/compact large language models for edge security analytics: }
A practical trend in AI-driven IoE security analytics is the deployment of distilled or compressed language models at the network edge. Knowledge Distillation (KD) techniques enable large language models (LLMs) to transfer their knowledge into compact architectures (e.g., Mini Language Model), while model compression methods such as quantization and pruning significantly reduce inference latency, memory footprint, and energy consumption \cite{wang2020minilm,han2015deep_compression}. These techniques make it feasible to deploy efficient language models directly on edge nodes and MEC infrastructures. In such architectures, compact LLMs running at the edge can perform near-real-time security analytics, including log analysis, anomaly detection, and automated threat triage, while larger cloud-based LLMs provide higher-level reasoning capabilities such as rule generation, threat intelligence synthesis, and automated response orchestration \cite{llm_ioe_security_2026}. This hybrid edge–cloud AI architecture supports scalable security monitoring in large-scale IoE deployments by combining low-latency local inference with powerful centralized
model capabilities.\vspace{.2cm}

\noindent\textbf{KD-enhanced semantic communications under multi-user interference: }
Beyond conventional bit-level transmission, \emph{semantic communications}
(SemCom) aim to convey task-relevant meaning rather than exact bits,
which is particularly 
\begin{figure*}[t]
  \centering
  \includegraphics[width=\textwidth, height=7.5cm]{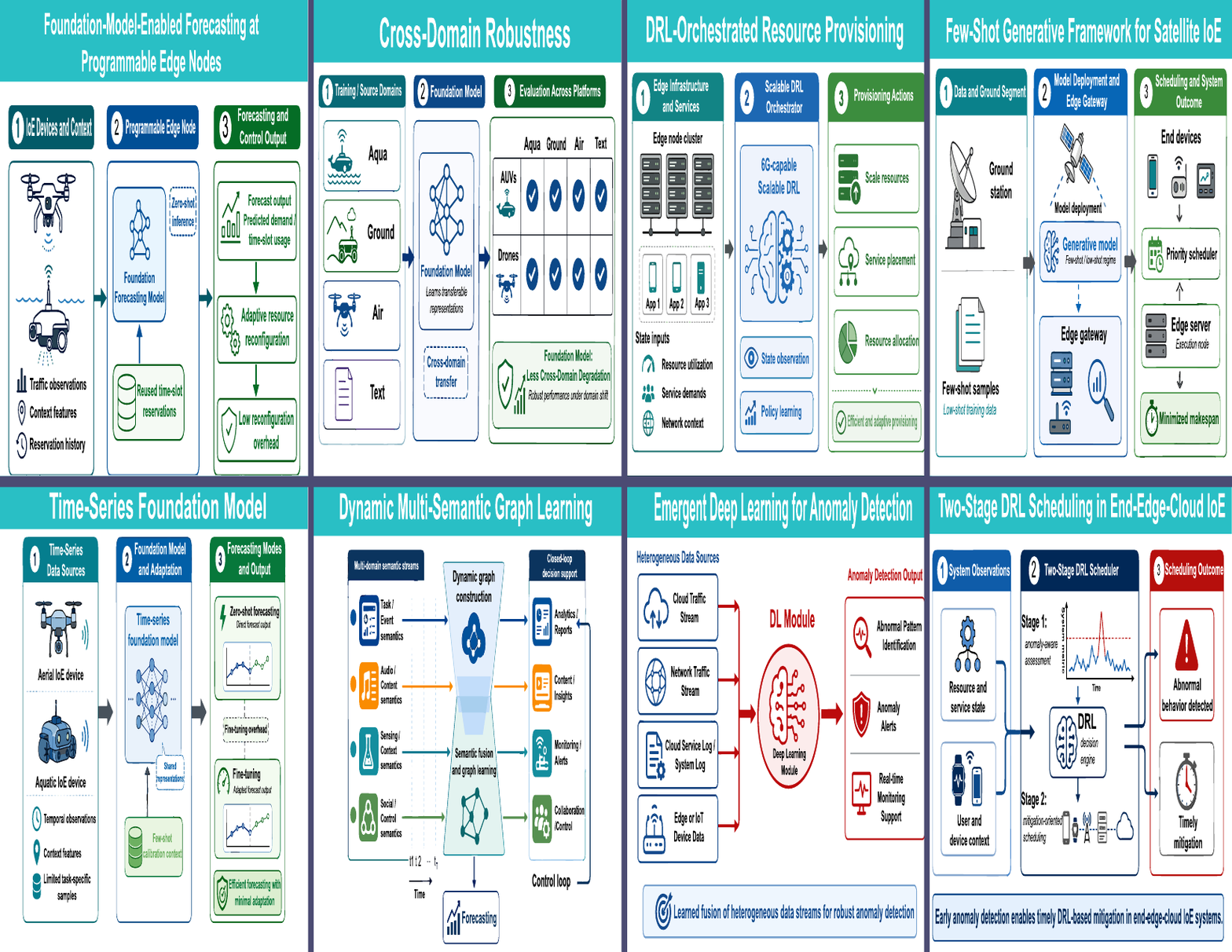}
  \vspace{-0.1cm}
  \caption{Foundation models show competitive accuracy, reliable uncertainty estimation, and low inference overhead-supporting dependable edge reconfiguration.}
  \label{fig:ai_orchestration_map}\vspace{-0.3cm}
\end{figure*}
attractive for massive IoE environments where bandwidth and energy resources are limited and transmitted content often exhibits redundancy \cite{strinati2022semantic_comms}. By focusing on semantic relevance, SemCom systems can reduce communication overhead while preserving the effectiveness of downstream IoE tasks such as monitoring and control. However, dense IoE deployments introduce practical \emph{multi-user interference}, which can degrade the performance of learned SemCom models. Recent work on task-oriented SemCom shows that prioritizing semantically relevant features enables efficient multi-user transmission while reducing communication overhead \cite{kang2023_task_oriented_semcom_jsac}.
To address this challenge, KD-based architectures have been proposed for multi-user environments, where a high-capacity teacher model transfers knowledge to lightweight student models deployed at edge devices \cite{10345474}. In such designs, a Transformer-based semantic
encoder--decoder is combined with lightweight channel modules, and the
student model is trained using a joint objective integrating cross-entropy loss with a temperature-scaled teacher--student Kullback--Leibler divergence. An important methodological aspect is the adoption of a \emph{train-without-interference / test-with-interference} evaluation protocol combined with meaning-oriented semantic metrics. This setup better reflects the generalization requirements of dense IoE deployments, where models must remain robust to previously unseen interference conditions while preserving semantic fidelity rather than focusing solely on bit-level accuracy \cite{10345474}. Performance is evaluated using semantic fidelity metrics such as BLEU and sentence-level similarity rather than conventional bit-error rate, highlighting that communication quality should be assessed in terms of meaning preservation in IoE services \cite{xie2021_semantic_comms}.
Experimental results further demonstrate improved robustness under \emph{unseen interference}, confirming the potential of KD-enhanced SemCom for large-scale IoE systems.

\vspace{.2cm}
\subsubsection{Forecasting, spatiotemporal learning, and DRL-driven IoE orchestration}
Here, we review a unified AI-native IoE framework comprising (i) foundation-model-based forecasting at programmable edge nodes, (ii) spatiotemporal graph learning for dynamic urban sensing, (iii) DRL-driven orchestration for proactive scaling, placement, and scheduling, and (iv) learning-based reliability monitoring across the end--edge--cloud continuum. For ease of navigation, Fig.~\ref{fig:ai_orchestration_map} shows these building blocks and their data/control flow, highlighting how predictions feed closed-loop resource decisions while monitoring modules provide feedback for robust, low-latency IoE service delivery.\vspace{0.2cm}

\noindent\textbf{AI-assisted access and forecasting at the edge:}
AI-native IoE systems increasingly rely on learning models deployed directly at the network edge to support both connectivity management and predictive network operation. In particular, ML can assist massive device access by rapidly identifying active devices in dense IoE environments, while forecasting models enable edge nodes to
anticipate traffic dynamics, channel conditions, and service demand. Together, these capabilities support proactive resource allocation and low-latency decision making across large-scale IoE systems.\vspace{0.2cm}

\noindent\textbf{User Activity Detection Network: }
Learning-based activity detection is a key enabler for massive sporadic access in large-scale IoE systems. A fundamental challenge in massive machine-type communications is to identify the subset of active devices among a large population of potential users before allocating limited pilot and channel resources. Classical approaches based on sparse signal recovery provide a theoretical foundation for User Activity Detection (UAD) in grant-free access schemes \cite{liu2018massive_access_jsac}. More recently, deep learning–based detectors have been proposed to exploit spatial and temporal correlations in received signals across distributed APs. In wireless-powered cell-free massive MIMO systems, Convolutional Neural Network (CNN)-based detectors can infer the active-user set directly from received identifier tensors, improving detection accuracy while maintaining low online inference latency (on the order of $\sim$0.1\,s) \cite{chen2020wirelesspowered}. Such learning-assisted UAD approaches are particularly well suited to dense IoE deployments, where rapid access decisions are required before allocating scarce pilot resources. Further studies show that deep neural networks can significantly enhance activity detection performance in grant-free random access by capturing complex correlations between received pilot signals and active device patterns, achieving high detection accuracy even under dense connectivity conditions \cite{elkeshawy2024_dl_uad_cellfree}. Overall, these results highlight the potential of integrating ML with cell-free massive MIMO architectures to enable scalable and low-latency access in future 6G-enabled IoE networks.\vspace{0.2cm}

\noindent\textbf{Foundation forecasting with cross-domain robustness:}
Recent advances in \emph{time-series foundation models} position them as key enablers of AI-native IoE systems by supporting \emph{general-purpose forecasting} directly at \emph{programmable edge nodes}. Unlike
task-specific models, these pretrained generative architectures can be
adapted across heterogeneous IoE environments with minimal
reconfiguration overhead. For instance, Marchetti \emph{et al.} propose
integrating a generative forecasting module within next-generation edge
nodes to enable seamless operation across diverse communication domains
(e.g., aqua--ground--air), thereby mitigating data scarcity and avoiding the need for training models from scratch
\cite{marchetti2025foundation_forecasting_ioe}. State-of-the-art models such as Chronos and Time-Series Foundation Model further demonstrate that large pretrained time-series models can achieve competitive forecasting performance in \emph{zero-shot} settings and can be efficiently adapted via lightweight \emph{fine-tuning} using limited data samples \cite{ansari2024chronos,Das2023TimesFM}. This capability allows edge nodes to dynamically adapt to evolving network conditions while maintaining low training and data collection overhead.
A defining advantage of foundation forecasting lies in its inherent
\emph{cross-domain robustness}. By leveraging diverse pretraining
corpora, these models generalize effectively across heterogeneous
environments, in contrast to conventional supervised approaches that
require domain-specific retraining. Empirical results show that, when
fine-tuned on a given domain, foundation models exhibit significantly
\emph{reduced performance degradation} when evaluated on unseen domains
\cite{marchetti2025foundation_forecasting_ioe,wu2022_ai_native_6g}. In
particular, cross-dataset evaluations spanning underwater, terrestrial,
and aerial communication scenarios confirm that zero-shot forecasting can
adapt to distribution shifts without domain-specific retraining, whereas
classical models often fail under such conditions.
From a networking perspective, this paradigm aligns with emerging
\emph{telecom foundation models}, where large pretrained models are
trained on massive network telemetry and subsequently adapted to diverse
operational tasks such as traffic prediction, anomaly detection, and
resource management \cite{zanouda2024telecom_foundation_models,qin2019deep_learning_phy}. 
More broadly, it supports the vision of \emph{AI-native 6G networks}, in
which intelligence is deeply embedded within the communication stack to
enable predictive, adaptive, and self-optimizing network operation across
heterogeneous IoE deployments
\cite{samdanis2023_ai_ml_service_enablers_6g}.
\vspace{.2cm}
\paragraph{Spatiotemporal learning for urban IoE prediction}
Large-scale IoE deployments generate massive streams of heterogeneous sensor data describing traffic conditions, environmental variables, and infrastructure states. Effectively exploiting these data requires learning models capable of capturing complex spatial and temporal dependencies across distributed sensing infrastructures.
Spatiotemporal learning frameworks (e.g., graph-based models) have therefore emerged as powerful tools for predicting dynamic urban phenomena and supporting proactive control in intelligent IoE systems.\vspace{0.2cm}

\noindent\textbf{Graph neural networks:}
In smart-city IoE environments, sensor streams often exhibit complex
spatiotemporal dependencies that evolve over time. Traffic sensors,
environmental monitoring devices, and urban infrastructure systems
generate data that are correlated both spatially (across locations)
and temporally (across time steps). Graph neural network architectures have recently emerged as powerful tools for modeling such dependencies by representing IoE sensing infrastructures as graphs where nodes correspond to sensors and edges capture spatial
or functional relationships \cite{yu2018stgcn}. Dynamic multisemantic graph attention models extend this idea by constructing multiple complementary graph views, such as spatial
adjacency graphs, dynamically learned dependency graphs, and
time-series similarity graphs. Attention mechanisms are then used
to fuse these heterogeneous graph representations, enabling the
model to capture evolving spatiotemporal interactions in
non-stationary urban environments. Recent studies on
spatiotemporal graph neural networks for smart-city IoT systems
demonstrate that attention-based residual graph architectures can
effectively model complex sensor relationships and significantly
improve traffic flow forecasting accuracy in large-scale urban
deployments \cite{zhang2023_st_residual_gat_iotj}. Such predictive models provide reliable short-term forecasts that can feed downstream control loops, such as traffic signal optimization and adaptive urban management systems
\cite{guo2021attention_traffic,smart_ioe_traffic_control}.\vspace{0.2cm}

\noindent\textbf{Energy-efficient optimization for IoE-6G resource management:}
Complementary to learning-based orchestration approaches (e.g., DRL at the edge), optimization-driven techniques provide powerful tools to address non-convex and multi-objective resource allocation problems under stringent energy, latency, and coverage constraints. 
In this context, hybrid metaheuristic methods such as the \emph{Energy-Efficient Hybrid Evolutionary Algorithm (EEHEA)}, which combines leader-based optimization with adaptive differential evolution, enable an effective balance between exploration and exploitation in dynamic environments \cite{singh2024eehea}. Reported results demonstrate that such approaches can simultaneously improve multiple performance metrics, including reduced energy consumption, enhanced latency, improved coverage, and better localization accuracy. 
These findings highlight the importance of integrating hybrid evolutionary optimization techniques alongside AI-based orchestration to achieve green, scalable, and high-performance IoE-6G systems.

\vspace{.2cm}
\paragraph{DRL-driven orchestration and scheduling}
Efficient operation of large-scale IoE infrastructures requires
intelligent orchestration mechanisms capable of dynamically managing computing resources, service placement, and task scheduling across distributed device--edge--cloud environments. Deep reinforcement learning has recently emerged as a promising approach to address these challenges, as it enables autonomous and adaptive decision-making under time-varying network conditions and heterogeneous resource constraints. By interacting with the environment, DRL agents can learn scalable policies that jointly optimize resource allocation, computation offloading, and service deployment strategies. In particular, both single-agent and multi-agent DRL frameworks have demonstrated strong capability in improving resource utilization, reducing latency, and enhancing service reliability in large-scale IoE systems
\cite{chen2018_drl_offloading_iot,ali2024_edgebus_kubernetes,
yang2025_beyond_edge_rl_mec,wang2023_marl_edge_scheduling}. These
capabilities make DRL a key enabler for closed-loop, AI-native
orchestration in future 6G-enabled IoE networks.\vspace{0.2cm}

\noindent\textbf{IScaler (orchestration):}
DRL has recently emerged as a promising approach to automate
orchestration decisions in MEC environments supporting large-scale
IoE services. In such settings, orchestration mechanisms must jointly
determine resource scaling and service placement across distributed
edge nodes under time-varying workloads. 
To address this challenge, Sami \emph{et al.} propose \emph{IScaler}, a
model-free DRL agent formulated as a Markov Decision Process (MDP), which enables proactive scaling and service placement while maintaining scalability in large state spaces and heterogeneous edge infrastructures
\cite{sami2021_ai_resource_provisioning_ioe_6g}.\vspace{0.2cm}

\noindent\textbf{Deep Reinforcement Learning–based Time Slot Scheduling: }
Many IoE services operate over hierarchical device--edge--cloud
pipelines, where raw data generated at devices are partially
processed at nearby edge nodes before being forwarded to the cloud
for further analysis \cite{shi2016edge,mao2017survey_mec}. In this
context, \cite{zhou2023_drl_tss} formulates resource allocation as an
NP-hard two-stage scheduling problem and proposes \emph{Deep Reinforcement Learning–based Time Slot},
which integrates Johnson's-rule-based presorting with deep
reinforcement learning to derive scheduling policies that minimize
the overall makespan under heterogeneous computing and communication
resources.\vspace{0.2cm}

\paragraph{Reliability monitoring and IoE security}
Ensuring reliability and security is a key requirement for
AI-native IoE infrastructures, where large numbers of distributed
devices continuously interact across heterogeneous networks.
Learning-based monitoring mechanisms are increasingly used to detect
abnormal behaviors, authenticate devices, and evaluate the reliability
of predictive models deployed at the edge.\vspace{0.2cm}

\noindent\textbf{Anomaly detection:}
In addition to control and orchestration, AI plays a crucial role in
\emph{reliability monitoring} for large-scale IoE deployments.
Operational IoE infrastructures continuously generate heterogeneous
telemetry streams originating from sensors, network devices, and
edge/cloud services, making manual monitoring increasingly challenging.
Learning-based anomaly detection approaches can automatically
identify abnormal behaviors, faults, or cyber-attacks by learning
patterns from historical system data. \cite{djenouri2023_emergent_dl_anomaly_ioe} introduces an emergent deep learning framework for anomaly detection in IoE environments, showing that neural models can effectively capture complex dependencies across heterogeneous data streams and support timely mitigation in safety- and
mission-critical deployments .
More broadly, deep learning–based anomaly detection systems are
increasingly used in IoT infrastructures to analyze network traffic,
sensor data, and operational telemetry in order to detect anomalies
and security threats in real time \cite{ferrag2020dl_ids_iot}. \cite{shen2024_federated_iot_ids} further shows that
FL can enable collaborative anomaly detection
across distributed IoT nodes without sharing raw data, improving
privacy preservation while maintaining high detection accuracy
in heterogeneous~IoT.\vspace{0.2cm}

\noindent\textbf{RF fingerprinting / satellite IoE:}
In particular, Radio-Frequency Fingerprinting (RFF) has emerged as an effective physical-layer identification mechanism for IoT devices by exploiting hardware-induced signal impairments that uniquely characterize transmitters \cite{xie2024_rff_iot_survey}. 
However, satellite IoE systems often operate under severe data constraints, where only a limited number of labeled signal samples are available for training reliable device identification models. This scarcity of data significantly limits the performance and generalization capability of conventional learning-based approaches.
To address this challenge, recent works have explored few-shot learning and data-generation techniques to enable robust identification under limited supervision. Building on the RFF paradigm, Zhao \emph{et al.} propose a ground--satellite generative learning framework in which a Specific Emitter Identification (SEI) model is trained on the ground and deployed on Low Earth Orbit (LEO) satellites for real-time device identification. Their approach converts RF signals into Gramian angular field representations and leverages generative data augmentation to improve fingerprint separability in low-shot regimes. Experimental results demonstrate significant accuracy gains, highlighting the effectiveness of generative few-shot learning for data-scarce satellite IoE environments \cite{zhao2026_fewshot_sei_satellite_ioe}.
More broadly, RFF-based techniques are increasingly investigated for device authentication and security in IoT and space communication systems, where unique hardware signatures can enable reliable transmitter identification even in dynamic wireless environments \cite{7239531}. Recent studies further show that deep-learning-based RFF methods can support secure access control in large-scale satellite IoT networks \cite{zhang2023_rff_iot_authentication}.\vspace{0.2cm}

\noindent\textbf{Evaluation metrics}
Following the adoption of foundation models for time-series forecasting
in IoE systems, it becomes essential to assess their performance not
only in terms of prediction accuracy but also with respect to
uncertainty awareness and deployment efficiency. For IoE control loops, forecasting quality cannot be assessed solely through point prediction accuracy; it must also consider \emph{uncertainty estimation} and \emph{runtime feasibility} at the edge.
In large-scale IoE systems, forecasting models are often integrated
into real-time decision pipelines where inaccurate or overconfident
predictions may lead to unstable control actions. Consequently,
recent studies emphasize probabilistic forecasting techniques that
provide prediction intervals and calibrated uncertainty estimates in
addition to point forecasts \cite{salinas2020deepar_o}.
Metrics such as Mean Absolute Error (MAE) evaluate point accuracy,
while measures such as Multi-Quantile Loss (MQL) enable the assessment
of distributional forecasts and confidence intervals, which are
particularly useful for confidence-aware resource allocation and
control decisions. Marchetti \emph{et al.} evaluate channel-state
forecasting using MAE and MQL to capture both deterministic and
probabilistic prediction performance in heterogeneous IoE 
settings \cite{marchetti2025foundation_forecasting_ioe}.
Beyond prediction quality, they also profile runtime characteristics
relevant to edge deployment, including inference latency, Central Processing Unit (CPU) utilization, peak memory consumption, and power usage. Their results highlight that time-series foundation models can operate effectively
in \emph{zero-shot} mode, eliminating additional training overhead while maintaining competitive forecasting performance. Such characteristics make foundation forecasting particularly attractive for programmable edge nodes, where limited compute and energy budgets require models that balance predictive accuracy with operational
efficiency \cite{ansari2024chronos}.\vspace{0.2cm}

\noindent\emph{\textbf{Insight:}} 
AI/ML enables IoE autonomy, but scalable deployment requires (i) edge-feasible models (compression/quantization),
(ii) robust learning under dynamics and interference, and (iii) closed-loop integration between prediction, orchestration, and monitoring.

\subsection{Edge/Fog/Cloud Computing}
\label{subsec:Edge/Fog}

Edge and fog computing complement the cloud by placing compute, storage, and control closer to IoE devices, reducing
latency and alleviating bandwidth usage \cite{yu2017edge_iot,9046806}. Beyond the classical edge--fog--cloud layering,
recent work frames IoE deployment as an \emph{IoE--Edge--Cloud continuum}, emphasizing end-to-end orchestration, placement,
and lifecycle management of services under heterogeneous resources and mobility \cite{gkonis2023continuum_survey}. Architectural
surveys highlight multi-tier designs, orchestration, and resource management as core mechanisms to support scalable IoE services
\cite{9046806}. Practical studies also demonstrate that edge computing can improve responsiveness for interactive IoT
workloads and reduce congestion compared to cloud-centric processing \cite{premsankar2018edge_case}. However, pushing computation
and data aggregation toward the edge also expands the attack surface; comprehensive surveys of edge-computing-assisted IoT
underline the need for security-by-design across edge nodes, APIs, and distributed control loops
\cite{alwarafy2021_ec_iot_security_privacy}. Energy-aware orchestration is increasingly treated as a first-class objective in
beyond-5G/6G systems, motivating cross-layer designs that expose energy information
and optimize placement and networking decisions accordingly
\cite{mao2022ai_green_6g}.

\vspace{0.2cm}

\subsubsection{VNF/CNF Placement in Cloud--Edge 5G/6G IoE Systems}
As 5G/6G infrastructures evolve toward service-based and geographically distributed architectures, network functions are increasingly deployed as virtualized components, either VM-based Virtual Network Functions (VNFs) or containerized Cloud-Native Network Functions (CNFs). These functions must be strategically placed across access/edge, aggregation, and core/cloud sites in order to satisfy strict latency,
throughput, and reliability requirements of emerging IoE services \cite{herrera2016_vnf_survey,han2015network}. The shift toward cloud-native networking further introduces microservice-based implementations and container orchestration frameworks (e.g., Kubernetes), enabling lightweight deployment and elastic scaling of network services across the device--edge--cloud continuum. However, coexistence of VNFs and CNFs significantly increases orchestration complexity, as operators must manage heterogeneous virtualization environments, service chains, and dynamic resource allocation across distributed infrastructures \cite{attaoui2023vnfcnf}. These challenges make efficient placement, scaling, and lifecycle management of network functions a central research problem in 5G/6G IoE systems, where orchestration frameworks must jointly optimize latency, resource utilization, and operational cost while maintaining service-level guarantees \cite{taleb2017mec_orchestration,mao2017survey_mec}.\vspace{0.2cm}

\noindent\textbf{Common optimization objectives:}
VNF/CNF placement is typically formulated as a single- or multi-objective optimization problem, where adding more
objectives increases decision complexity and forces explicit trade-offs. Common objectives include minimizing end-to-end
latency, reducing energy consumption, limiting the number of active nodes/instances, and improving resource utilization,
subject to QoS and Service Level Agreements (SLAs) constraints \cite{attaoui2023vnfcnf}.

\vspace{0.2cm}

\subsubsection{VNF/CNF Placement: Solutions and Design Challenges}

\begin{figure*}[!t]
  \centering
  \includegraphics[width=\textwidth, height=11cm]{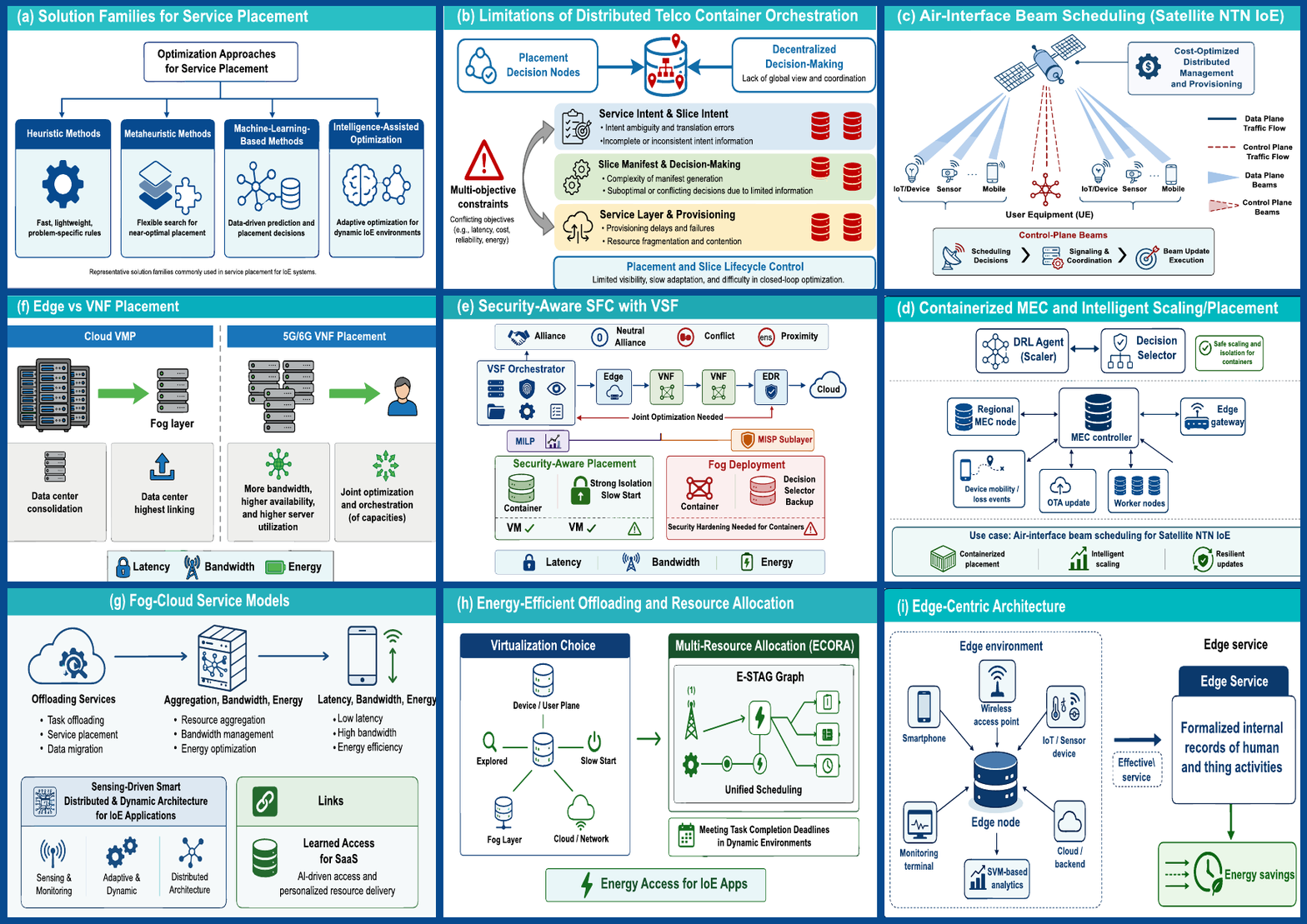}
  \caption{Taxonomy of VM/VNF/CNF placement solution families: heuristics, meta-heuristics, and learning-based methods.}
  \label{fig:placement_solution_families}\vspace{-0.3cm}
\end{figure*}

Building on the above formulation, we now review the main
solution approaches and associated design challenges for
VNF/CNF placement in large-scale IoE systems.
Fig.~\ref{fig:placement_solution_families} summarizes the main solution families used for VNF/CNF placement. Most placement formulations are Non-deterministic Polynomial-time hard; consequently, the literature is commonly grouped into: (i) \emph{heuristics} (fast, problem-specific),
(ii) \emph{meta-heuristics} (high-level, flexible, near-optimal search), and (iii) \emph{machine-learning-based} approaches (including deep and reinforcement learning for adaptive decisions) \cite{attaoui2023vnfcnf}.\vspace{0.2cm}

\noindent\textbf{CNF orchestration limits in distributed 5G/6G: }
Although Kubernetes enables container orchestration, it is not designed by default for widely distributed telco sites with stringent latency/performance constraints. This motivates placement/scheduling enhancements that incorporate network-awareness, decentralized decision-making, and multi-objective policies (energy, QoS, SLA) \cite{attaoui2023vnfcnf}.\vspace{0.2cm}

\noindent\textbf{Placement as a slicing control problem: }
Today, VM placement decisions become tightly coupled with 5G/6G network slicing, where operators instantiate isolated logical networks on shared infrastructure. A slicing viewpoint decomposes the system into (i) a business layer (marketplace/slice intent), (ii) a service layer (configuration, management, scaling based on a slice manifest), and (iii) an infrastructure layer (real-time reconfigurable virtualized resources), highlighting that VM/VNF placement is effectively a life-cycle management function inside the slice control loop \cite{vmp_multicriteria_survey}.\vspace{0.2cm}

\noindent\textbf{Edge-NFV orchestration tradeoff: bandwidth vs. VNF reuse under latency constraints: }
When IoE services are implemented as Network Function Virtualization (NFV)-based Service Function Chains (SFC) at the edge, orchestration faces a fundamental tradeoff: selecting short paths reduces network bandwidth consumption, but may require launching additional VNF instances (higher server usage), while selecting longer paths can reuse already-deployed VNFs (saving server resources) at the cost of higher bandwidth usage and potentially larger end-to-end delay. Joint optimization of routing and VNF assignment is therefore essential for scalable multi-service IoE at the edge, especially under strict latency constraints \cite{infocom20_edge_nfv}.\vspace{0.2cm}

\noindent\textbf{Cloud VMP vs. VNF placement in 5G/6G:}
While classical VM placement emphasizes consolidation and cost/energy tradeoffs in data centers, telecom-grade VNF placement additionally targets efficient resource expenditure and service trustworthiness under slice constraints, linking placement decisions directly to end-to-end service guarantees in 5G/6G IoE \cite{vmp_multicriteria_survey}.\vspace{0.2cm}

\noindent\textbf{Security-aware service chaining with Virtual Security Functions: }
Beyond elastic scaling and placement, large-scale IoE services increasingly rely on \emph{NFV/SDN-based service function chaining} to deploy application functions together with \emph{Virtual Security Functions} (VSFs) (e.g., firewall, deep packet inspection, Intrusion Detection System (IDS)) along the device--edge--cloud continuum. A key insight is that security cannot be treated as an afterthought: VSFs must be placed \emph{jointly} with VNFs under latency/SLA constraints while also respecting VSF operational rules that govern correct security behavior. For example, Tamim \emph{et al.} introduce four practical constraints---\emph{alliance}, \emph{conflict}, \emph{redundancy}, and \emph{proximity}---to capture dependency/anti-affinity and delay-sensitive relations among VSFs and VNFs during SFC placement, and formulate a latency-minimizing Mixed-Integer Linear Programming (MILP) model that enforces these security rules during deployment \cite{tamim2020vsf}. Moreover, realistic edge orchestration must account for access-side congestion: queueing delay at access points can be modeled (e.g., M/M/1) and included in the end-to-end latency budget, which can materially change feasible routing/placement decisions under bursty IoE traffic \cite{infocom20_edge_nfv}.\vspace{0.2cm}

\noindent\textbf{Why this matters for low-latency IoE?}
In a representative virtualized Evolved Packet Core use case, the security-aware MILP is compared against a latency-agnostic greedy placement that ignores latency and security rules, showing the benefit of incorporating both delay requirements and VSF constraints when deploying SFCs. This supports the broader IoE survey takeaway that orchestration in MEC/NFV should be \emph{policy- and security-aware at placement time}, especially for mission-critical IoE services that are simultaneously latency-constrained and attack-exposed \cite{tamim2020vsf}.

\vspace{.2cm}
\subsubsection{Service orchestration models in IoE--Fog--Cloud systems}
Beyond the algorithmic aspects of placement and scheduling, IoE service orchestration is also shaped by the architectural models used to deliver computation, storage, and sensing capabilities across the device--edge--cloud continuum. Different service models determine how workloads are distributed between fog nodes, edge infrastructure, and centralized cloud platforms, while virtualization technologies enable flexible deployment and scaling of services. Summarizes key service models and architectural abstractions that guide orchestration decisions in IoE-enabled fog and cloud systems.\vspace{0.2cm}

\noindent\textbf{Service models in the IoE--Fog--Cloud ecosystem:}
From a workload perspective, fog-enabled IoE systems are often categorized into complementary service models: (i) \emph{offloading}, where fog performs near-source processing while heavy tasks are delegated to the cloud; (ii) \emph{aggregation}, where fog/gateways fuse or compress streams to reduce redundancy and backhaul load; and (iii) \emph{Peer-to-Peer} (P2P) cooperation, where devices exchange data or partial results through local connectivity (e.g., Device-to-Device (D2D)) \cite{foe2017_fog_of_everything}. This taxonomy helps interpret orchestration decisions as selecting among (or combining) these models under latency, bandwidth, and energy constraints.\vspace{0.2cm}

\noindent\textbf{Virtualization strategy, containers vs. VMs:}
Virtualization is central to scalable fog/MEC hosting, yet the choice of strategy impacts deployment speed, resource efficiency, and isolation. Container-based virtualization can enable faster instantiation and higher density than full VMs, which is attractive when many lightweight IoE service instances must be deployed and scaled rapidly at the edge \cite{foe2017_fog_of_everything}. However, because containers share the host operating system, isolation can be weaker than in VM-based designs, motivating careful security hardening and policy enforcement in multi-tenant IoE deployments.\vspace{0.2cm}

\noindent\textbf{Cloud-assisted sensing service models:}
Cloud-Assisted Remote Sensing (CARS) positions remote sensing as a key enabler of IoE by leveraging cloud capabilities such as global data and resource sharing, real-time remote access and analytics, elastic provisioning, and pay-as-you-go service models \cite{abdelwahab2014cars}. This paradigm enables the abstraction of sensing and remote-sensing capabilities through cloud-based service models.
In particular, CARS frameworks support (i) elastic data storage and processing capabilities (akin to Infrastructure as a Service and Platform as a Service), and (ii) higher-level sensing and analytics functionalities delivered as services to applications (SaaS-like delivery). This service-oriented model naturally leads to \emph{Sensing-as-a-Service} (SenaaS), where sensing infrastructures and data can be virtualized, leased, and reused across multiple domains, thereby supporting scalable, interoperable, and multi-stakeholder IoE ecosystems.
Such cloud-based sensing abstractions enable on-demand access to large-scale sensing data and analytics pipelines, facilitating flexible and data-driven IoE service provisioning across heterogeneous environments.\vspace{.2cm}

\noindent\textbf{Energy-efficient computation offloading:}
IoE computation-intensive applications (e.g., AR and smart grids) often exceed the capabilities of resource-constrained devices. While cloud computing provides abundant resources, cloud-only offloading may introduce excessive latency, increased transmission energy, and higher data-exposure risks. To address these limitations, fog computing enables near-source execution, reducing both delay and energy consumption. However, limited fog capacity under high device density motivates \emph{fog--cloud complementarity} and joint orchestration across the continuum.
In heterogeneous access settings, devices must decide \emph{whether} to offload computation and \emph{which AP} to select for uplink transmission. An illustrative approach is the Energy-Efficient Computation Offloading and Resource Allocation (ECORA) scheme, which adopts a three-layer architecture (device, fog, and cloud) and jointly optimizes offloading decisions, transmit power, and computation-resource allocation to minimize system cost. By decoupling the problem into subproblems and solving them iteratively, ECORA demonstrates that joint optimization can significantly reduce system cost (up to 50\%) compared with conventional approaches, particularly under limited fog resources or high device density \cite{li2019_ecora_fogcloud_ioe}.

\subsubsection{Intelligent orchestration architectures for IoE}
AN increasing number of research efforts explore intelligent orchestration architectures that combine edge computing, AI-driven decision-making, and adaptive resource coordination to support large-scale IoE services. These architectures aim to dynamically manage computation, communication, and storage resources across the device--edge--cloud continuum while satisfying latency, energy, and reliability constraints. The following examples illustrate representative frameworks ranging from containerized MEC orchestration and edge-centric resource controllers to emerging paradigms such as cybertwin-assisted service provisioning and dynamic multi-resource scheduling in Non-Terrestrial Network (NTN)-enabled IoE systems.\vspace{0.2cm}

\noindent\textbf{Containerized MEC and intelligent scaling/placement:}
Container-based hosting and orchestration are widely adopted to support rapid updates and elastic management of IoE services at the MEC layer. In a representative MEC cluster, worker nodes host services as containers, while a master orchestrator load-balances requests and issues placement/scaling decisions \cite{sami2021_ai_resource_provisioning_ioe_6g}. To reduce the risk of poor actions during early DRL learning phases, an \emph{Intelligent Scaling and Placement} sublayer can combine a DRL agent (IScaler) with a heuristic optimizer as a bootstrapper/backup, and a switching component to select safer decisions when needed.\vspace{0.2cm}

\noindent\textbf{Example of edge-centric resource distribution via a web-server controller:}
An illustrative energy-aware edge architecture for autonomous IoE uses a \emph{localized web server} that connects IoE devices to integrated Internet servers at the network edge and allocates tasks using a resource-allocation mechanism targeting reduced delay and energy utilization \cite{babbar2023_massiveiot_ioe}. In this framework, a \emph{Boltzmann-machine-based} strategy is used to model smart and scalable distributed resource decisions under dynamic QoS conditions, by mapping resource-state observations to workload-dispersion actions and learning a policy that improves energy utilization over time \cite{babbar2023_massiveiot_ioe}.\vspace{0.2cm}

\noindent\textbf{Cybertwin-driven service provisioning for massive IoE at the edge:}
Beyond classical MEC orchestration, \emph{cybertwin} extends the digital-twin concept into a network-assisted entity that maintains digital records of activities of \emph{humans and things} and can act as a contact hub for service delivery in 6G-enabled edge networks \cite{adhikari2022_cybertwin_tii}. A representative cybertwin-driven framework combines collaborative edge--cloud resources and uses \emph{deep reinforcement learning} to distribute incoming IoE tasks according to dynamic service requirements, targeting simultaneous delay reduction and energy savings \cite{adhikari2022_cybertwin_tii}. In addition, the same framework integrates edge-level AI analytics (e.g., support vector machine-based classification with feature-extraction) to improve data-processing accuracy while limiting edge overhead. These results motivate cybertwin-assisted orchestration as a scalable mechanism for multi-application IoE provisioning in 6G edge.\vspace{0.2cm}

\noindent\textbf{Beam management for initial access in LEO-integrated IoE:}
At the air-interface level, resource coordination challenges also arise in NTN-enabled IoE systems. In LEO-integrated 5G NR networks, satellites must efficiently manage beam resources for initial access, where signaling beams are used to broadcast synchronization signal blocks and system information to ground users. This process involves beam sweeping, beam selection, and access coordination under highly dynamic channel conditions and satellite mobility. \cite{zhang2026_iabm_leo_5gnr} proposes an initial access beam management framework that optimizes beam scanning and access procedures to improve coverage and access reliability in LEO satellite networks. Such mechanisms highlight the importance of efficient beam-level coordination for supporting scalable and reliable IoE connectivity in NTN environments.\vspace{0.2cm}

\noindent\textbf{Multi-resource scheduling under dynamic connectivity (satellite/NTN IoE): }
MEC orchestration problems typically involve joint decisions across \emph{communication}, \emph{compute}, and sometimes \emph{storage} resources; this coupling becomes even more pronounced in disruption-prone infrastructures such as satellite/NTN networks supporting IoE applications. FlexSatIoE introduces an \emph{Enhanced Storage Time-Aggregated Graph} abstraction that jointly represents time-varying transmission, buffering, and compute resources, enabling unified scheduling across these dimensions rather than optimizing routing or placement in isolation \cite{flexsatioe2026}. This perspective extends terrestrial MEC orchestration by emphasizing that buffering and routing policies are integral to satisfying application-level completion delay constraints in dynamic IoE environments.\vspace{0.2cm}

\noindent\emph{\textbf{Insight:}} 
The edge/fog/cloud continuum reduces latency and backhaul pressure, but pushes complexity into orchestration: Placement must become multi-objective (QoS/SLA, energy, security policies) and robust to heterogeneity and dynamics.

\subsection{Energy-as-a-Service for RF-Powered IoE systems}
Large-scale IoE deployments face a critical sustainability barrier: the operational burden of monitoring, replacing, or recharging batteries across massive device populations. A promising 6G-era solution is \emph{Energy-as-a-Service (EaaS)}, where dedicated \emph{Energy Stations} (ESs) provide WET to energy-harvesting IoE devices, enabling battery-less or low-maintenance operation. In this platform, ESs act as shared infrastructure (analogous to cellular base stations), serving heterogeneous IoE operators and applications aim to reduce operational expenditure while optimizing Capital Expenditure (CAPEX). This study introduces a tractable theoretical framework for \emph{EaaS-enabled connectivity} in RF-powered IoE systems by leveraging stochastic geometry and continuum percolation theory. ESs and IoE devices are modeled as independent Poisson Point Processes (PPPs). 
We construct a \emph{WET-aware Random Geometric Graph (WET-RGG)}, where a device is considered \emph{active} only if it lies within at least one ES charging region (i.e., a WET coverage disk). D2D communication links are established exclusively between active devices whose pairwise distance does not exceed a predefined communication range. 
By applying percolation theory to the WET-RGG, this work characterizes the \emph{phase transition} behavior and derives \emph{necessary/sufficient conditions} for the emergence of large-scale multi-hop D2D connectivity under EaaS \cite{lin2026_eaas_rf_ioe_percolation}. Importantly, the framework yields \emph{deployment guidelines} in terms of the \emph{critical ES density} needed to ensure large-scale connectivity, and proposes practical approximations that connect directly to CAPEX-aware planning. Two complementary regimes are emphasized: (i) a sparse-device regime where active devices can be approximated as a thinned PPP (inner-city style approximation), and (ii) a dense-device regime where each ES induces a connected local cluster and inter-cluster connectivity depends primarily on ES spacing (Gilbert-disk style approximation). This modeling viewpoint is valuable, it links \emph{energy provisioning} (WET range, ES density) to \emph{network scalability} (percolation/connectivity) and provides a rigorous foundation for “energy infrastructure planning” as a first-class design dimension in 6G-enabled IoE.\vspace{0.2cm}

\noindent\emph{\textbf{Insight:}}  EaaS transforms energy from a device-level constraint into a network-level infrastructure planning problem, where connectivity and scalability can be engineered via ES deployment density, WET-zone size, and D2D range.

\subsection{Big Data Analytics and Data Mining}
\label{subsec:Big_Data}

IoE generates massive, heterogeneous, and high-velocity data, requiring analytics pipelines capable of extracting knowledge from
distributed sources \cite{alfuqaha2015iot}. Recent edge analytics surveys emphasize that practical IoE pipelines must cope with
limited memory, intermittent connectivity, and power constraints, motivating lightweight streaming analytics and adaptive partitioning
between edge and cloud \cite{nayak2024_edge_analytics_review}. Data mining techniques for IoT/IoE have been surveyed extensively,
covering challenges such as streaming analytics, noisy data, semantic interpretation, and scalable learning over heterogeneous sources
\cite{tsai2014datamining_iot}.\vspace{0.2cm}

\noindent\textbf{From raw streams to graph-structured knowledge: }
A practical pattern in urban IoE analytics is to transform multi-source streams into dynamic graphs and learn spatiotemporal representations via attention, enabling scalable forecasting and decision-making (e.g., mobility prediction feeding adaptive traffic control) \cite{smart_ioe_traffic_control}.\vspace{0.2cm}

\noindent\textbf{Edge-level feature retrieval to reduce redundancy: }
A practical bottleneck in IoE analytics is that raw device streams often contain redundant and irrelevant features, which increases edge computation complexity and may degrade predictive performance. An edge-centric approach is to incorporate lightweight \emph{feature retrieval} at localized edge devices---for example, using Tone Groups to extract only the most informative characteristics from IoE application data—thereby reducing duplicate/unnecessary attributes before model inference \cite{babbar2023_massiveiot_ioe}. Such pruning is especially useful when edge storage is limited, so only critical retrieved features are processed locally while the remaining data can be retained for remote-server analytics.\vspace{0.2cm}

\noindent\textbf{Scalable stream processing and edge-side reduction: }
In massive IoE/IoT, the data layer must scale under high-speed streams and limited device resources. Practical directions include \emph{edge-side compression and aggregation} to reduce redundancy and network load, as well as \emph{stream processing} pipelines that support near real-time analytics at scale (e.g., message-queue and streaming frameworks) \cite{IEEE9667513}. Such mechanisms decrease communication overhead and energy consumption while enabling timely extraction of actionable knowledge from distributed sources.\vspace{0.2cm}

\noindent\emph{\textbf{Insight:}} 
IoE data pipelines must be streaming-first and edge-aware: reduce redundancy early, preserve semantics, and manage privacy constraints while still enabling timely analytics for control loops.

\subsection{Connectivity Layer: 5G/6G and Beyond}
\label{subsec:Connectivity_Layer}
The connectivity layer constitutes the communication backbone of IoE systems, enabling seamless data exchange across massive numbers of heterogeneous devices and distributed infrastructures. As IoE applications evolve toward ultra-dense deployments and mission-critical services, 5G and emerging 6G networks are expected to provide scalable, reliable, and low-latency connectivity through advanced radio technologies, new spectrum utilization, and AI-native network design. Here, we review key connectivity enablers and design principles that support large-scale IoE deployments, spanning access mechanisms, spectrum evolution, non-terrestrial integration, and programmable wireless environments.
\begin{figure}[t]
  \centering
  \includegraphics[width=\columnwidth, height=6.5cm]{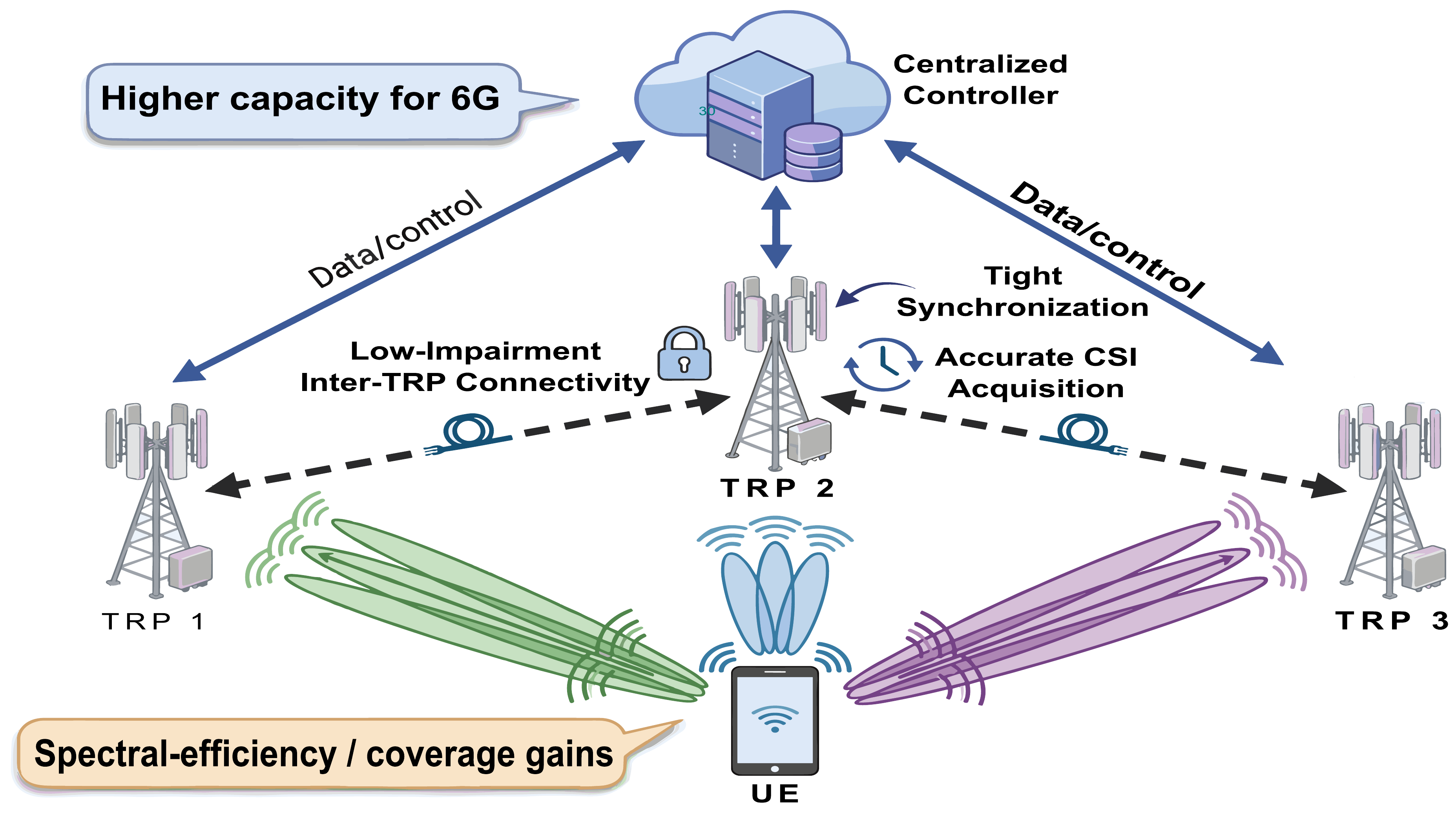}
  \vspace{-.5cm}\caption{A coherent joint transmission scenario with centralized coordination and multiple TRPs jointly serving a UE.}
  \label{fig:cjt_scenario}\vspace{-0.3cm}
\end{figure}

\vspace{0.2cm}

\subsubsection{5G/6G Connectivity for IoE at Scale}
Connectivity remains foundational for IoE, especially as systems scale to massive numbers of devices and support mission-critical services.
6G visions emphasize ultra-low latency, high reliability, massive connectivity, and AI-native networking to enable next-generation IoE
applications \cite{saad2020vision6g}. Widely cited 6G vision work also formalizes the shift toward AI-native, data-driven networking and
defines KPIs and service classes that map directly to large-scale IoE requirements. From a physical-layer perspective, key challenges (e.g., cell-free architectures, hardware constraints, and new spectrum regimes) are consolidated in influential 6G roadmapping work \cite{matthaiou2021_road_to_6g}. Comprehensive surveys further detail 6G requirements, enabling technologies, and testbeds
that shape the communication substrate for future large-scale IoE deployments \cite{wang2023road6g}. For massive access and device density, surveys specifically targeting massive IoT over 6G provide architectural viewpoints and technology enablers that complement general 6G roadmaps
\cite{guo2021_massive_iot_6g_survey}.\vspace{0.2cm}

\noindent\textbf{Massive access mechanisms for dense IoE: }
At the access layer, Massive Internet of Things (MIoT) scalability depends on reducing control overhead and collisions under extreme device density. Grant-free access is widely discussed as a key enabler because it reduces scheduling signaling and can lower latency for sporadic transmissions, improving scalability in dense IoE settings \cite{IEEE9667513}. In parallel, non-orthogonal multiple access techniques such as sparse code multiple access support higher connection density by mapping symbols to sparse codewords and enabling multi-user detection, with recent trends exploring learning-assisted decoding to further improve performance in challenging conditions. Prior evaluations also indicate that cell-free operation can outperform small-cell baselines in per-user spectral-efficiency distributions, with macro-diversity (more APs) often providing larger gains than simply adding antennas per AP \cite{chen2020wirelesspowered}.\vspace{0.2cm}

\noindent\textbf{Routing and buffering optimization for delay-sensitive satellite IoE:}
Efficient routing and buffering mechanisms are essential for delay-sensitive IoE applications over NTN, where intermittent connectivity and dynamic topology significantly impact performance. Recent works have explored optimization frameworks that jointly consider routing, buffering, and transmission dynamics to improve application-level performance under such conditions. In particular, flexible buffering and delay-aware routing strategies have been shown to significantly reduce end-to-end delay and improve task completion rates in disruption-prone satellite IoE environments \cite{flexsatioe2026}.\vspace{0.2cm}

\noindent\textbf{Decoupling UAD from CSI acquisition: }
In massive IoE, pilot scarcity makes it difficult to simultaneously achieve massive connectivity and high spectral efficiency. A useful design principle is to decouple UAD from channel estimation: devices first transmit short, possibly non-orthogonal identifiers for activity discovery, and only the detected active set is then assigned pilots for CSI acquisition during subsequent coherence intervals. This reduces overhead wasted on inactive devices and improves scalability under sporadic traffic \cite{chen2020wirelesspowered}.\vspace{0.2cm}

\noindent\textbf{Multi-band spectrum evolution and spectrum management: }
From a spectrum viewpoint, 5G deployments largely relied on sub-6~GHz and mmWave bands together with enabling PHY techniques such as massive MIMO and beamforming \cite{sharma2025_6g_overview}. As IoE services expand toward higher connection density and more diverse service classes, 6G is expected to operate across a broader spectrum range (including cmWave and THz), which motivates spectrum-management mechanisms that can adapt to heterogeneous propagation conditions and application requirements. In addition, semantic communications is emerging as an AI-native Physical Layer and Medium Access Control Layer complement for IoE, and recent KD-based multi-user SemCom designs specifically study robustness under co-channel interference while reducing model size, which is relevant for dense 6G IoE deployments \cite{10345474}.

\vspace{0.2cm}

\subsubsection{Coherent Joint Transmission in Multi-Transmission and Reception Point Systems}
\label{subsubsec:cjt}
Coherent Joint Transmission (CJT) extends multi-TRP cooperation by enabling \emph{coherent} beamforming across multiple Transmission--Reception Points (TRPs), so that spatially distributed antennas act as a larger effective array and can transmit the same set of data layers to a User Equipment (UE). This is a high-capacity enabler for 6G, especially in the FR3 band (roughly 7--24~GHz), where channels become more line-of-sight and the per-TRP spatial multiplexing rank can be limited; CJT mitigates this limitation by exploiting additional spatial diversity across TRPs. Fig.~\ref{fig:cjt_scenario} illustrates a typical CJT setup with centralized coordination and multiple TRPs jointly serving the UE, which highlights the need for low-impairment inter-TRP connectivity, tight synchronization, and accurate CSI acquisition to unlock the promised spectral-efficiency and coverage gains \cite{SamsungCJT2026}.

\vspace{0.2cm}

\subsubsection{NTNs and Satellite IoE}

NTN are emerging as a key extension of terrestrial 5G/6G systems to support IoE services beyond traditional coverage areas. By leveraging satellite constellations, particularly LEO systems, NTN enables wide-area connectivity for sensing, monitoring, and distributed intelligence across remote and unconnected regions. However, the characteristics of satellite communications, including high mobility, intermittent links, and limited onboard resources, introduce new challenges in access, resource management, and end-to-end service delivery.\vspace{.2cm}

\noindent\textbf{Initial access and beam management in LEO:}
As IoE extends toward NTN, \emph{initial access} becomes a practical bottleneck due to high link dynamics, large propagation losses, and limited payload/antenna resources on LEO satellites. In LEO-aided 5G NR NTN, conventional initial-access beam management based on repeated measurements and reporting can yield \emph{outdated} beam measurements and increased access delay. 
To address these challenges, recent work proposes a framework that dynamically coordinates \emph{signaling beams} (for access) and \emph{service beams} (for traffic), aiming to balance coverage capability and service efficiency under limited beam resources \cite{zhang2026_iabm_leo_5gnr}. The framework models signaling--service beam coordination using a queueing-based formulation and derives an online Lyapunov-based algorithm to adapt beam allocation under time-varying LEO dynamics with low computational complexity. 
Reported results show that the proposed approach improves access performance while reducing computational time complexity by more than 90\% compared with traditional offline methods. More broadly, this design illustrates the effectiveness of combining queueing-theoretic modeling with Lyapunov-based online optimization for dynamic resource management in NTN-enabled IoE systems. Building on these access-layer challenges, routing and buffering become critical at the network layer for delay-sensitive applications.\vspace{0.2cm}

\noindent\textbf{Delay-sensitive IoE over LEO (flexible routing and buffering): }
LEO satellite constellations are increasingly relevant to IoE because they provide wide-area connectivity for sensing, monitoring, and distributed intelligence beyond terrestrial coverage. However, satellite links are intermittent and disruption-prone, and many IoE tasks require \emph{complete data reception before computation}. To capture this requirement, recent work defines the \emph{application completion delay} as the time from when application data leaves the source until the computation is finished at the destination; violating delay bounds directly reduces the task success ratio in delay-constrained IoE services \cite{flexsatioe2026}. Importantly, minimizing this delay may require \emph{delay-sensitive bundle transfer} (sending packets as a whole) rather than per-packet fairness scheduling, and this bundle behavior can violate the classical \emph{optimal substructure} assumption behind shortest-path routing. To address this, FlexSatIoE proposes flexible buffering along satellite routes to collaboratively utilize storage across nodes, achieving improved delay and completion performance in realistic satellite settings.\vspace{0.2cm}

\noindent\emph{\textbf{Insight:}}  NTN-enabled IoE requires joint design across access (beam management) and network-layer control (routing/buffering) to meet delay and reliability constraints under intermittent connectivity.

\vspace{0.2cm}

\subsubsection{Visible Light Communication for 6G IoE}
Visible Light Communication (VLC) is a promising complementary access technology for enabling 6G-era IoE applications, leveraging the visible-light spectrum and existing lighting infrastructure. Compared with RF communications, VLC offers advantages such as high bandwidth, enhanced security (limited leakage beyond illuminated regions), reduced interference, and potentially improved energy efficiency in some deployment settings \cite{10609788}. However, VLC links also exhibit distinctive propagation properties that strongly affect system design and evaluation: weak penetration and diffraction, predominantly line-of-sight behavior with relatively weak multipath dispersion, and sensitivity to environmental conditions (e.g., sunlight and weather in outdoor scenarios). Consequently, accurate channel characterization and modeling are essential for performance evaluation and optimization of VLC-IoE systems. Recent survey work systematizes VLC channel modeling methods (deterministic and statistical approaches), typical IoE scenarios (indoor, industrial, outdoor, underground, underwater), and emerging 6G combinations such as RIS-assisted VLC and ISAC, while outlining open research directions for future VLC-IoE deployments \cite{10609788}.

\vspace{0.2cm}

\subsubsection{Reconfigurable Intelligent Surfaces and Smart Radio Environments}
RIS is emerging as a key 6G enabler for IoE by shaping the propagation environment through software-controlled electromagnetic responses, thereby improving coverage and reliability in challenging deployments (e.g., dense indoor environments). In healthcare IoE scenarios, RIS is particularly attractive due to its low power consumption, high energy efficiency, and its ability to mitigate blockage by converting non-line-of-sight conditions into more favorable links \cite{wu2025_ris_nf_ff_ioe_healthcare}. Importantly, as carrier frequencies increase and RIS apertures become large, the operating regime may transition from conventional FF assumptions to NF conditions, which impacts both communication performance and localization accuracy. \cite{wu2025_ris_nf_ff_ioe_healthcare} proposes hybrid NF--FF channel models for RIS-assisted IoE that adaptively balance NF components (supporting high-precision positioning) and FF components (supporting wider monitoring coverage), enabling more robust system design across heterogeneous IoE service requirements.\vspace{0.2cm}

\noindent\textbf{Energy-efficient programmable surfaces:}
Beyond the general RIS concept, \emph{scalable intelligent/reflecting surfaces} have been highlighted as a practical 6G approach to improve link performance with \emph{minimal interference} and attractive energy properties. In particular, by keeping most elements effectively in a low-activity (sleep-like) mode and activating only a small subset when needed, these surfaces can deliver large-array gains while supporting energy-efficient operation---an important requirement for massive-scale IoE deployments \cite{babbar2023_massiveiot_ioe}.\vspace{0.2cm}

\noindent\emph{\textbf{Insight:}} 
6G connectivity for IoE is not only higher rate: scalability depends on access efficiency (grant-free, Non-Orthogonal Multiple Access (NOMA)), new propagation regimes (NF/FF), and integrated communication--compute control under stringent reliability constraints.

\subsection{Molecular Communication and IoBNT}
\label{subsec:IoBNT}

MC is an emerging paradigm in which information is conveyed through molecules, enabling communication at micro/nano scales and extending networked systems toward biological domains. Recent survey connects MC-driven IoBNT with the broader IoE vision and highlights bacterial systems as particularly relevant building blocks due to their intrinsic sensing, communication, and adaptation mechanisms \cite{10654347}.\vspace{0.2cm}

\noindent\textbf{Bacteria as bio-inspired IoE nodes and integrated sensing, communication, and computing:}
A useful bio-inspired perspective draws an analogy between a bacterium and an IoT device: pili can be viewed as transceiver-like structures supporting interaction/transfer, ribosomes resemble processing/memory units, bacterial receptors function as sensors, flagella act as actuators for mobility/control, and genetic material (genome/plasmids) resembles a control unit governing system behavior \cite{10654347}. As discussed in this work, integrated sensing, communication, and computing using bacteria can improve communication rate, reliability, and system performance, thereby opening new IoE research directions.\vspace{0.2cm}

\noindent\textbf{Communication primitives from bacterial mechanisms:}
Key bacterial mechanisms can be mapped to communication-network design: (i) Quorum Sensing (QS) provides threshold-based collective coordination using signaling molecules (autoinducers), suggesting bio-inspired protocol and signal-processing designs; (ii) chemotaxis exhibits run-and-tumble navigation under noisy gradients, inspiring adaptive routing/navigation strategies; and (iii) biofilm-mediated electrochemical signaling and conductive structures (e.g., nanowires) support more noise-resilient and potentially faster information transfer than diffusion-only models \cite{10654347}.\vspace{0.2cm}

\noindent\textbf{Open issues for scalable bio-nano IoE:}
\cite{10654347} also highlights open directions that align with IoE concerns: realistic modeling beyond oversimplified assumptions, protocol design inspired by QS, multimodal signaling models, energy-efficient communication strategies, and the need for reproducible testbeds and microfluidic experimental platforms.

\subsection{Federated Learning for Distributed Intelligence}
\label{subsec:FL_DI}

\noindent\textbf{Why FL fits massive IoE?}
FL is particularly well suited for massive IoE/MIoT deployments, as it enables collaborative intelligence while avoiding raw-data exchange, thereby aligning with privacy requirements and energy/bandwidth constraints at the network edge \cite{IEEE9667513,lim2020fl_men}. In this paradigm, multiple devices or edge nodes jointly train models without sharing local datasets, making FL highly attractive for privacy-sensitive and bandwidth-constrained IoE environments \cite{lim2020fl_men,fl6g_survey}. 
However, despite its advantages, FL introduces several system-level challenges. Its privacy guarantees can be compromised by inference and poisoning attacks, motivating the integration of defenses such as secure aggregation, differential privacy, and robust aggregation mechanisms \cite{mothukuri2021_fl_security_privacy}. Moreover, IoE environments are inherently heterogeneous, involving intermittent device participation, diverse communication conditions, and unequal computational capabilities, which complicate scalable deployment \cite{nguyen2021_fl_iot_survey}. In mobile edge settings, additional considerations such as system architecture design, communication efficiency, personalization, and robustness further influence FL performance \cite{lim2020fl_men}. 
Beyond the standard centralized FL paradigm, decentralized and cooperative FL approaches—based on D2D collaboration and consensus mechanisms—have been proposed to improve scalability and reduce single points of failure in massive IoE systems \cite{savazzi2020fl_cooperating}. Recent 6G-oriented perspectives further position FL as a key enabler of distributed intelligence across device--edge--cloud infrastructures \cite{fl6g_survey,saad2020vision6g}.\vspace{0.2cm}

\noindent\textbf{Federated learning vs.\ classical distributed learning: }
Unlike classical distributed learning, which typically assumes homogeneous resources and independent and identically distributed (i.i.d.)\ data distributions, FL must operate under realistic industrial conditions characterized by non-i.i.d.\ and imbalanced datasets, intermittent participation, and heterogeneous device/network capabilities. These factors directly affect convergence behavior, communication overhead, and overall system reliability in large-scale IoE deployments \cite{gulhane2025intro_fl_industry}. 
In industrial IoE scenarios—such as smart factories, healthcare systems, and energy networks—data are generated across diverse entities with different reliability levels, communication constraints, and feature distributions. To address this complexity, several FL taxonomies have been proposed to characterize deployment settings and data-partition structures, thereby guiding the design of suitable FL architectures for real-world IoE systems \cite{gulhane2025intro_fl_industry}.\vspace{0.2cm}

\noindent\textbf{Cross-device vs.\ cross-silo FL:}
From a deployment perspective, FL can be categorized into \emph{cross-device} FL (large populations of edge devices with sporadic participation) and \emph{cross-silo} FL (a smaller number of reliable entities such as organizations or institutions). This distinction is particularly relevant in IoE, as it reflects differences in scale, reliability, and communication constraints \cite{gulhane2025intro_fl_industry}.\vspace{0.2cm}

\noindent\textbf{Horizontal vs.\ vertical vs.\ federated transfer learning:}
From a data-partition perspective, \emph{horizontal FL} applies when clients share the same feature space but have different data samples, while \emph{vertical FL} applies when clients share the same sample space but have different feature sets. \emph{Federated transfer learning} further generalizes these settings by supporting scenarios where both samples and features differ across clients. This taxonomy is essential for mapping FL techniques to practical IoE scenarios involving multi-organization collaboration and heterogeneous sensing modalities \cite{gulhane2025intro_fl_industry}.\vspace{0.2cm}

\noindent\textbf{Canonical FL objective (global model from local objectives):}
A common formulation of FL aims to learn a shared parameter vector $w$ by minimizing an aggregate of local objectives, i.e., $ \min_{w} \sum_{k=1}^{K} p_k F_k(w)$, where $F_k$ denotes the local loss at client $k$ and $p_k$ reflects its relative contribution. This formulation highlights the importance of accounting for data heterogeneity, sampling bias, and client reliability in both optimization and orchestration within IoE systems \cite{gulhane2025intro_fl_industry}.\vspace{0.2cm}

\noindent\textbf{Communication overhead and heterogeneity in FL-aided IoE:}
At scale, FL becomes communication-intensive in bandwidth-constrained IoE environments, motivating the development of communication-efficient techniques such as model compression, reduced communication rounds, and split or federated split learning variants. In parallel, statistical heterogeneity (non-i.i.d.\ and imbalanced data) and system heterogeneity (unequal compute, energy, and connectivity across devices) can significantly slow convergence and destabilize training. Consequently, practical FL deployment in IoE requires robust aggregation methods, adaptive orchestration strategies, and intelligent client selection mechanisms to ensure scalability and reliability \cite{gulhane2025intro_fl_industry}.

\subsection{Blockchain / Distributed Ledger Technologies}
\label{subsec:Blockchain_DLT}

Trust, integrity, and accountability are important in multi-stakeholder IoE ecosystems. Recent studies \cite{ali2019blockchain_iot} show that blockchain and Distributed Ledger Technology (DLT) can support tamper-evident logging, decentralized authentication, and auditable coordination for IoT/IoE services, while also introducing overhead and scalability trade-offs.\vspace{0.2cm}

\noindent\textbf{What blockchain changes in IoE security (five concern areas):}
When blockchain is used to decentralize IoE, security concerns extend beyond classical IoT issues and can be grouped into: (i) communication and network security in P2P systems, (ii) identity management and authentication at massive scale, (iii) reliable distributed consensus under adversaries, (iv) decentralized cooperation and trust establishment, and (v) transaction-data privacy protection \cite{wei2019_ioe_blockchain}.\vspace{0.2cm}

\noindent\textbf{IoE-compatibility: full nodes vs lightweight nodes:}
Blockchain-based IoE deployments commonly differentiate \emph{full nodes} (store and synchronize the full ledger) from \emph{lightweight nodes} (store only block headers and verify transactions by querying neighbors), which aligns naturally with heterogeneous IoE devices with limited compute/storage \cite{wei2019_ioe_blockchain}.\vspace{0.2cm}

\noindent\textbf{Hardware-assisted blockchain for sustainable IoE security:}
A key limitation of blockchain in massive IoE is that classical consensus can be too heavy for constrained devices. PUFchain addresses this challenge by integrating the hardware security primitive Physical Unclonable Function (PUF) with a lightweight consensus mechanism, namely Proof of Possession (PoP). PUFs provide inherently unclonable device identities without requiring the storage of long-term secret keys, while PoP facilitates fast and energy-efficient block validation, rendering the framework suitable for IoE-scale authentication and data integrity \cite{pufchain2020}.

\subsection{Standards and Interoperability Mechanisms}
\label{subsec:SIM}
Large-scale IoE ecosystems involve heterogeneous devices, communication protocols, and service platforms deployed across multiple administrative domains. Ensuring secure and reliable operation therefore requires not only technical defense mechanisms but also standardized frameworks that enable interoperability, policy enforcement, and coordinated security management across vendors and infrastructures. 
As summarized in Table~\ref{tab:hybrid_vs_autonomous}, different security strategy families present complementary trade-offs between scalability, automation, and contextual decision-making capabilities.\vspace{0.2cm}

\noindent\textbf{Hybrid (human--AI collaborative) defenses:}
Hybrid IoE security combines automated monitoring and ML-based analytics with human supervision for contextual interpretation, escalation handling, and ethically sensitive decisions. This human-in-the-loop design can reduce false alarms and improve response quality when alerts are ambiguous or trade-offs (e.g., safety vs. availability) require expert judgment \cite{corradini2025ioe_cybersecurity_overview}.\vspace{0.2cm}

\noindent\textbf{Fully autonomous defenses:}
Fully autonomous IoE security aims to detect and mitigate threats with minimal human intervention by using adaptive ML (e.g., reinforcement learning) and closed-loop response policies. Decentralized components such as blockchain-like logging can further improve integrity and resilience by reducing single points of failure; however, autonomy raises concerns about computational overhead and handling unforeseen, ethically complex scenarios \cite{corradini2025ioe_cybersecurity_overview}. Interoperability remains critical due to device and protocol heterogeneity. Surveys \cite{lee2021standards_iot,alfuqaha2015iot} on standards for interoperability and security highlight the role of international standardization bodies, protocol stacks, and security frameworks that enable cross-vendor integration and systematic security assurance.\vspace{0.2cm}

\noindent\emph{\textbf{Insight:}} 
Trust mechanisms (DLT, SPbD, standards) can harden IoE ecosystems, but must be engineered for constrained nodes: lightweight identity, policy-aware orchestration, and interoperability-by-design are crucial for deployability.
\begin{table}[t]
\centering
\caption{Security strategy families in IoE: strengths vs. limitations.}
\label{tab:hybrid_vs_autonomous}
\setlength{\tabcolsep}{4pt}
\renewcommand{\arraystretch}{1.15}
\begin{tabularx}{\columnwidth}{|p{0.28\columnwidth}X|}
\hline
\rowcolor{Emerald!50}
\textbf{Family} & \textbf{Typical strengths / limitations} \\
\hline
\rowcolor{Emerald!10}
Hybrid human--AI & + Better contextual decisions and governance;\newline
-- operational burden and scalability limits when human review is heavy. \\
\rowcolor{Emerald!25}
Fully autonomous & + Fast response and scalable automation;\newline
-- ethical/context gaps and compute/resource overhead in constrained IoE nodes. \\
\hline
\end{tabularx}
\vspace{-0.3cm}
\end{table}

\begin{figure*}[t]
  \centering
  \includegraphics[height=10cm,width=2\columnwidth]{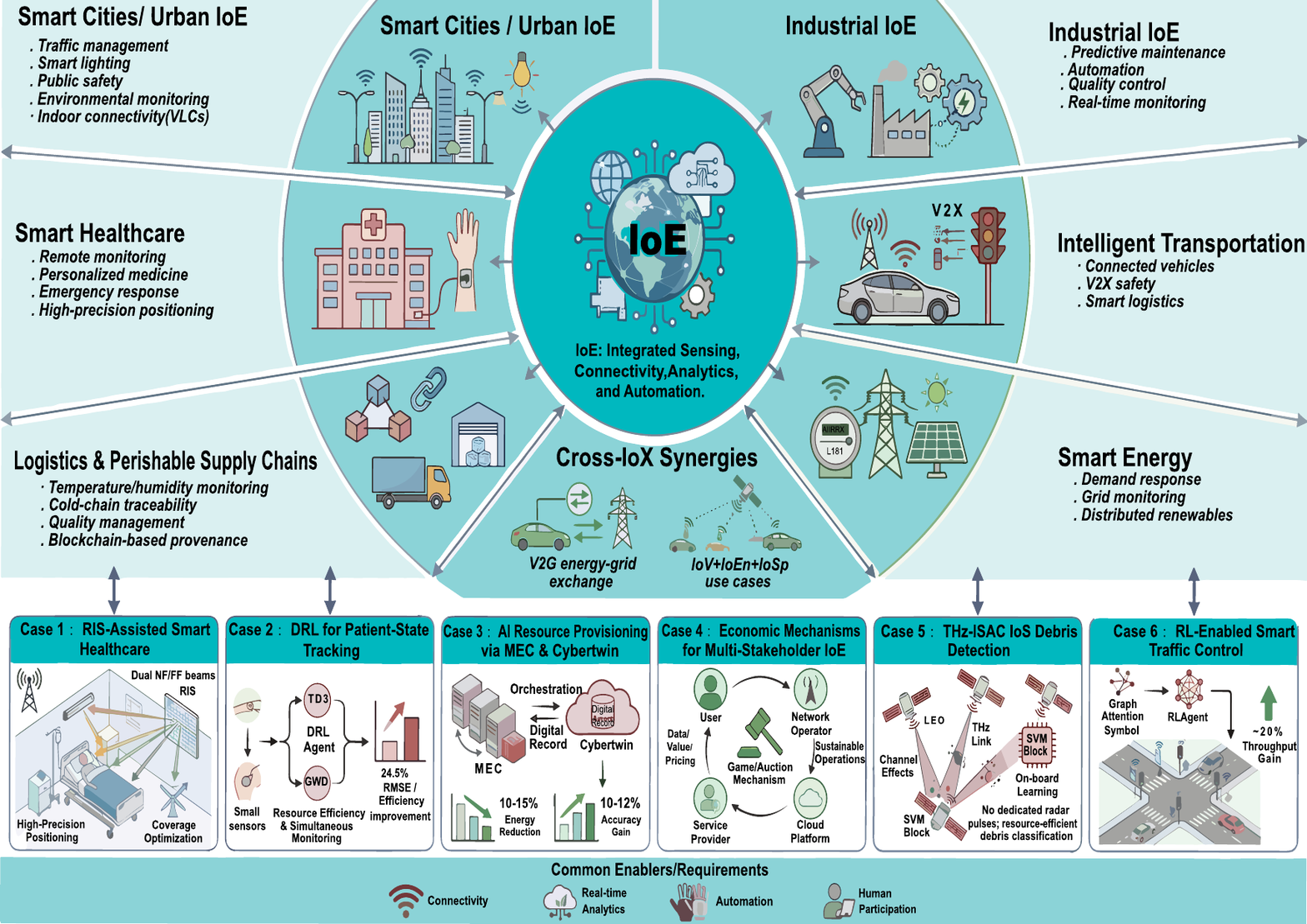}
  \caption{Main IoE smart applications and use cases.}
  \label{fig:IoE_Apps}\vspace{-0.3cm}
\end{figure*}

\section{IoE Applications and Use Cases}
\label{sec:apps}
IoE enables a broad range of applications, including smart cities, smart healthcare, industrial IoE, intelligent transportation, and smart energy. These applications benefit from a tight integration of sensing, connectivity, analytics, automation, and human participation \cite{cisco2012ioe,alfuqaha2015iot,saad2020vision6g}, as illustrated in Fig.~\ref{fig:IoE_Apps}. \vspace{-.1cm}

\subsection{Representative Application Domains}
\label{subsec:RAD}
\noindent The IoE paradigm spans a wide-spectrum of application domains, where heterogeneous entities—ranging from devices and infrastructure to humans and intelligent services—interact to enable data-driven, context-aware, and scalable solutions. These application domains illustrate how IoE integrates sensing, communication, computation, and intelligence across diverse environments, including urban, industrial, healthcare, and agricultural ecosystems. In this section, we present representative IoE application domains, highlighting their key functionalities, enabling technologies, and associated challenges in large-scale, real-world deployments.\vspace{0.2cm}

\noindent \textbf{Smart Cities (Indoor/Urban IoE) and Intelligent Transportation:} traffic management, smart lighting, public safety, environmental monitoring, and indoor connectivity. In addition to RF, optical wireless communications (e.g., VLC) are promising 6G enablers for dense indoor IoE due to high bandwidth, reduced electromagnetic interference, and spatial confinement benefits \cite{10609788}. Intelligent Transportation represents a core application domain of the IoE, enabling the seamless interconnection of vehicles, roadside infrastructure, cloud--edge platforms, and human stakeholders to support safe, efficient, and autonomous mobility services. IoE-enabled transportation systems facilitate connected and cooperative vehicles through Vehicle-to-Everything communications and cooperative perception mechanisms that fuse multi-source sensor data, including computer vision and positioning information, to enhance situational awareness and road safety. Advanced intelligence techniques, such as federated graph learning, further enable scalable and privacy-preserving knowledge sharing across distributed transportation entities under non-i.i.d. data conditions. In addition, IoE supports smart logistics, intelligent fleet management, and the integration of autonomous transportation within smart city ecosystems through edge--cloud orchestration. Despite these advantages, intelligent transportation IoE systems face challenges related to stringent latency requirements, data heterogeneity, security, and large-scale coordination across cyber--physical infrastructures~\cite{10291222,11015835,9305204}.\vspace{.2cm}

\noindent \textbf{Smart Healthcare:} It is a prominent application domain of the IoE, where heterogeneous entities, including medical sensors, wearable and implantable devices, patients, healthcare professionals, and intelligent platforms, are interconnected to enable data-driven healthcare services. IoE-based smart healthcare systems support continuous and context-aware remote patient monitoring, personalized medicine through real-time physiological data analytics, and proactive emergency response enabled by low-latency and high-reliability communications. Furthermore, high-precision positioning and monitoring services facilitate patient localization, medical asset tracking, and situation-aware clinical decision-making. These capabilities collectively enhance service reliability, operational efficiency, and quality of care, while introducing challenges related to scalability, interoperability, energy efficiency, and data security in large-scale smart healthcare ecosystems~\cite{10307968}.\vspace{.2cm}
    
\noindent \textbf{Industrial IoE (IIoE):} It constitutes a key application domain of the IoE, where industrial machines, robotic systems, enterprise workflows, and human operators are interconnected through intelligent automation and data-driven services. By integrating IoE with robotic process automation and generative AI, industrial environments enable predictive maintenance based on real-time equipment condition monitoring, automated and adaptive production processes, and digitalized quality control across manufacturing pipelines. Furthermore, real-time industrial monitoring supports resource optimization, waste reduction, and productivity enhancement through continuous sensing, intelligent decision-making, and closed-loop control mechanisms. Despite these advantages, Industrial IoE systems introduce challenges related to interoperability, ethical deployment of AI-driven automation, standardization, and secure orchestration of large-scale industrial services~\cite{10488437}.\vspace{.2cm}

\noindent \textbf{Smart Energy:} A fundamental application domain of the IoE, it enables the intelligent interconnection of energy generation units, storage systems, smart meters, buildings, grid infrastructures, and end users within smart city ecosystems. IoE-based smart energy systems support advanced demand response mechanisms, real-time grid monitoring, and predictive energy management through continuous data acquisition and analytics at the edge and cloud levels. The integration of distributed renewable energy sources is facilitated by IoE-driven coordination, semantic metadata management, and ML-based optimization. Furthermore, P2P energy trading enabled by blockchain and smart contracts allows decentralized, transparent, and energy-aware optimization of energy distribution and consumption. While these benefits are significant, smart energy IoE systems remain subject to important challenges related to interoperability, scalability, data integration, security, and the reliable operation of heterogeneous cyber--physical energy infrastructures~\cite{10803796,8539513,10205731,9803076}.\vspace{.2cm}
    
\begin{figure}[t]
  \centering
  \includegraphics[width=\columnwidth, height=6cm]{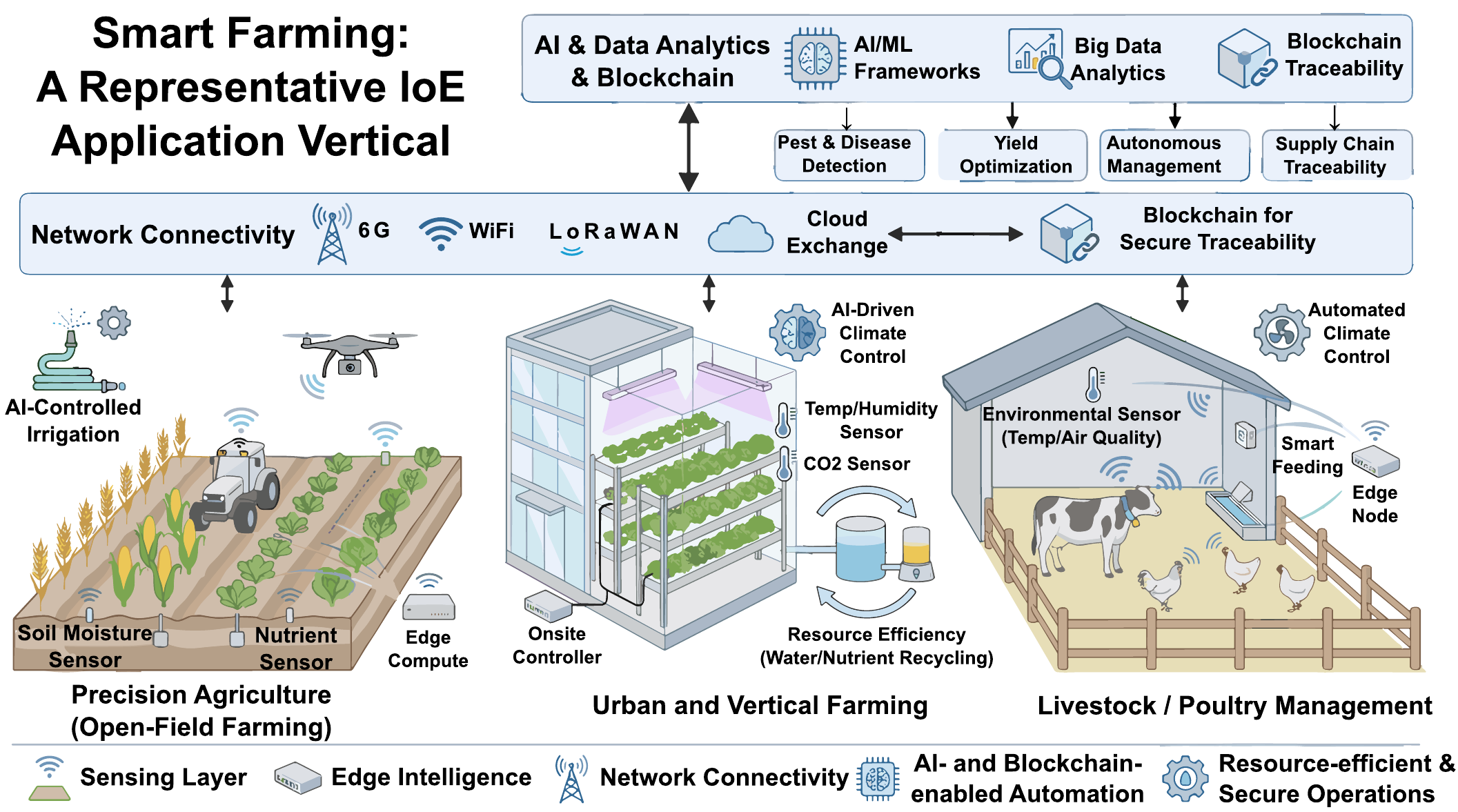}
  \caption{IoE-enabled smart farming ecosystem integrating AI, edge intelligence, 5G/6G connectivity, and blockchain for intelligent, automated, and secure agricultural operations. }
  \label{fig:IoE-smart-farming}\vspace{-0.3cm}
\end{figure}

\noindent \textbf{Smart Farming:} As illustrated in Fig.~\ref{fig:IoE-smart-farming}, smart farming constitutes a representative IoE application vertical where sensing, communication, intelligence, trust, and sustainability layers converge in real-world, large-scale deployments. Comprehensive surveys highlight how IoT-enabled sensing infrastructures, embedded systems, and wireless sensor networks enable real-time monitoring of soil moisture, microclimate conditions, irrigation processes, and crop health, forming the technological backbone of precision agriculture ecosystems \cite{8784034,8883163,9374808,9122412}. Low-cost IoT architectures further demonstrate the feasibility of scalable and energy-efficient agricultural monitoring systems \cite{9148455}. Beyond basic sensing, AI-driven automation frameworks increasingly integrate ML, deep learning, and computer vision techniques for intelligent irrigation scheduling, pest and disease detection, yield prediction, and land-cover classification \cite{9311735,10908635}. Edge-IoT-UAV integration with 3D-LiDAR point clouds further enhances precision agriculture through high-resolution crop monitoring and biomass estimation, enabling distributed edge intelligence across large farmlands \cite{10680916}. Emerging soilless and urban farming paradigms leverage IoT-based precision control to address food security challenges and resource efficiency in metropolitan environments \cite{10879001}.
Secure and trustworthy smart farming infrastructures are equally critical. Blockchain-enabled and FL-based architectures have been proposed to strengthen intrusion detection, privacy preservation, and data integrity in distributed agricultural systems \cite{10714358}. Broader security and privacy challenges in smart farming ecosystems, spanning edge-cloud architectures and cyber-physical integration, are extensively analyzed in \cite{9003290}. In parallel, long-range connectivity technologies such as Long Range facilitate sustainable, low-power wide-area agricultural communications aligned with Agriculture~4.0 visions \cite{10735371}.
Looking ahead, the convergence of IoT with digital twins and metaverse-enabled cyber-physical integration further expands the scope of agricultural IoE systems, enabling immersive monitoring, remote actuation, and data-driven decision support across physical and virtual domains \cite{10002946}. Collectively, these advances demonstrate how compute-continuum architectures, edge intelligence, secure distributed learning, resilient connectivity, and sustainable resource management must cohesively operate to support geographically distributed, energy-constrained, and economically sensitive agricultural environments. As such, smart farming serves as a concrete microcosm of the broader 6G-enabled IoE vision.\vspace{.2cm}

\noindent \textbf{Smart Logistics and Perishable Supply Chains:} It constitutes a critical application domain of the IoE by enabling robust connectivity among distributed sensors, transportation assets, edge--cloud platforms, and multiple supply chain stakeholders. IoE--enabled systems support real-time temperature and humidity monitoring across cold chains, maintaining required environmental conditions for perishable goods. Quality management platforms that integrate edge--cloud blockchain architectures provide authenticated, tamper-resistant provenance and trustworthy sharing of product quality data, enhancing transparency and stakeholder confidence throughout the logistics network. Furthermore, advanced IoE and digital twin service platforms facilitate cold chain traceability, anomaly detection, indoor localization, and efficient information integration to improve operational efficiency and end--to--end visibility across logistics processes. These integrated capabilities help optimize resource allocation, reduce spoilage, and strengthen collaboration across transportation, storage, and distribution tiers, while also posing challenges in data heterogeneity, latency management, and scalable interoperability~\cite{WU2023100443,yang2022_edgecloud_blockchain_supply_chain}.\vspace{.2cm}

\noindent \textbf{Cross-IoX applications enabled by synergies:}
IoE enables a new class of cross-domain applications emerging from the interaction among heterogeneous IoXs, where the resulting functionality goes beyond the capabilities of individual domains. For instance, the convergence of the IoV, IoEn, and Internet of Money can enable P2P vehicle charging supported by blockchain-based energy trading using digital tokens. In such systems, vehicles can dynamically exchange energy while leveraging decentralized financial mechanisms, improving energy efficiency and reducing infrastructure pressure. These interactions can be further enhanced by integrating the Internet of Space (IoSp), which provides global connectivity and real-time tracking of vehicle states (e.g., location and battery levels) to optimize matching between energy providers and consumers \cite{akan2023_ioe_molecules_universe}.
Beyond energy and transportation, cross-IoX synergies also enable novel applications spanning multiple scales and domains, such as bio-cyber interfaces combining the IoNT and Internet of People and Senses for intrabody health monitoring and human augmentation, as well as precision agriculture scenarios integrating Internet of Agricultural Things, IoV, and IoSp for large-scale monitoring, automation, and decision making. These examples highlight that IoE applications are fundamentally driven by the seamless cooperation of heterogeneous systems, where the interaction itself becomes the primary source of innovation and added value. This paradigm shift from isolated IoT verticals to synergistic IoX interactions is a defining characteristic of the IoE vision.

\subsection{Deep-Dive IoE Case Studies}
\label{subsec:Deep-Dive}
To further illustrate the practical realization of the IoE vision, this section presents a set of representative case studies spanning multiple domains. These case studies highlight how heterogeneous IoX components interact through sensing, communication, intelligence, and economic mechanisms to enable advanced, real-world applications. Rather than focusing on isolated use cases, the discussion is structured along key functional dimensions of IoE systems, including sensing and localization, edge intelligence and orchestration, economic optimization, networking, cyber--physical control, and security.

\subsubsection{Sensing, Localization, and Healthcare Intelligence}
IoE-enabled smart healthcare systems exemplify the tight integration of sensing, communication, and intelligence, where accurate localization and continuous monitoring must be jointly optimized under dynamic and resource-limited environments.\vspace{.2cm}

\noindent\textbf{RIS-Assisted Monitoring and Precision Positioning in Smart Healthcare: } 
Smart healthcare illustrates the dual-service nature of IoE: (i) \emph{high-precision positioning} for medical instruments and devices and (ii) \emph{reliable monitoring coverage} for patients across wards. With increasing carrier frequencies and larger reconfigurable intelligent surface apertures, deployments may operate in both NF and FF regimes, motivating hybrid NF--FF channel models with adaptive weighting to balance positioning accuracy and coverage performance \cite{wu2025_ris_nf_ff_ioe_healthcare}. This case study highlights how smart radio environments can jointly support connectivity and sensing/localization requirements in dense indoor IoE healthcare settings.\vspace{.2cm}

\noindent\textbf{DRL-Driven Patient-State Tracking and Resource Management in NF IoE Healthcare Systems: }
NF IoE healthcare monitoring introduces a tightly coupled sensing--communication--control problem, where maintaining accurate patient-state tracking requires continuous adaptation of sensing strategies and resource allocation under dynamic NF propagation conditions and multi-patient interactions. A representative framework, termed \emph{PPM-TD3-GWD}, the proposed framework integrates probabilistic Patient State Prediction Modeling (PPM) with a Twin-Delayed Deep Deterministic Policy Gradient (TD3) agent, wherein the reward function is defined based on the Gaussian Wasserstein Distance (GWD) between prior and posterior state distributions. This design quantifies information gain and guides sensing and positioning decisions toward the most informative measurements while constraining resource usage \cite{wang2025_drl_nf_ioe_healthcare}. Experimental results (including a diabetes monitoring scenario) demonstrate significant performance improvements in tracking accuracy, resource efficiency, and event detection compared to baseline approaches. Moreover, the framework scales effectively to multi-patient environments while maintaining robustness under network congestion \cite{wang2025_drl_nf_ioe_healthcare}.\vspace{.2cm}

\subsubsection{Edge Intelligence and Resource Orchestration}

Edge intelligence is a cornerstone of IoE systems, enabling real-time decision making and scalable service delivery by distributing computation and learning capabilities across the device--edge--cloud continuum. In such environments, efficient resource orchestration is critical to handle dynamic workloads, heterogeneous applications, and stringent QoS/QoE needs.\vspace{.2cm}

\noindent\textbf{AI-Based Resource Provisioning for Multi-Application IoE Services on MEC: } 
A fundamental challenge in large-scale IoE systems is the \emph{elastic provisioning} of edge resources for multiple concurrent applications with time-varying demands. In MEC clusters, services are typically deployed as containers across distributed worker nodes, requiring the orchestrator to jointly optimize \emph{service placement} and \emph{resource scaling} to prevent QoS/QoE degradation. A representative approach introduces an \emph{Intelligent Scaling and Placement} layer, where a DRL agent (IScaler) proactively determines scaling and placement decisions, while a heuristic optimizer acts as a bootstrapper and fallback mechanism during early learning stages \cite{sami2021_ai_resource_provisioning_ioe_6g}. The underlying MDP formulation is designed to ensure scalability by handling large input spaces with reduced memory overhead, and is evaluated using realistic traces from the Google Cluster Usage dataset to capture dynamic workload and resource conditions \cite{sami2021_ai_resource_provisioning_ioe_6g}.\vspace{.2cm}

\noindent\textbf{Cybertwin-Assisted Edge Provisioning: }
Cybertwin extends the digital-twin paradigm by introducing a cyber representation that acts as both a \emph{contact hub} and a persistent \emph{digital record} for IoE entities at the network edge. This abstraction facilitates intelligent task distribution and coordinated resource provisioning across edge--cloud infrastructures, addressing scalability and QoS challenges in 6G-enabled IoE systems \cite{adhikari2022_cybertwin_tii}. By enabling context-aware and adaptive orchestration, cybertwin-based frameworks can significantly reduce latency and improve energy efficiency. Reported results indicate energy consumption reductions of approximately 10--15\% compared to baseline methods, alongside prediction accuracy improvements of around 10--12\% in representative scenarios.\vspace{0.2cm}

\noindent\emph{\textbf{Insight:}} IoE orchestration requires tightly coupled optimization of computation, communication, and learning processes, where AI-driven mechanisms enable scalable and adaptive resource management across distributed edge infrastructures.

\subsubsection{Economic and Multi-Stakeholder Optimization}

IoE systems inherently involve multiple interacting stakeholders with heterogeneous objectives, making purely technical optimization insufficient. Instead, economic and game-theoretic frameworks play a central role in coordinating resource allocation, incentivizing participation, and ensuring sustainable system operation across the IoE ecosystem.\vspace{.2cm}

\noindent\textbf{Economic Mechanisms for Sustainable Multi-Stakeholder IoE Ecosystems: } 
Large-scale IoE deployments typically include sensing/data owners, connectivity providers, edge/cloud operators, and application service providers, each with distinct incentives and constraints. In such environments, economic mechanisms based on \emph{pricing}, \emph{incentive design}, and \emph{auction theory} can be leveraged to align individual stakeholder behavior with global system objectives, such as maximizing coverage, minimizing latency, and improving data fusion quality and service reliability \cite{ding2025_economic_ioe}. 
Recent studies structure these interactions along the IoE information pipeline, including \emph{information collection}, \emph{transmission}, \emph{fusion}, and \emph{service application}, where each stage introduces specific economic challenges and design requirements. To address these challenges, both \emph{game-theoretic approaches} (e.g., Stackelberg games, contract theory, and auction-based models) and \emph{learning-based mechanisms} (e.g., multi-agent reinforcement learning for dynamic pricing and incentive adaptation) have been proposed to operate under uncertainty and time-varying system conditions \cite{ding2025_economic_ioe}.\vspace{.2cm}

\noindent\emph{\textbf{Insight:}} IoE resource management is fundamentally a joint economic--technical optimization problem, where incentive-compatible and learning-driven mechanisms are essential to achieve efficient, fair, and scalable operation in multi-stakeholder ecosystems.

\vspace{0.2cm}

\subsubsection{Space and Non-Terrestrial IoE Systems}

The evolution of IoE toward 6G and beyond naturally extends connectivity and intelligence to non-terrestrial domains, giving rise to an IoSp where satellite systems jointly support communication, sensing, and global coordination across vast and heterogeneous environments.\vspace{.2cm}

\noindent\textbf{THz-ISAC for IoSp---Debris Detection and Classification: } 
As LEO satellite constellations become a key component of future IoE infrastructures, ensuring space situational awareness is critical for reliable and safe operations. In this context, ISAC has emerged as a promising paradigm to unify communication and environmental sensing functionalities within the same system. A representative framework, \emph{DebriSense-THz}, leverages THz communication signals to simultaneously perform data transmission and space debris detection, eliminating the need for dedicated radar systems \cite{dong2025_debrisense_thz_isac_twc}. 
The proposed approach employs an electromagnetic multi-ray channel model that captures key propagation phenomena, including \emph{reflection, scattering, and diffraction}, within a hybrid ray-tracing and Rician fading framework. By exploiting CSI variations under different debris densities and MIMO configurations, the system enables efficient debris detection and classification using communication signals alone \cite{dong2025_debrisense_thz_isac_twc}. This case study highlights how IoE systems can extend beyond terrestrial networks to enable joint communication--sensing capabilities in space environments.\vspace{.2cm}

\noindent\emph{\textbf{Insight:}} This example demonstrates that future IoE systems will operate over integrated terrestrial and non-terrestrial infrastructures, where ISAC-based designs enable efficient reuse of communication resources for sensing, thereby improving scalability, energy efficiency, and system sustainability.

\vspace{0.2cm}

\subsubsection{Smart Cities and Cyber-Physical Control}

Smart cities represent a key application domain of IoE, where large-scale cyber--physical systems integrate sensing, communication, and control to enable real-time, data-driven decision making in complex urban environments.\vspace{.2cm}

\noindent\textbf{Smart IoE-Integrated Traffic Control via Dynamic Graph Attention and Reinforcement Learning: } 
Traffic control in smart cities is a representative closed-loop IoE application, where heterogeneous roadside sensors and connected vehicles continuously generate real-time measurements that must be translated into timely control actions, such as adaptive traffic signal timing. A recent Smart IoE framework addresses this challenge by combining (i) traffic flow prediction using a dynamic multisemantic graph attention network and (ii) traffic signal control using an improved multi-agent Proximal Policy Optimization (PPO) algorithm, enabling prediction-aware and adaptive intersection management under time-varying urban conditions \cite{smart_ioe_traffic_control}. 
By jointly leveraging graph-based learning for spatial--temporal traffic modeling and reinforcement learning for control optimization, the framework effectively captures complex traffic dynamics and enables coordinated decision making across intersections. Reported evaluations demonstrate consistent performance gains over baseline methods, with throughput improvements on the order of $\sim$19--23\% under varying traffic conditions, highlighting the practical benefits of integrating IoE sensing, graph learning, and reinforcement learning for urban mobility optimization \cite{smart_ioe_traffic_control}.\vspace{.2cm}

\noindent\emph{\textbf{Insight:}} This case study illustrates that smart-city IoE systems operate as closed-loop cyber--physical systems, where the tight integration of sensing, prediction, and control through AI-driven mechanisms is essential to achieve efficient, adaptive, and scalable urban services.

\vspace{0.2cm}

\subsubsection{Security and Trust in IoE Systems}

Security and trust are fundamental enablers of IoE systems, where large-scale, heterogeneous, and data-intensive interactions across cyber--physical infrastructures significantly expand the attack surface and introduce new vulnerabilities.\vspace{.2cm}

\noindent\textbf{Illustrative Security Case Studies in IoE: }
Recent IoE security studies highlight the evolution of intelligent, scalable, and adaptive security mechanisms tailored to heterogeneous cyber--physical environments. These approaches span multiple layers of the IoE stack. First, \emph{ML-assisted IDSs} have been widely adopted in IoT/IoE-enabled critical infrastructures, such as smart grids and industrial systems, where deep learning models enable the detection of sophisticated cyber--physical attacks beyond the capabilities of traditional signature-based methods \cite{vinayakumar2020dl_ids_iot}. 
Second, \emph{forensic analysis and attack reconstruction} techniques leverage emulation platforms and realistic testbeds to capture the diversity of IoE devices, communication protocols, and traffic patterns, enabling detailed investigation and mitigation of complex attack scenarios \cite{corradini2025ioe_cybersecurity_overview}. Third, \emph{large-scale anomaly detection frameworks} are designed to process massive IoE telemetry streams in 6G environments, where edge-based and FL approaches enable distributed, privacy-preserving, and low-latency detection mechanisms across geographically dispersed systems \cite{nguyen2020federated_iot_security}. 
By combining edge intelligence, FL, and data-driven security analytics, these frameworks significantly enhance the resilience of IoE infrastructures against evolving threats while maintaining scalability and efficiency under high data volume and system heterogeneity.\vspace{.2cm}

\noindent\emph{\textbf{Insight:}} These case studies demonstrate a paradigm shift from static, rule-based security toward adaptive, learning-enabled, and distributed trust frameworks, which are essential to secure large-scale IoE systems operating under stringent latency, privacy, and scalability constraints. 


\section{ Key Trends Shaping Future IoE Architectures}
\label{sec:challenges}

\noindent
This section synthesizes the main \emph{cross-layer} challenges that arise when IoE systems scale from IoT-style connectivity
to heterogeneous, multi-stakeholder and intelligence-driven ecosystems. In particular, we highlight how scalability,
data management, security/privacy, energy sustainability, and ultra-low-latency reliability constraints become
\emph{mutually coupled}: improving one dimension often introduces new constraints or costs in the others. We detail these challenges through the subsections below, moving from general intensification/coupling to
security/trust, cloud federation and portability, scalability/heterogeneity, orchestration, sustainability, and
latency/reliability.

\subsection{Artificial General Intelligence for Autonomous IoE}

Future IoE systems are expected to incorporate increasingly autonomous
decision-making capabilities enabled by AI-native architectures.
Unlike traditional IoT analytics pipelines, which primarily rely on
centralized processing and task-specific models, emerging IoE
infrastructures require distributed intelligence capable of adapting
to dynamic environments, heterogeneous data streams, and large-scale
coordination across devices, edge nodes, and cloud services.
Recent 6G visions emphasize the role of AI-native networking and
learning-driven control loops to support real-time orchestration,
resource allocation, and service optimization in highly dynamic
environments \cite{saad2020vision6g}. 
At the architectural level, this evolution is closely tied to the
integration of edge intelligence and programmable network infrastructures,
where AI models are deployed across distributed nodes to enable low-latency,
context-aware decision-making while preserving data privacy and security
\cite{9144755}. In parallel, advances in communication technologies such as
grant-free access schemes, NOMA-based resource allocation, and integrated
satellite--aerial--terrestrial networks provide the scalable and reliable
connectivity substrate required for large-scale autonomous IoE operation
\cite{9906897,11176044}. These networking innovations are essential to
support continuous learning, coordination, and adaptation across highly
distributed environments.

In this context, the convergence of advanced ML techniques—
including graph learning, reinforcement learning, multi-agent systems, and foundation models—marks a transition toward more general-purpose intelligence within IoE ecosystems. Recent studies on foundation models for IoE demonstrate how generative and pre-trained architectures can enable unified forecasting and decision-making across heterogeneous
network telemetry, paving the way for more flexible and transferable intelligence at the edge \cite{marchetti2025foundation_forecasting_ioe}. Moreover, emerging paradigms such as human-centric cybertwins and composable digital twins introduce a semantic and contextual layer that allows IoE systems to model, predict, and interact with both physical and human environments in a more holistic manner \cite{9720181,10621154}. These developments collectively point toward a progressive shift from
narrow, task-specific intelligence to more general and adaptive decision mechanisms. While fully realized Artificial General Intelligence remains an open research challenge, its conceptual integration into IoE provides a useful abstraction for designing systems capable of reasoning, knowledge transfer, and cross-domain adaptation. In practice, this
translates into self-adaptive, self-optimizing, and resilient IoE infrastructures, where robustness, security, and trustworthiness are critical design requirements \cite{11165493}. Ultimately, such systems may evolve toward self-governing IoE ecosystems operating at a planetary
scale, supported by continuous learning and autonomous coordination across the device--edge--cloud continuum.

\vspace{0.2cm}

\paragraph{AI-Native Networking and Autonomous Networks}
Recent studies emphasize that future communication infrastructures are expected to evolve toward \emph{AI-native networking}, where AI becomes an integral component of the network architecture rather than an external optimization tool. In this paradigm, learning models are embedded across multiple layers of the communication stack to enable intelligent sensing, adaptive resource management, and closed-loop network control capable of operating in highly dynamic environments \cite{cui2025ai6g}. 
Such architectures support a wide range of autonomous network functions including traffic prediction, spectrum allocation, and dynamic service orchestration across distributed device--edge--cloud infrastructures. Earlier research has already demonstrated the potential of ML techniques to enable intelligent network management and self-optimizing communication systems \cite{jiang2017machine_learning_wireless}. In particular, deep reinforcement learning has been widely investigated for autonomous resource allocation and adaptive control in wireless networks \cite{ye2019deep_rl_v2v}. 
Furthermore, distributed decision-making frameworks based on multi-agent reinforcement learning provide a promising approach for coordinating large-scale network entities operating under heterogeneous and dynamic conditions \cite{zhang2021multiagent_rl_survey}. Combined with edge computing architectures that bring intelligence closer to data sources \cite{mao2017survey_mec}, these developments indicate a gradual evolution toward increasingly autonomous IoE infrastructures capable of self-configuration, self-optimization, and adaptive service orchestration across 
complex cyber--physical environments.

\vspace{0.2cm}

\paragraph{Toward Self-Running and AI-Native Communications}
Recent research increasingly envisions future communication infrastructures as fully autonomous systems capable of operating with minimal human intervention. The concept of \emph{self-running networks} has been proposed as an architectural paradigm in which intelligent control loops, data-driven analytics, and automated orchestration mechanisms continuously monitor, optimize, and adapt network behavior across complex environments \cite{shajarian2025self_running_networks}. In parallel, the emergence of \emph{AI-native network architectures} aims to embed AI directly into the design of communication systems, enabling learning-driven
decision mechanisms that dynamically manage resources, services, and network operations according to task-specific objectives \cite{wang2025ainative_architecture}. 
At the radio access level, these ideas are further extended through the vision of \emph{autonomous RAN}, where AI-powered agents perform tasks such as traffic prediction, spectrum management, and network configuration without continuous human supervision \cite{liu2025autonomous_ran_6g}.
Together, these developments indicate a broader transition toward self-evolving communication infrastructures in which distributed intelligence coordinates sensing, communication, and computing resources across large-scale IoE ecosystems.

\vspace{0.2cm}

\paragraph{Foundation Models and Large AI Models for Autonomous Networking}
Recent work, such as \cite{10599304}, suggests that the emergence of large-scale AI models and
foundation models may significantly transform the design and operation
of future communication networks. Unlike traditional ML
systems tailored to specific networking tasks, foundation models are trained on massive datasets and can generalize across multiple applications such as traffic prediction, anomaly detection, resource allocation, and network optimization. Recent surveys highlight how large AI models may serve as general-purpose intelligence layers for
future communication infrastructures, enabling networks to reason across heterogeneous data sources and dynamically adapt their operation in complex environments \cite{jiang2025large_ai_models_communication}.
In parallel, the concept of telecom or network foundation models has been proposed to leverage large-scale network telemetry data and pretraining techniques in order to support diverse networking tasks through a unified learning framework \cite{zanouda2024telecom_foundation_models}. \cite{aboulfotouh2024wireless_foundation_models} also explores the development of \emph{radio foundation models} based on transformer architectures that learn general representations of wireless channels and communication behaviors, potentially enabling adaptive and intelligent wireless systems in future 6G networks. Together, these developments indicate a transition toward more general-purpose and self-learning communication infrastructures, bringing networking systems closer to the vision of highly autonomous IoE ecosystems.

\subsection{Global Connectivity as the IoE Backbone}
Global connectivity forms the fundamental infrastructure enabling large-scale IoE ecosystems. 
Supporting massive-scale IoE requires coping with heterogeneous devices, communication protocols, 
and diverse data modalities while ensuring reliable and low-latency communication across highly 
distributed environments. Classic IoT surveys emphasize that heterogeneity and scale complicate 
system management, integration, and service composition \cite{atzori2010iot,alfuqaha2015iot}. 
Looking forward, 6G visions highlight that future networks must support massive connectivity, 
ultra-low latency, and high reliability simultaneously, requiring new architectures and AI-native 
control mechanisms capable of orchestrating communication resources at unprecedented scale 
\cite{saad2020vision6g,wang2023road6g}.

\vspace{.2cm}
\paragraph{Scalable initial access and bounded per-AP coordination in cell-free RANs}
Supporting massive device connectivity is a fundamental requirement of IoE systems, particularly in 6G environments where ultra-dense deployments and heterogeneous devices must be efficiently managed. In this context, scalable cell-free Radio Access Network (RAN) architectures have emerged as a promising solution to provide uniform service quality and flexible resource allocation for massive IoE access.
Scalable cell-free operation requires that per-AP processing and per-AP served-user cardinality remain bounded as the number of potential devices grows. A practical initial-access approach jointly assigns (i) a pilot and (ii) a serving-AP subset using a competition rule based on large-scale fading, under the constraint that each AP serves at most one UE per pilot. A blacklisting mechanism can prevent repeated conflicts and help ensure that even users in weak channel conditions are eventually assigned resources, which is important for fairness and reliable massive access \cite{chen2020wirelesspowered}.

\vspace{0.2cm}

\paragraph{Scalable power control for wireless-powered massive access (efficiency vs. fairness)}
In wireless-powered massive access, energy sustainability and throughput depend strongly on how downlink energy-transfer power and uplink transmit power are controlled. A scalable design pattern is user-centric operation where each AP serves only a bounded subset of devices and applies simple local power-control rules, keeping fronthaul and signal-processing complexity under control. Uplink fractional power control further enables a tunable fairness--efficiency tradeoff, which is useful when massive IoE devices experience highly uneven large-scale fading   \cite{chen2020wirelesspowered}.

\subsection{Edge Computing and Decentralized Intelligence}
Large-scale IoE deployments require coordinated orchestration across widely distributed device--edge--cloud infrastructures under strict QoS/SLA constraints. Compared with classical cloud scheduling, IoE orchestration must jointly handle (i) heterogeneous resources and workloads, (ii) tight end-to-end latency budgets, and (iii) limited backhaul capacity, making placement and scaling decisions significantly more complex at scale \cite{attaoui2023vnfcnf,vmp_multicriteria_survey,FGCS2024_ref}. Many IoE applications, including industrial automation, real-time control, and immersive services, demand ultra-low latency and high reliability. The tactile Internet literature highlights that achieving such responsiveness requires careful co-design across communication and computing infrastructures \cite{promwongsa2020tactile_survey}. Edge and fog computing architectures therefore play a critical role in reducing end-to-end delays compared with cloud-only processing \cite{premsankar2018edge_case,yu2017edge_iot}. In parallel, 6G roadmaps emphasize new physical-layer and network solutions capable of supporting URLLC-like operation at massive scale \cite{wang2023road6g,saad2020vision6g}. In the specific context of cloud-assisted remote sensing and sensing-as-a-service, additional system-level challenges arise beyond general connectivity and device heterogeneity. These include QoS/SLA-aware resource allocation for sensing workloads, scalability and dimensioning under fluctuating sensing demand, sensor-network virtualization and embedding (treating sensing sources as allocatable cloud resources), and inter-cloud interoperability and portability for migration and cross-cloud service composition \cite{abdelwahab2014cars}.\vspace{.2cm}
\begin{figure*}[t]
  \centering  \includegraphics[height=9cm,width=2\columnwidth]{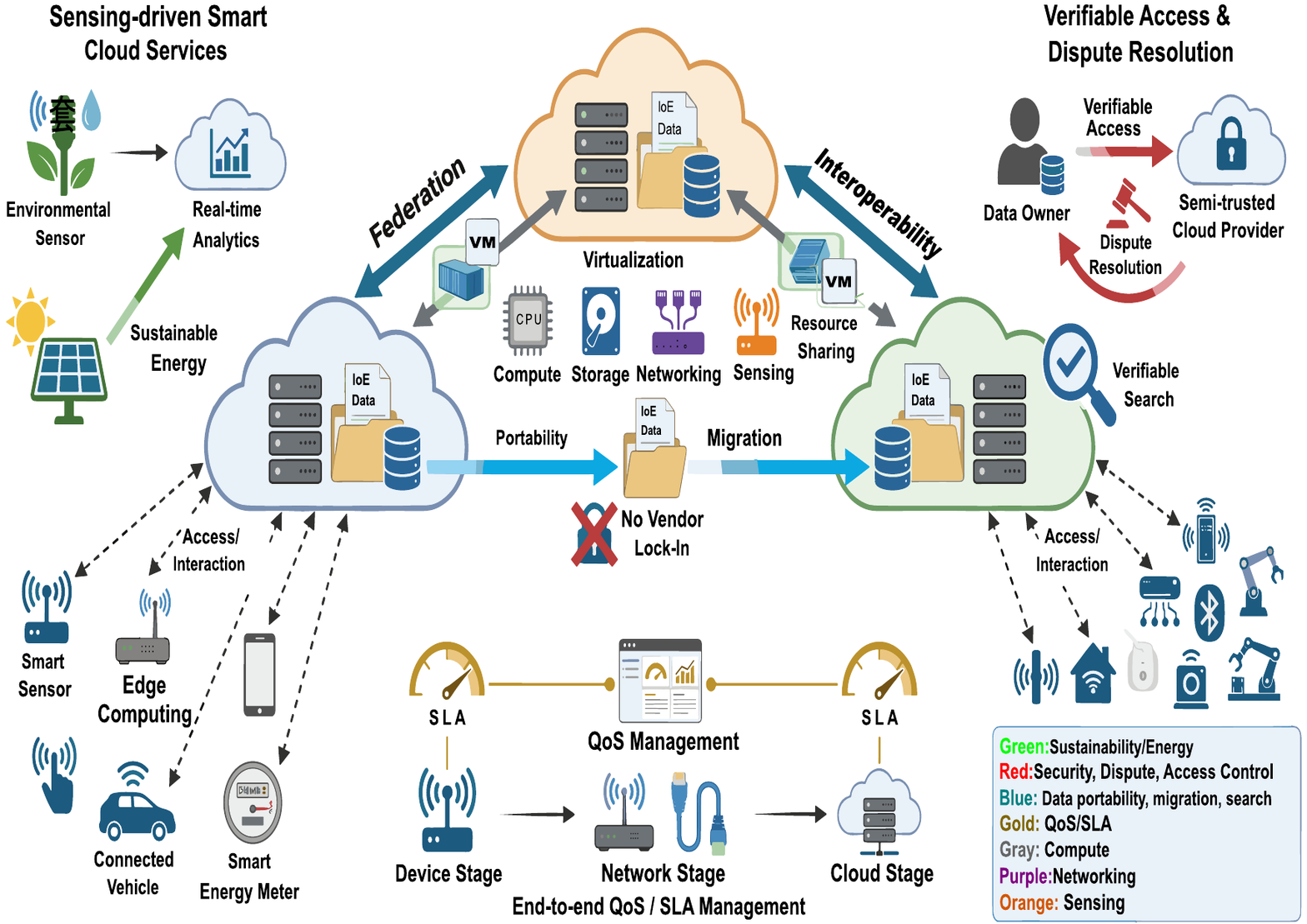}
  \caption{Reference multi-cloud federation architecture for IoE, highlighting inter-cloud portability, resource virtualization, and cross-layer QoS/SLA concerns.}
  \label{fig:ioe_multicloud_arch}\vspace{-0.35cm}
\end{figure*}

\paragraph{Why challenges intensify in IoE beyond IoT?}
The transition from the IoT to the IoE substantially amplifies many classical challenges. While IoT primarily focuses on connecting heterogeneous devices and sensors, IoE integrates a broader ecosystem of interconnected entities spanning people, processes, data, and diverse IoX verticals operating across multiple domains \cite{alfuqaha2015iot,zanella2014iot}. These heterogeneous systems differ not only in device technologies but also in their spatiotemporal scale, operating environments, and communication modalities—ranging from conventional electromagnetic wireless communications to emerging paradigms such as molecular and acoustic communications—and even in the material nature of connected entities (biotic vs. abiotic) \cite{akan2023_ioe_molecules_universe}.
Such heterogeneity significantly complicates interoperability, as semantic and syntactic inconsistencies arise across different IoX verticals, communication protocols, and data models. Recent studies emphasize that large-scale IoT/IoE environments exhibit substantial heterogeneity in devices, services, and data formats, creating major system-integration and interoperability challenges across distributed platforms \cite{iot_heterogeneity_survey_2023}. In parallel, the proliferation of interconnected sensors, services, and AI-enabled applications generates massive volumes of heterogeneous data streams that must be processed across the device–edge–cloud continuum. This trend increases the complexity of big-data management, context-aware analytics, and real-time decision-making in large-scale cyber–physical ecosystems, motivating the adoption of edge computing architectures that bring computation closer to data sources \cite{satyanarayanan2017edge}. Consequently, resource management becomes significantly more challenging, as future IoE infrastructures must dynamically coordinate heterogeneous communication, computing, and energy resources across distributed architectures in order to maintain QoS and scalability for cross-domain applications \cite{shi2016edge,iot_edge_cloud_data_2024}.
\vspace{0.2cm}

\paragraph{System-level requirements driving scalability}
Beyond protocol heterogeneity, massive-scale IoE is constrained by stringent latency requirements (often from milliseconds to tens of milliseconds), limited device energy/resources, and the risk of bandwidth/backbone congestion when raw data are streamed toward centralized clouds \cite{foe2017_fog_of_everything}. These constraints motivate distributed self-organization and cooperative processing (edge/fog and D2D), as well as adaptive workload placement to meet application QoS under dynamic connectivity \cite{foe2017_fog_of_everything}.\vspace{-0.05cm}

\subsection{Multi-Cloud Federation and Resource Virtualization in IoE}

The increasing scale and heterogeneity of IoE ecosystems require cloud infrastructures capable of supporting distributed services, resource sharing, and cross-domain interoperability. In this context, multi-cloud federation and resource virtualization emerge as key architectural principles that enable scalable orchestration across geographically distributed device--edge--cloud infrastructures.\vspace{.2cm}

Large-scale IoE systems increasingly rely on federated multi-cloud ecosystems, where multiple cloud providers collaboratively support distributed services and cross-domain applications. Fig.~\ref{fig:ioe_multicloud_arch} illustrates a representative federated architecture and highlights key cross-layer interactions across the device--edge--cloud continuum.
However, practical deployment introduces several challenges beyond local edge orchestration. First, \emph{inter-cloud interoperability} and \emph{data/service portability} are critical to avoid siloed IoE data and vendor lock-in. Second, the \emph{virtualization of heterogeneous IoE resources}—including compute, storage, networking, and sensing—is necessary to enable flexible resource sharing and service composition across providers. Third, ensuring end-to-end \emph{QoS/SLA management} across distributed infrastructures remains a major challenge, as performance depends on tightly coupled interactions between device, network, and cloud layers \cite{abdelwahab2014cars,FGCS2024_ref}.
These challenges are transversal across multiple IoE verticals. For instance, \emph{sensing-driven smart cloud services} require tight coordination between data acquisition at the edge and elastic processing in the cloud, where both data quality and system responsiveness jointly determine service performance \cite{abdelwahab2014cars}. Furthermore, cross-domain data sharing raises trust and accountability concerns, motivating mechanisms that enable verifiable access (e.g., verifiable search) and dispute resolution between data owners and semi-trusted cloud providers in cloud-assisted IoE environments \cite{miao2019fairdyndsf}.\vspace{.2cm}

\noindent\emph{\textbf{Insight:}} Multi-cloud federation in IoE is not only a scalability enabler but also a key challenge, requiring unified frameworks for interoperability, resource abstraction, and trust-aware service management across heterogeneous domains.

\subsection{Edge-Orchestration and Scalability Challenges}
Beyond architectural considerations such as multi-cloud federation and resource virtualization, IoE infrastructures must also address the practical challenges of orchestrating services across highly distributed edge environments. In particular, the large scale, heterogeneity, and strict QoS/SLA requirements of IoE applications introduce significant complexity in service placement, resource allocation, and orchestration mechanisms.

\vspace{0.2cm}

\paragraph{Computational hardness of edge orchestration}
Many edge-NFV orchestration problems that jointly decide routing and VNF placement with latency constraints are NP-hard, motivating heuristic and approximation-based solvers rather than exact optimization at scale. For example, practical approaches use a two-stage design: first generating feasible paths under constraints, then assigning VNFs along the path to maximize instance reuse while controlling bandwidth and delay, with provable performance bounds and topology-based evaluations \cite{infocom20_edge_nfv}.

\vspace{0.2cm}

\paragraph{VNF/CNF coexistence and multi-objective trade-offs}
In 5G/6G IoE infrastructures, network functions may be deployed as VM-based VNFs or containerized CNFs, and supporting their coexistence increases lifecycle and placement complexity. Moreover, placement is typically multi-objective, balancing latency, energy consumption, number of active instances/nodes, and resource utilization under QoS/SLA constraints, which forces explicit trade-offs \cite{attaoui2023vnfcnf,vmp_multicriteria_survey}.

\vspace{0.2cm}

\paragraph{Scalability and orchestration platforms}
Most placement formulations are NP-hard, so exact optimization does not scale; practical systems rely on heuristics, meta-heuristics, or learning-based controllers, each with limitations under dynamics and uncertainty. In addition, while Kubernetes supports CNF management, it is not designed by default for widely distributed telco edge sites where network-awareness and strict latency/SLA enforcement must be first-class inputs, motivating enhanced placement and decentralized control mechanisms \cite{attaoui2023vnfcnf}.

\subsection{Sustainability and Green IoE Architectures}

Sustainability and energy efficiency have become fundamental design objectives for large-scale IoE deployments. 
Many IoE devices operate under strict energy constraints due to limited battery capacity or energy-harvesting 
capabilities, making long-term autonomous operation a critical challenge. Surveys on energy-harvesting IoT systems 
highlight sensing, computing, and communication strategies aimed at extending device lifetime and reducing 
maintenance requirements, while also revealing trade-offs associated with harvesting uncertainty, workload 
scheduling, and protocol design \cite{ma2020eh_iot}. In parallel, emerging 6G visions emphasize energy efficiency 
as a key requirement for supporting massive-scale intelligent services in a sustainable manner 
\cite{saad2020vision6g}.

\vspace{0.2cm}

\paragraph{Multi-metric optimization in IoE-6G} Recent MIoT literature emphasizes that key system dimensions—including \emph{network scalability}, \emph{data management}, \emph{security and privacy}, and \emph{energy efficiency}—are tightly coupled in real deployments; improving one dimension often introduces constraints or costs in others. This interdependence motivates holistic, cross-layer solutions rather than siloed optimizations \cite{IEEE9667513,FGCS2024_ref}. In IoE-enabled 6G deployments, energy efficiency is further intertwined with other system-level objectives such as latency, coverage, and localization accuracy. Hybrid evolutionary optimization approaches can address such multi-objective trade-offs in complex resource-allocation settings. For instance, the solution in \cite{singh2024eehea} combines leader-based optimization with adaptive differential evolution to improve energy-aware
allocation while demonstrating gains across energy expenditure, latency, coverage, and localization metrics.

\vspace{0.2cm}

\paragraph{Freshness--energy tradeoff via Age--Energy Efficiency}
Classical energy-efficiency metrics include battery lifetime, Joules/bit, and energy-per-inference, but they do not capture the timeliness of status information. Beyond classical energy-per-bit metrics (e.g., Joules/bit, battery lifetime), real-time IoE and IIoE systems must also preserve
\emph{information freshness}, commonly captured by the Age of Information (AoI). Recent work studies the \emph{Long-Term Age--Energy Efficiency (AEE)} in wireless-powered industrial IoE status-update networks, explicitly optimizing the tradeoff between timely updates and energy consumption in dynamic wireless environments
\cite{zheng2024_aee_dldqn}.

\vspace{0.2cm}
\paragraph{Renewable energy and smart-grid-enabled IoE}
Early visions of the IoT already highlighted the importance of energy-efficient device operation and scalable sensing infrastructures to support large numbers of interconnected devices \cite{borgia2014_iot_vision}.  Energy harvesting technologies—including solar, vibration, thermal, and RF harvesting—have been widely explored to enable long-term autonomous operation of distributed IoE devices without frequent battery replacement, particularly in large-scale sensing deployments and industrial environments \cite{ma2020eh_iot,sanislav2021energy_harvesting_iot}. Numerous late studies further investigate the practical feasibility of energy-harvesting-powered IoT systems and quantify the available energy resources in real deployment scenarios, providing models and empirical analyses for designing self-powered sensing infrastructures \cite{vanleemput2023eh_iot,beach2023kinetic_eh_iot}. At the infrastructure level, smart grids play a critical role in supporting green IoE ecosystems by enabling bidirectional energy management, demand-response mechanisms, and intelligent coordination between distributed energy resources and communication networks \cite{sangoleye2021energy_harvesting_iot}. Through the integration of sensing, communication, and data analytics, smart grids allow IoE systems to dynamically optimize energy production, storage, and consumption across heterogeneous cyber--physical infrastructures. These developments contribute to the broader vision of \emph{green IoE architectures}, where renewable energy integration, energy-aware resource allocation, and intelligent workload placement jointly minimize the environmental footprint of large-scale digital infrastructures while maintaining performance and reliability for future 6G services.

\subsection{IoE--Metaverse Convergence: Physical--Digital Continuum}

The convergence of IoE infrastructures and metaverse platforms is
expected to create tightly coupled physical--digital ecosystems in
which real-world sensing, actuation, and control are continuously
synchronized with immersive virtual environments. From now on, IoE devices and edge platforms provide real-time data streams that feed digital twins and immersive interfaces, enabling applications such as remote industrial control, collaborative robotics, smart-city visualization, and immersive training systems. Achieving this physical--digital continuum requires integrating ultra-low-latency communication, edge computing, scalable data
pipelines, and semantic-aware communication mechanisms capable of
maintaining consistency between the physical environment and its
virtual representation \cite{li2025_6g_isci_review}. Henceforth,
emerging digital twin systems, AI-driven edge architectures,
and semantic communication frameworks are increasingly viewed as key
enabling technologies for linking large-scale IoE infrastructures with
immersive metaverse environments. Consequently, IoE–metaverse
integration represents a promising direction for future cyber–physical
services supported by 6G infrastructures.

\vspace{.2cm}
\paragraph{Digital Twin for IoE-Enabled Cyber--Physical Systems}
Recent research highlights digital twin networks as a key enabling technology for tightly coupling physical infrastructures with their virtual counterparts in IoE systems. In the IoE context, digital twin networks integrate heterogeneous data streams originating from distributed IoX entities—including sensors, devices, and cyber--physical assets—to replicate the behavior, state, and operational dynamics of physical systems in real time \cite{lin2022digital_twin_6g,ahmadi2022digital_twin_6g,duong2023digital_twin_metaverse}. 
By combining sensing data, communication networks, and computing resources across the device--edge--cloud continuum, these architectures create synchronized digital replicas of IoE environments. This enables real-time monitoring, predictive analytics, and adaptive control of large-scale IoE cyber--physical systems. Consequently, operators can simulate network conditions, anticipate faults, and optimize resource allocation before applying changes to the physical system, thereby improving reliability, efficiency, and operational resilience.

\vspace{0.2cm}
\paragraph{AI-Driven Digital Twin for Intelligent IoE Services}
Beyond basic digital replication, recent studies emphasize the integration of artificial intelligence and edge computing into digital twin architectures to enable intelligent and autonomous IoE systems. In this context, Digital Twin Edge Networks extend the digital twin concept to distributed IoE edge infrastructures, where localized digital replicas of network entities support real-time decision making and resource optimization closer to data sources \cite{tang2022diten_6g}. 
AI-enabled digital twin networks further enhance IoE operations by leveraging ML models to analyze large-scale IoE telemetry, predict system behavior, and dynamically adapt communication and computing resources across heterogeneous environments \cite{sheraz2024ai_digital_twin_6g}. These capabilities are particularly relevant for IoE ecosystems integrating physical and virtual domains, where continuous interaction between sensing, analytics, and control enables adaptive and scalable service provisioning.

\vspace{0.2cm}
\paragraph{Semantic and Task-Oriented Communication for IoE}
Semantic communication is emerging as a key paradigm for future IoE systems, where massive volumes of heterogeneous data are generated by distributed IoX entities. Unlike conventional communication approaches that focus on transmitting raw bits, semantic communication prioritizes the transmission of task-relevant information, thereby improving efficiency and reducing communication overhead \cite{lu2023rethinking_semantic_comm}.  In IoE environments, task-oriented communication frameworks enable intelligent information exchange by transmitting only the data required to support specific applications, such as control, monitoring, or decision making \cite{shi2023task_oriented_comm}. These approaches are particularly beneficial for large-scale IoE deployments, where bandwidth, latency, and energy constraints require efficient data representation and transmission. By integrating semantic reasoning with sensing and AI-driven decision making, IoE systems can support complex cyber--physical applications while maintaining scalability and efficiency \cite{sagduyu2024semantic_6g,park2023semantic_metaverse}.  Together, digital twin infrastructures, AI-driven edge intelligence, and semantic communication frameworks constitute key building blocks of future IoE systems, enabling tight integration between physical and digital environments through real-time sensing, learning, and control across distributed IoX domains.

\subsection{Security and Privacy-by-Design in IoE Systems}
\begin{table}[t]
\centering
\caption{IoE cybersecurity challenges frequently reported in the literature.}
\label{tab:ioe_cyber_challenges}
\setlength{\tabcolsep}{4pt}
\renewcommand{\arraystretch}{1.15}
\begin{tabularx}{\columnwidth}{|p{0.3\columnwidth}X|}
\hline
\rowcolor{Emerald!60}
\textbf{Challenge} & \textbf{Why it is hard in massive IoE} \\
\hline
\rowcolor{Emerald!10}
Scalable authentication & Heterogeneous endpoints and constrained devices make strong authentication expensive at scale. \\

\rowcolor{Emerald!25}
Security-aware resource allocation & Schedulers must incorporate threat intelligence while meeting latency/QoS constraints. \\
\rowcolor{Emerald!10}
Outdated network architectures & Legacy stacks struggle with multipath routing, scalability, and latency requirements. \\
\rowcolor{Emerald!25}
Intrusion detection at scale & Large attack surface (e.g., smart grids) motivates ML-based detection with context (node roles/workload). \\
\rowcolor{Emerald!10}
Forensics in dynamic environments & Device/protocol diversity complicates evidence collection and post-incident analysis. \\
\rowcolor{Emerald!25}
Real-time constraints & Security must operate under low latency and limited compute/energy budgets. \\
\rowcolor{Emerald!10}
Ethical/context limits of automation & Fully automated defenses may miss contextual/ethical nuances without human oversight. \\
\hline
\end{tabularx}
\vspace{-2mm}
\end{table}
The above discussion of IoE security challenges follows common themes highlighted in recent IoE cybersecurity syntheses \cite{corradini2025ioe_cybersecurity_overview}. IoE enlarges the attack surface by connecting massive numbers of heterogeneous and resource-constrained devices. A widely cited survey \cite{sicari2015iot_security} identifies persistent challenges spanning authentication, access control, confidentiality, key management, and trust establishment, and emphasizes that security must be addressed end-to-end across devices, networks, and services \cite{sicari2015iot_security}. 
These challenges are further summarized in Table~\ref{tab:ioe_cyber_challenges}, which highlights key scalability, architectural, and operational security issues in massive IoE systems. Standardization efforts also remain essential to ensure interoperability and security alignment across heterogeneous ecosystems \cite{lee2021standards_iot}. 
When IoE systems rely on distributed intelligence mechanisms such as FL, which are inherently privacy-presing by design, additional challenges may still arise. In particular, information leakage from model updates, poisoning attacks, and unreliable participants can compromise both privacy and system robustness \cite{lim2020fl_men,fl6g_survey}. Finally, learning-based anomaly detection is increasingly adopted as a complementary defense layer to identify abnormal behaviors and cyber threats in large-scale IoE environments \cite{djenouri2023_emergent_dl_anomaly_ioe}.\vspace{0.2cm}

\noindent\textbf{Why cybersecurity challenges intensify in IoE? }
In IoE, cybersecurity becomes more complex than in classical IoT because the ecosystem couples heterogeneous devices and protocols with data-driven processes and human-in-the-loop interactions, expanding the attack surface and stressing real-time operation under tight resource constraints.
As a result, conventional security mechanisms often fail to scale, and IoE deployments require  security-aware architectures and adaptive defenses that remain effective under dynamic workloads \cite{corradini2025ioe_cybersecurity_overview,IEEE11358823,IEEE9667513}.
\cite{corradini2025ioe_cybersecurity_overview} synthesizes IoE-specific cybersecurity challenges and organizes solutions into two families: hybrid human--AI collaborative defenses and fully autonomous security frameworks, while also highlighting open gaps in scalability, real-time constraints, and ethical governance.\vspace{0.2cm}

\noindent\textbf{Privacy leakage from model updates and practical solutions: }
Although FL avoids sharing raw data, private information can still be inferred from shared gradients/updates in some settings. Common mitigation directions include differential privacy (adding controlled noise with an accuracy--privacy tradeoff) and cryptographic protections such as homomorphic encryption or secure multi-party computation for secure aggregation, at the cost of additional computation/communication overhead. Hence, privacy in FL-enabled IoE is best treated as an end-to-end system design problem rather than a guaranteed byproduct of data locality \cite{gulhane2025intro_fl_industry}.\vspace{0.2cm}

\noindent\textbf{Governance and compliance challenges in industrial FL/IoE: }
Beyond technical security, industrial FL deployments raise governance questions related to data ownership, consent,
data sovereignty, regulatory compliance, and accountability when multiple stakeholders collaborate. These issues are amplified
in IoE because stakeholders (device owners, operators, and service providers) may have conflicting objectives, and attacks such
as poisoning or privacy leakage can create both safety and legal consequences \cite{gulhane2025intro_fl_industry}.\vspace{0.2cm}

\begin{figure*}[t]
  \centering
  \includegraphics[width=1.6\columnwidth]{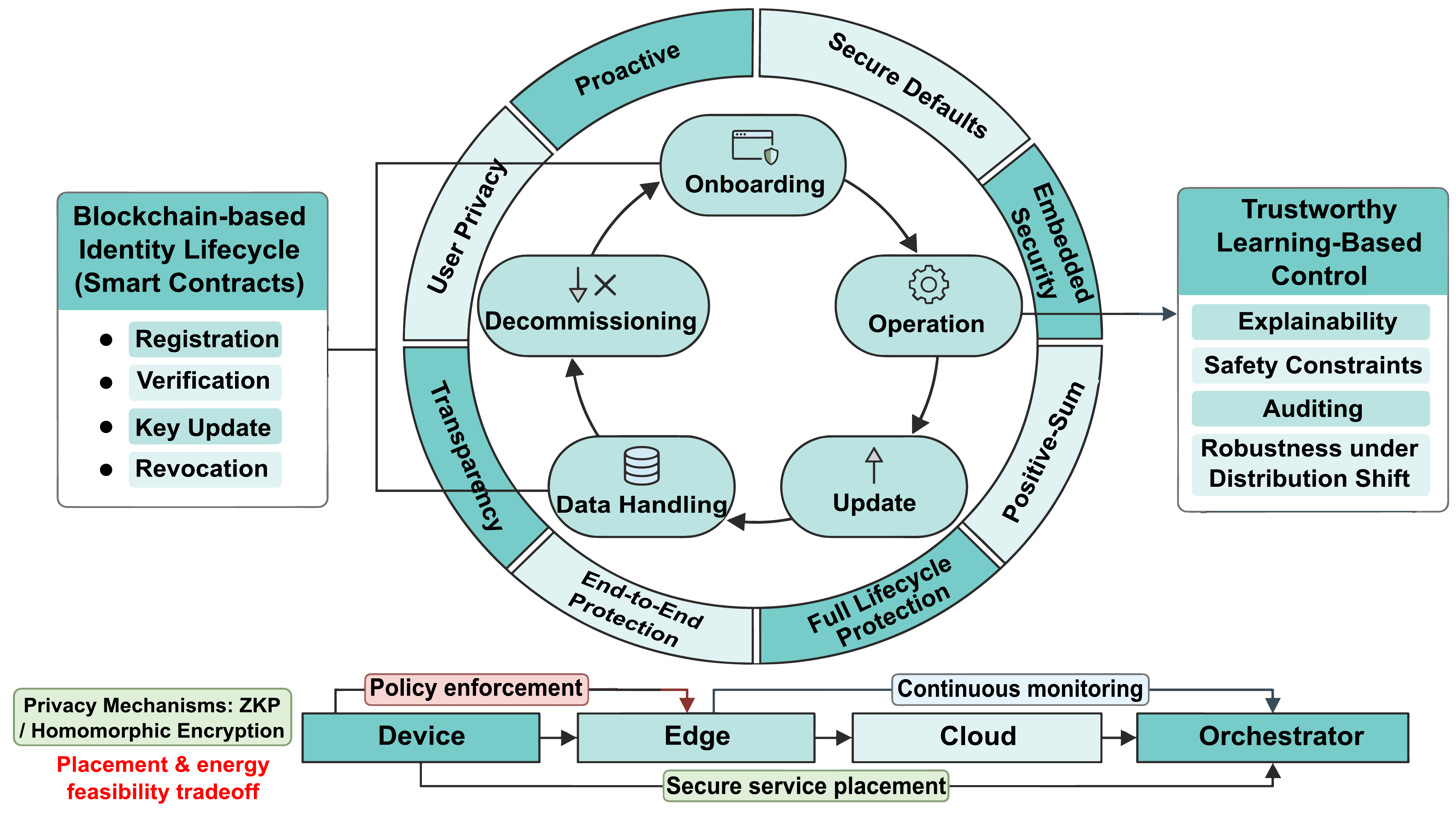}
  \caption{Security and Privacy by Design across the IoE lifecycle and the device--edge--cloud continuum, illustrating lifecycle-integrated controls, policy-aware orchestration, identity management, and trustworthiness mechanisms for learning-driven systems.}
  \label{fig:spbd_ioe_arch}\vspace{-.4cm}
\end{figure*}

\subsubsection{Security and Privacy by Design for IoE}

A widely accepted principle in IoE system design is that security and privacy should be integrated from the outset rather than introduced as an afterthought \cite{spbd_ioe}. The SPbD/PSbD paradigm promotes 
proactive and embedded protection mechanisms throughout the system lifecycle. Its key principles 
\begin{table}[t]
\centering
\caption{Security risks and typical defenses in blockchain-enabled IoE (condensed).}
\label{tab:blockchain_ioe_risks}
\small
\setlength{\tabcolsep}{3pt}
\begin{tabular}{|p{1.75cm}|p{3.2cm}|p{3.2cm}|}
\hline
\rowcolor{Emerald!50}
\textbf{Layer} & \textbf{Attack examples} & \textbf{Typical defenses} \\
\hline
\rowcolor{Emerald!10}
Physical & jamming, device capture, firmware replacement & state detection, recovery mechanisms \\
\hline
\rowcolor{Emerald!25}
Network or Transport & eavesdropping, spoofing, DoS, byzantine behaviors & lightweight crypto, access control, ZKP/HE where feasible \\
\hline
\rowcolor{Emerald!10}
Application & sybil/forgery, selfish behaviors & admission control, lightweight auth, reputation based on ledger data \\
\hline
\end{tabular}
\end{table}
include: (i) proactive rather than reactive security, 
(ii) secure and privacy-preserving default configurations, 
(iii) security embedded directly into system architecture, 
(iv) positive-sum design that avoids trade-offs between functionality and protection, 
(v) end-to-end lifecycle protection, 
(vi) visibility and transparency, and 
(vii) respect for user privacy \cite{spbd_ioe}. 
Fig.~\ref{fig:spbd_ioe_arch} illustrates how these principles can be mapped 
across the IoE lifecycle and the device--edge--cloud continuum. 
In parallel, Table~\ref{tab:blockchain_ioe_risks} summarizes representative 
security risks across IoE layers along with typical defense mechanisms, 
highlighting the need for cross-layer protection strategies.\vspace{0.2cm}

\noindent\textbf{SPbD mapped to the IoE lifecycle: }
For IoE, SPbD is most actionable when mapped to lifecycle phases: onboarding (secure identity and keying), operation (least privilege and continuous monitoring), update (secure patching and rollback), data handling (minimization and purpose limitation), and end-of-life (secure decommissioning and data deletion). As summarized in Table~\ref{tab:spbd_lifecycle}, this lifecycle mapping clarifies where security controls must exist along the device--edge--cloud continuum rather than only at endpoints \cite{spbd_ioe}.\vspace{0.2cm}

\noindent\textbf{Trustworthiness and accountability of learning-driven control: }
When IoE decisions are produced by graph learning and RL (e.g., traffic signal control), practical deployment also requires explainability, safety constraints, and auditing to ensure that learned policies remain robust under distribution shifts and do not violate operational policies \cite{smart_ioe_traffic_control,spbd_ioe}.\vspace{0.2cm}

\begin{table}[t]
\centering
\caption{Operationalizing SPbD across the IoE lifecycle.}
\label{tab:spbd_lifecycle}
\small
\setlength{\tabcolsep}{3pt}
\renewcommand{\arraystretch}{1.15}
\begin{tabularx}{\columnwidth}{|p{0.28\columnwidth}X|}
\hline
\rowcolor{Emerald!50}
\textbf{Lifecycle phase} & \textbf{Concrete SPbD actions (use cases)}\\
\hline
\rowcolor{Emerald!10}
Onboarding & secure identity provisioning; mutual authentication; secure defaults; least-privilege roles\\
\rowcolor{Emerald!25}
Operation & continuous monitoring/anomaly detection; auditable logging; policy enforcement at edge/orchestrator\\
\rowcolor{Emerald!10}
Update & secure OTA updates; signed firmware; rollback support; vulnerability response playbooks\\
\rowcolor{Emerald!25}
Data handling & data minimization; encryption in transit/at rest; access control; privacy-aware analytics/aggregation\\
\rowcolor{Emerald!10}
Decommissioning & secure wipe; credential revocation; data deletion verification; reuse-safe device reset\\
\hline
\end{tabularx}
\end{table}

\noindent\textbf{How to operationalize SPbD in 6G-scale IoE: }
In practice, these principles translate into secure-by-default onboarding/authentication, least-privilege access control, secure updates and decommissioning, auditable logging, and data-minimization in edge--cloud pipelines, alongside policy-aware orchestration that enforces security constraints during service placement and scaling \cite{spbd_ioe}.\vspace{0.2cm}

\noindent\textbf{Operationalizing SPbD with decentralized identity: }
A practical SPbD mechanism in blockchain-enabled IoE is to manage the \emph{device identity lifecycle} via smart contracts, covering registration (type/manufacturer/expiration/public key), verification, secure updates (e.g., firmware/keys), loss reporting, and obsolescence. This leverages blockchain immutability and automated execution to reduce identity forgery and lower trust-establishment costs compared with purely CA-based approaches \cite{wei2019_ioe_blockchain}. \vspace{0.2cm}

\noindent\textbf{Privacy vs. feasibility on constrained nodes: }
While zero-knowledge proofs and homomorphic encryption can protect transaction privacy in public ledgers, their extra computation overhead may be prohibitive for resource-constrained IoE devices, making feasibility
(and where to place these functions: device vs. edge/gateway vs. cloud) a key design decision \cite{wei2019_ioe_blockchain}.\vspace{0.2cm}

\subsubsection{LLM-Driven Security for IoE/IoEn (Edge--Cloud IDS)}

\begin{figure}[t]
  \centering
  \includegraphics[width=\columnwidth, height=6.3cm]{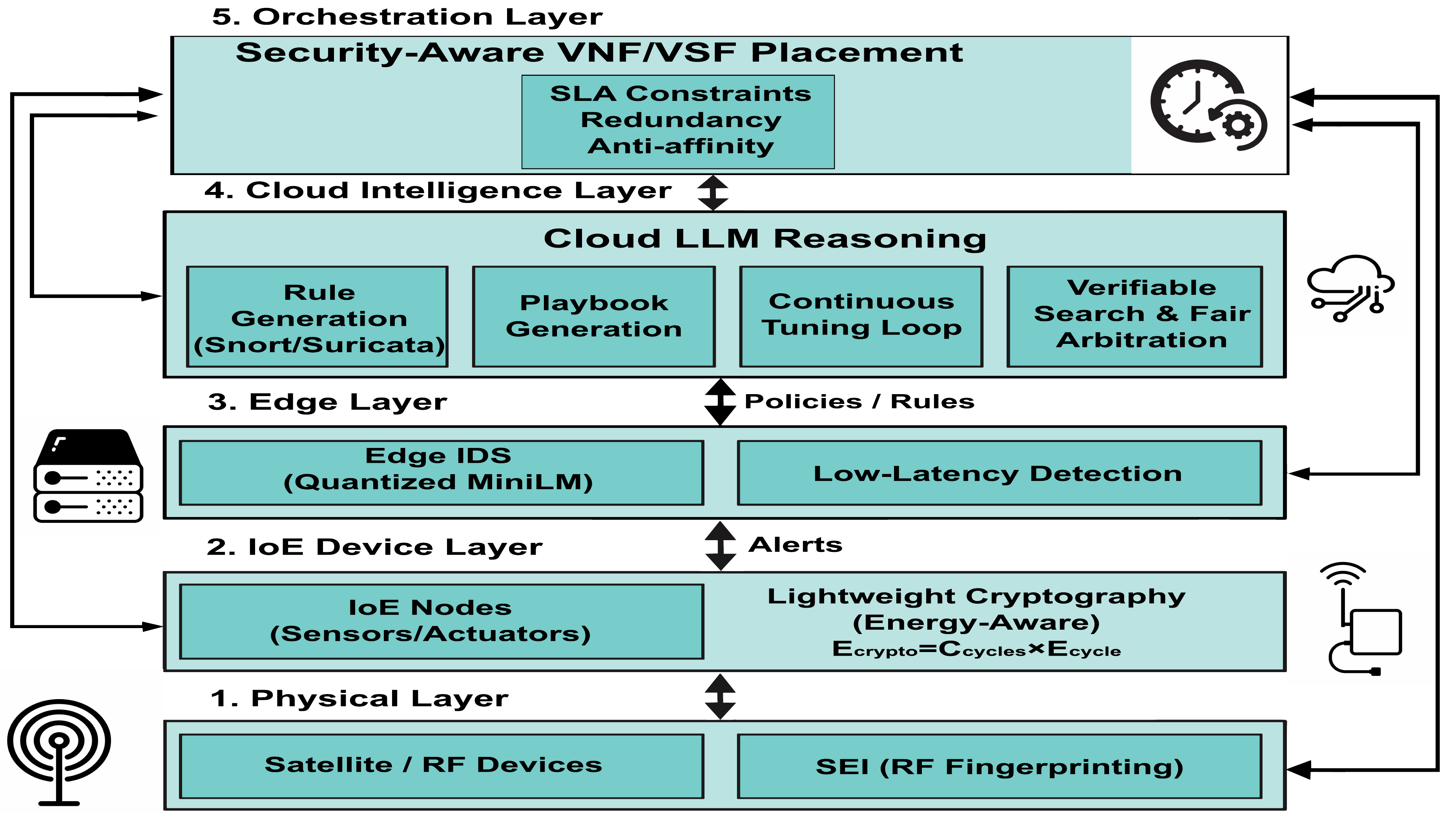} 
  \caption{Multi-cloud federation architecture for IoE, highlighting inter-cloud portability, resource virtualization, and cross-layer QoS/SLA concerns.}
  \label{fig:llm_ioe_security_arch}\vspace{-0.3cm}
\end{figure}

Fig.~\ref{fig:llm_ioe_security_arch} illustrates the multi-layer adaptive security architecture discussed in this subsection, spanning physical-layer authentication, energy-aware cryptography, orchestration-level constraints, and LLM-driven edge--cloud intrusion detection.\vspace{0.2cm}

\noindent\textbf{Edge--cloud architecture and operational motivation: }
Recent work proposes an LLM-driven adaptive intrusion-detection architecture for IoE deployments that combines a lightweight edge detector with cloud-level reasoning. In this design, a compact MiniLM model runs at the edge, while a cloud LLM (e.g., GPT-4) generates actionable security artifacts such as IDS rules (Snort/Suricata) and human-readable response playbooks from edge alerts, improving explainability and incident-response workflows \cite{llm_ioe_security_2026}.\vspace{0.2cm}

\noindent\textbf{Deployability under edge constraints and operational impact :}
To meet tight latency budgets on commodity IoE hardware, the authors report that the 8-bit integer numerical format (INT8) quantization reduces MiniLM inference latency (e.g., 92\,ms to 54\,ms on Raspberry Pi~4) while preserving detection quality. Human-in-the-loop evaluation indicates a 41\% reduction in mean time-to-contain when using cloud-generated playbooks, and a continual tuning loop reduces false positives during pilot deployment (31\% to 26\%), highlighting the value of adaptive policy generation and feedback-driven tuning for evolving IoE threats \cite{llm_ioe_security_2026}.\vspace{0.2cm}

\noindent\textbf{Limitations and robustness considerations: }
It was shown that robustness against adversarial manipulation (e.g., evasion attacks and prompt/log injection affecting the cloud tier) remains an open issue, motivating stronger stress-testing and defenses for trustworthy LLM-assisted IoE security \cite{llm_ioe_security_2026}. In addition to cryptographic protection and authentication, IoE security must also be enforced at the \emph{orchestration} layer, where VNF/VSF placement decisions in NFV-based service chains should satisfy both SLA latency bounds and VSF operational rules (e.g., proximity, redundancy, and anti-affinity) \cite{tamim2020vsf}.\vspace{0.2cm}

\noindent\textbf{Security--energy coupling on constrained devices: }
Security mechanisms are not ``free'' in massive IoE: for resource-constrained nodes, cryptographic protection can be approximated by an energy model $E_{\text{crypto}} = C_{\text{cycles}} \times E_{\text{cycle}}$, where $C_{\text{cycles}}$ is the number of CPU cycles consumed and $E_{\text{cycle}}$ is the energy per cycle \cite{IEEE9667513}. Reported comparisons indicate that moving from lightweight ciphers (on the order of $10$--$30$k cycles) toward stronger primitives such as AES-256/ECC-256 (on the order of $150$--$350$k cycles) can increase energy consumption by roughly $5$--$10\times$ on typical embedded processors, motivating lightweight and hardware-accelerated security designs for massive IoE \cite{IEEE9667513}.\vspace{0.2cm}

\noindent\textbf{Fair and verifiable data sharing in cloud-assisted IoE: }
When IoE data are outsourced to the cloud, ensuring confidentiality is not sufficient: practical deployments also require (i) \emph{verifiable search} over encrypted data (to detect incomplete/incorrect results) and (ii) \emph{fair arbitration} mechanisms to resolve disputes between data owners and semi-trusted cloud servers, especially under \emph{dynamic updates} (insert/modify/delete) \cite{miao2019fairdyndsf}.\vspace{0.2cm}

\noindent\textbf{Physical-layer authentication for satellite IoE: }
In satellite IoE, relying only on upper-layer authentication can remain vulnerable to forgery, replay, and deception attacks in open environments. SEI provides a complementary physical-layer authentication mechanism by exploiting RFFs caused by intrinsic hardware imperfections, which are difficult to forge or replicate. Recent work further shows that few-shot and generative-learning-based SEI can be deployed in a ground--satellite pipeline to improve identification accuracy even under severely limited labeled samples \cite{zhao2026_fewshot_sei_satellite_ioe}.\vspace{0.2cm}

\noindent\textbf{From challenges to directions: }
Taken together, these challenges suggest that future IoE systems must be designed as closed-loop, cross-layer infrastructures: (i) AI-native control to manage complexity, (ii) security/privacy enforced at lifecycle and orchestration time, (iii) economic incentives for sustainable multi-stakeholder participation, and (iv) energy-aware operation that also preserves timeliness (AoI).
These observations motivate the research directions summarized next.

\section{Key Research Frontiers in 6G-Enabled IoE}
\label{sec:future}

\noindent
Future research on 6G-enabled IoE is expected to converge toward a set of
interrelated high-impact directions centered on intelligence, scalability,
trustworthiness, and sustainability. In particular, the transition toward
AI-native and semantic-aware communication paradigms is anticipated to
play a central role in enabling scalable and context-aware autonomy across
distributed IoE systems, as highlighted in recent 6G visions and semantic
communication frameworks \cite{saad2020vision6g,strinati20216g}. At the same
time, economic-aware mechanisms, including pricing and incentive design,
will be essential to support cooperation and resource sharing in
multi-stakeholder environments, as emphasized in emerging works on
resource economics and network optimization for beyond-5G systems
\cite{wen2024incentive6g,hexa2025value}. From a systems perspective, safe
and elastic orchestration across the device--edge--cloud continuum,
combined with zero-touch management supported by digital twins, will be
critical to achieve fully autonomous and self-optimizing infrastructures
\cite{akyildiz2020sixg,amadeo2024digitaltwins}. 
In parallel, communication-efficient federated intelligence will be needed
to ensure scalable learning under resource constraints, while maintaining
robustness against heterogeneity and adversarial threats, as widely
discussed in recent FL and edge intelligence surveys \cite{kairouz2021advances_fl,lim2020fl_men}. Security and privacy must be addressed as end-to-end design principles, embedded into system
architectures through lifecycle-aware and policy-driven mechanisms \cite{yan2020survey6g_security}. Moreover,
sustainability considerations will drive the development of green IoE frameworks that jointly optimize energy consumption and information freshness, in line with recent research on energy-efficient 6G systems and age-of-information-aware optimization \cite{akyildiz2020sixg,kaul2012aoi}.
Finally, emerging challenges related to dense deployments—such as multi-TRP coordination, synchronization, and feedback overhead—highlight the need for lightweight and scalable coordination mechanisms in next-generation wireless infrastructures \cite{SamsungCJT2026}.

\subsection{AI-Native and Semantic-Aware Orchestration for 6G IoE}
\label{subsec:AI-Native_Orchestration}

\noindent
Future 6G-enabled IoE systems are expected to evolve toward
AI-native architectures that tightly integrate communication,
computation, and control. In this paradigm, intelligence is embedded
across the entire system stack, enabling scalable autonomy, adaptive
resource management, and context-aware service provisioning in
highly dynamic environments.\vspace{.2cm}

\noindent\textbf{AI-native communication and orchestration: }
AI-native IoE systems rely on learning-driven mechanisms to optimize network behavior and service delivery in real time. In particular, semantic- and goal-oriented communication has been identified as a key paradigm for future 6G-enabled IoE systems, enabling networks to prioritize task-relevant information rather than raw bit transmission, thereby improving communication efficiency and scalability \cite{saad2020vision6g,wang2023road6g}. Research on FL for 6G \cite{fl6g_survey} further reinforce this vision by enabling distributed and communication-efficient intelligence across large-scale IoE environments. Building on this paradigm, AI-driven orchestration emerges as a fundamental requirement to manage heterogeneous IoE services across the device--edge--cloud continuum. In this context, recent studies model orchestration as a sequential decision-making problem and leverage DRL to enable adaptive and closed-loop resource management under dynamic workloads \cite{sami2021_ai_resource_provisioning_ioe_6g}. More advanced survey works highlight the need for scalable and multi-dimensional orchestration strategies that jointly optimize communication, computation, and storage resources in edge intelligence systems \cite{zhang2023_multidimensional_orchestration_edge_intelligence_6g,yang2025_beyond_edge_rl_mec}. Such a development opens several important research directions for IoE systems, including the design of scalable DRL-based orchestration frameworks, the integration of semantic communication with learning-driven control, and the development of distributed and communication-efficient intelligence mechanisms for large-scale heterogeneous environments.\vspace{.2cm}

\noindent\textbf{Scalable optimization and distributed control: }
At a finer granularity, IoE orchestration extends to microservice-based architectures, where containerized service graphs must be dynamically placed, scaled, and migrated across distributed edge infrastructures. Recent work shows that combining reinforcement learning with structured representations, such as graph-based dependency modeling, enhances scalability and generalizability of placement strategies by capturing service interdependencies and network dynamics \cite{chen2024_adaptive_edge_microservice_placement_tmc}. However, the dynamic and uncertain nature of IoE environments raises serious reliability challenges, motivating the integration of safety-aware learning mechanisms, including constrained policy updates and fallback control strategies, to ensure stable operation under varying workloads and network conditions \cite{sami2021_ai_resource_provisioning_ioe_6g,yang2025_beyond_edge_rl_mec}. From an optimization perspective, IoE orchestration is inherently multi-objective, involving tradeoffs among energy efficiency, latency, operational cost, and QoS/SLA guarantees. Classical NFV and cloud resource allocation studies highlight the complexity of these tradeoffs and the need for flexible optimization frameworks capable of handling heterogeneous resources and service requirements \cite{herrera2016_vnf_survey,maia2024_integrated_edge_survey}. Here, heuristic and multi-criteria optimization approaches remain relevant for tackling large-scale and combinatorial placement problems \cite{vmp_multicriteria_survey}. 
Furthermore, the distributed and partially observable nature of IoE systems implies the use of learning-based and decentralized optimization techniques. Namely, belief-based models and distributed optimization methods
enable scalable coordination under uncertainty, while federated and multi-agent reinforcement learning approaches support collaborative and privacy-aware decision making across edge nodes \cite{asheralieva2022_aoi_security_ioe,seid2023_mafrl_iomt}.  These developments point toward several key research directions, including the design of scalable graph-based orchestration models, the integration of safety constraints into learning-driven control, and the development of distributed and multi-agent optimization frameworks for large-scale IoE.\vspace{0.2cm}

\noindent\textbf{Security-oriented research directions: }
The integration of AI-native orchestration and distributed intelligence introduces new security and trust challenges in 6G IoE systems. Future research must focus on lightweight and scalable security mechanisms tailored to heterogeneous and resource-constrained devices, as well as real-time, adaptive anomaly detection frameworks capable of identifying sophisticated threats with minimal overhead \cite{corradini2025ioe_cybersecurity_overview}. In addition, security-aware orchestration strategies should explicitly incorporate threat models and risk assessment into resource allocation and service placement decisions. These directions build on extensive IoT security research emphasizing lightweight cryptographic primitives, privacy-preserving communication protocols, and scalable protection mechanisms for massive device ecosystems \cite{sicari2015iot_security,conti2018iot_security_survey}.
Recent advances in deep learning–based IDSs demonstrate strong potential for detecting complex attacks in large-scale IoT environments while maintaining practical efficiency \cite{ferrag2020dl_ids_iot}. Furthermore, FL–based
security frameworks enable collaborative and privacy-preserving threat detection across distributed IoE nodes without sharing raw data \cite{shen2024_federated_iot_ids}.

\subsection{Zero-Touch Management and Digital-Twin-Assisted
Operations}
As IoE systems evolve toward large-scale, multi-domain, and highly dynamic environments, managing network operations and services becomes increasingly complex. In this context, zero-touch management and digital-twin-assisted operation emerge as key enablers for achieving autonomous, scalable, and reliable IoE infrastructures. By combining AI-driven closed-loop control, virtual representations of physical systems, and decentralized coordination mechanisms, these paradigms aim to enable real-time monitoring, adaptive decision making, and efficient resource management across the device--edge--cloud continuum.\vspace{.2cm}

\noindent\textbf{Trustworthy AI for Zero-Touch IoE: }
\label{subsec:Zero-Touch}
As IoE scales to massive device densities and multi-domain services,
manual operation becomes increasingly impractical. A key research
direction for future 6G systems is therefore the adoption of
\emph{Zero-Touch Network and Service Management (ZSM)}, where monitoring,
configuration, healing, and optimization processes are automated
end-to-end through closed-loop control mechanisms driven by AI and data analytics \cite{sharma2025_6g_overview,9749186}. While such automation can significantly improve scalability and
operational efficiency, it also raises important challenges related to decision transparency, reliability, and trust. In highly dynamic IoE environments, autonomous control systems must not only optimize network performance but also provide interpretable and verifiable decision-making mechanisms. These requirements motivate the development of trustworthy AI approaches, including explainable AI and hybrid learning--reasoning frameworks, to support reliable closed-loop operation in future zero-touch IoE management systems.\vspace{.2cm}

\noindent\textbf{Neuro-symbolic Explainable AI twin for trustworthy zero-touch IoE: }
An important research gap for 6G-era ZSM is \emph{trustworthy} closed-loop operation, i.e., the ability to explain and correct orchestration decisions under time-varying network dynamics. Munir et al. \cite{neysy_xai_twin_zsm} propose a \emph{neuro-symbolic Explainable AI (XAI) twin} that couples (i) an \emph{implicit learner} (neural model) operating in the physical space to produce fast, data-driven decisions with (ii) an \emph{explicit learner} operating in the virtual space that performs symbolic reasoning via a Bayesian Belief Network (BBN) and corrects decisions via a Bayesian multi-armed bandit mechanism. This design targets explainable and reliable service execution decisions in closed-loop zero-touch IoE management.\vspace{.2cm}

\noindent\textbf{Design insight and explainability for extensible ZSM: }
The explicit learner uses a directed acyclic graph based representation inside the BBN to map contextual metrics to decisions, which supports extensibility (e.g., adding new metric nodes without redesigning the full controller) and enables evidence-based reasoning over network dynamics \cite{neysy_xai_twin_zsm}.\vspace{.2cm}

\noindent\textbf{Reported outcomes (accuracy vs. trust): }
The authors report that the proposed XAI twin achieves high decision accuracy (about 96.26\%) while improving a trust/reasoning score by roughly 18\%--44\% compared to learning-based baselines, illustrating the value of pairing neural decision-making with symbolic explanation in zero-touch IoE control loops \cite{neysy_xai_twin_zsm}.\vspace{.2cm}

\noindent\textbf{Cybertwin-Based Edge Management: }
In parallel with zero-touch management frameworks, \emph{digital twin}
technologies are emerging as a powerful tool for network management in
IoE environments. By maintaining virtual replicas of physical network
components, digital twins enable continuous monitoring of system
states, proactive fault diagnosis, and predictive optimization through
simulation-based what-if analysis before actions are deployed in the
real infrastructure \cite{khan2022_digital_twin_wireless}. Such virtual
representations allow operators to detect potential network issues,
test configuration changes, and evaluate system performance without
disrupting the physical IoE deployment \cite{empl2023_digital_twin_iot_security}. Building upon this concept, cybertwin architectures extend digital representation capabilities toward edge-based control frameworks, where virtual entities associated with devices or services are deployed closer to data sources. This approach enables faster decision-making, improved contextual awareness, and more efficient management of distributed IoE resources across edge computing
environments.\vspace{.2cm}

\noindent\textbf{Cybertwin-driven zero-touch resource allocation (digital-twin at the edge): }
Beyond classical digital twins, \emph{cybertwin} places a virtual representation of ``ends'' (things/people) at the \emph{edge cloud}, acting as a network function that supports tracking, logging, and control for IoE applications. This edge-resident twin can serve as a practical substrate for closed-loop, zero-touch management by maintaining up-to-date device/service context and enabling fast local decisions \cite{jain2022_mwba_rat}.\vspace{.2cm}

\noindent\textbf{Blockchain and Optimization for Resource Allocation: }
Managing resources efficiently in large-scale IoE systems remains
challenging due to device heterogeneity, dynamic workloads, and
stringent latency requirements. In such environments, decentralized
coordination mechanisms can play a key role in enabling scalable and
resilient network operation. Recent studies suggest that distributed
control strategies inspired by \emph{swarm networking} can improve
adaptability and scalability in dense IoE deployments by allowing
groups of nodes to cooperate through local interactions to achieve
global coordination \cite{dressler2023swarm_iot,sharma2025_6g_overview}. Complementing these decentralized approaches, blockchain-based frameworks and advanced optimization techniques are increasingly explored to support secure and efficient resource allocation in IoE infrastructures. By combining trustworthy resource monitoring with intelligent optimization algorithms, these solutions aim to enable scalable and dependable orchestration of communication and computing resources in future 6G-enabled IoE ecosystems.\vspace{.2cm}

 \noindent\textbf{Blockchain-assisted optimization for resource allocation: }
A concrete direction is to combine cybertwin-based control with \emph{blockchain-backed} monitoring/sharing of resources and \emph{metaheuristic} optimization for real-time resource allocation. The Multi-Weight/Blockchain-Aware optimization with Resource Allocation Techniques (MWBA-RAT) framework integrates blockchain for dependable resource management and proposes a quasi-oppositional search-and-rescue optimizer, further enhanced with graph clustering to improve the initial population quality, aiming to reduce overall communication cost between gateways and resources \cite{jain2022_mwba_rat}.\vspace{.2cm}

\noindent\textbf{Evidence of potential gains and open gaps: }
In reported experiments, MWBA-RAT achieves lower system average cost (e.g., 2.63 under 500 required computation resources) and lower power consumption (e.g., 0.012\,mW under 10 nodes) compared with baseline methods, suggesting that cybertwin-assisted, optimization-based RA can be effective for scalable 6G IoE operation. However, broader validation under realistic mobility, heterogeneous workloads, and strict real-time constraints remains an open direction for future ZSM-ready deployments \cite{jain2022_mwba_rat}.

\subsection{NF and RIS-Aided Localization for 6G IoE}
\label{subsec:Near-Field_RIS}
A concrete 6G research direction is the integration of \emph{NF communications} and \emph{RIS-assisted} smart radio environments to jointly support IoE connectivity, sensing, and high-precision localization. As wireless systems move toward higher carrier frequencies and extremely large antenna arrays, the classical FF planar-wave propagation assumption becomes insufficient, and many links operate in the radiating NF region where spherical-wave propagation effects must be considered \cite{cui2023nearfield_survey,bjornson2020ris_smart_radio_environment}. Now, IoE links may simultaneously span FF monitoring regimes and NF positioning regimes, motivating adaptive NF--FF channel models capable of smoothly balancing coverage and localization accuracy \cite{wu2025_ris_nf_ff_ioe_healthcare,cui2023nearfield_survey,zhi2024_xlmimo_jsac}. Other studies show that RIS-enabled smart radio environments can actively shape propagation conditions to improve signal focusing, coverage, and localization performance in dense IoE deployments \cite{basar2019ris_smart_env}.
Practical 6G IoE deployments generate additional system challenges. These include (i) low-overhead RIS configuration and calibration, (ii) energy-aware operation of large RIS arrays through techniques such as selective element activation, and (iii) robustness to complex indoor multipath environments typical of dense IoE scenarios (e.g., hospitals and smart buildings) \cite{wu2025_ris_nf_ff_ioe_healthcare}. Addressing these challenges requires cross-layer system design that combines RIS control, edge intelligence, and communication-efficient learning techniques to deliver reliable, sustainable, and accurate 6G-enabled IoE services. 

\subsection{Communication-Efficient Federated Intelligence} \label{subsec:CEFI} 
Scaling distributed intelligence in next-generation networks requires mitigating the severe communication bottleneck caused by frequent and massive model exchanges, while simultaneously improving robustness against device heterogeneity and non-i.i.d. data distributions. Foundational FL studies emphasize communication-efficient training strategies, including gradient compression, quantization, and update sparsification, as key mechanisms to reduce communication overhead without significantly degrading model convergence \cite{lim2020fl_men,mcmahan2017communication}. Furthermore, intelligent client selection and resource-aware scheduling have emerged as critical strategies to optimize communication rounds by prioritizing edge devices with favorable channel conditions and informative local updates \cite{nishio2019client}.

As a paradigm shift, future 6G will consider over-the-air (OTA) computation, which harnesses the waveform superposition property of wireless multiple-access channels for rapid, simultaneous gradient \cite{zhu2020broadband}. To ensure that such high-efficiency communication does not compromise global model fairness, recent frameworks like OTA-FFL formulate FL as a multi-objective minimization problem, utilizing a modified Chebyshev approach to balance average performance with Pareto-optimality across heterogeneous clients \cite{hamidi2025ota}. However, the performance of OTA-FL is often constrained by the "straggler effect" and signal aggregation errors. To address this, joint device selection and transmit power optimization frameworks have been proposed to manage the trade-off between prioritizing strong channels (to minimize noise) and including diverse datasets from power-constrained devices to avoid model bias \cite{liu2026age}. Additionally, knowledge-transfer methods like Federated Distillation (FD) remain a key strategy for ultra-dense networks by communicating lightweight soft predictions rather than full model parameters \cite{fl6g_survey, jeong2018communication}.

\subsection{Trustworthy and Incentivized Participation}
\label{subsec:TIP}

Future IoE systems must ensure trustworthy participation in distributed learning and service provisioning. Blockchain-based mechanisms can support auditability and decentralized coordination \cite{ali2019blockchain_iot}. Hardware-rooted identity (e.g., PUF-assisted lightweight consensus) can complement ledger-based trust by enabling device authentication and integrity with reduced computation/energy overhead on constrained IoE nodes \cite{pufchain2020}.
 Recent 6G FL directions further motivate combining incentive design (including game-theoretic ideas) with trust mechanisms to sustain participation and mitigate malicious behavior in large-scale IoE learning \cite{fl6g_survey}.

\subsection{Security and Privacy for 6G IoE}
\label{subsec:SPRI}
The need for security and privacy is intensifying as 6G integrates IoE distributed intelligence and critical infrastructure services. Prospective studies outline key threat surfaces and research needs spanning identity, trust, secure AI, and privacy-preserving networking for 6G ecosystems \cite{porambage2021roadmap_6g_security}. From a roadmap perspective, SPbD/PSbD provides a unifying checklist for building trustworthy 6G IoE systems, ensuring protections are proactive, embedded, and enforced end-to-end across the lifecycle \cite{spbd_ioe}. Explainability is also becoming a prerequisite for trustworthy automation: neuro-symbolic XAI-twin designs can quantify reasoning/trust and correct closed-loop decisions, complementing purely data-driven ZSM controllers \cite{neysy_xai_twin_zsm}.\vspace{0.2cm}
\begin{table*}[!t]
\centering
\caption{Comparison of Economic Models Across IoE Layers}
\label{tab:ioe_economic_models}
\renewcommand{\arraystretch}{1.2}
\begin{tabular}{|p{2.3cm}|p{3.8cm}|p{4.1cm}|p{3.8cm}|p{1.8cm}|}
\hline
\rowcolor{Emerald!50}
\textbf{IoE Layer} & \textbf{Economic Objective} & \textbf{Typical Models} & \textbf{Key Advantages} & \textbf{Representative Refs.} \\
\hline
\rowcolor{Emerald!10}

Data Collection 
& Incentivize participation and maximize sensing coverage under energy/privacy constraints 
& Reverse auctions, contract theory, Stackelberg games, incentive-compatible mechanisms 
& Efficient user participation, fairness-aware data acquisition, energy-aware compensation 
& \cite{wen2024incentive6g,yang2016crowdsensing,wickramasinghe2024auction} \\
\rowcolor{Emerald!25}

Data Transmission 
& Optimize resource allocation and spectrum usage while ensuring economic efficiency 
& Stackelberg games, coalition games, pricing-based resource allocation, hierarchical optimization 
& Improved power efficiency, reduced latency, incentive-compatible communication 
& \cite{6279592,8240666,9711524,bounaira2025trustworthy} \\
\rowcolor{Emerald!10}

Information Fusion 
& Ensure data quality, truthful reporting, and efficient edge intelligence processing 
& Trust-aware optimization, reputation systems, incentive-based learning, mechanism design 
& Enhanced robustness, higher data reliability, improved system efficiency 
& \cite{9075087,yan2014survey_trust_iot,8763885,computers14010010} \\
\rowcolor{Emerald!25}

Service Application 
& Maximize revenue and customer value through pricing and service differentiation 
& Usage-based pricing, service bundling, platform economics, value-chain optimization 
& Increased monetization opportunities, improved user satisfaction, scalable service deployment 
& \cite{oecd2024digital,mckinsey2025tech,hexa2025value,gowroju2025survey6g}\\
\rowcolor{Emerald!10}

Cross-Layer Ecosystem Coordination 
& Maximize total ecosystem profit and stakeholder utility across IoE value chains 
& Cooperative games, supply-chain coordination, multi-stakeholder optimization, revenue-sharing models 
& Higher global efficiency, stronger network effects, sustainable multi-provider collaboration 
& \cite{hexa2025value,gowroju2025survey6g} \\
\rowcolor{Emerald!25}
\hline
\end{tabular}\vspace{-0.5cm}
\end{table*}

\noindent\textbf{Freshness--integrity co-optimization as a security objective: } Beyond confidentiality and authentication, next-generation IoE must explicitly optimize the \emph{value of information}, which depends on \emph{freshness} (Age of Information) and on whether the received/processed information is \emph{correct} under adversarial or failure-prone edge computing. This motivates security designs that jointly control sampling/updating and resilient computation so that stale or corrupted updates do not propagate into IoE decision processes \cite{asheralieva2022_aoi_security_ioe}.\vspace{0.2cm}

\noindent\textbf{Resilient edge analytics via coded distributed computing: } For MEC-assisted IoE analytics, coded distributed computing (e.g., Lagrange coded computing) can provide \emph{A-security} against malicious edge devices and \emph{S-resilience} against stragglers by increasing redundant task assignments. In particular, the decoding condition links the required number of assigned tasks/nodes to the desired security and resilience levels, enabling system-level tuning of robustness under edge heterogeneity and adversarial behavior \cite{asheralieva2022_aoi_security_ioe}.

\subsection{Economic-Aware Mechanism Design for 6G-Scale IoE}
\label{subsec:Economic-Aware}

A concrete research direction is to combine \emph{AI-native orchestration} with \emph{economic mechanism design} to sustain large-scale IoE ecosystems. Recent economic IoE surveys emphasize that game-based methods (e.g., pricing-, incentive-, and auction-based mechanisms) can align stakeholder actions with system-level goals, while learning-based approaches can adapt these mechanisms under dynamic and uncertain conditions \cite{ding2025_economic_ioe}.
Economic mechanisms also arise naturally in Internet-of-Energy settings, where end users can trade energy and optimize ES and Electric Vehicle (EV) charging decisions under both market trading prices and utility real-time pricing, linking incentive design directly to measurable reductions in grid energy waste \cite{lin2017_ioe_energy_trading}. For example, economic incentives can promote collaborative resource sharing (e.g., via slicing-style sharing and cost distribution), reducing the burden on individual entities and improving long-term ecosystem durability \cite{ding2025_economic_ioe}. Moreover, privacy and security can be treated economically by rewarding high-quality participation (e.g., in FL) and using game-theoretic attacker--defender models that explicitly incorporate attack costs and defense rewards \cite{ding2025_economic_ioe}. Overall, integrating economics with 6G IoE control loops opens research opportunities in multi-objective utility design, market-aware resource allocation, and incentive-compatible privacy/security enforcement at scale.

\subsection{IoE Ecomonics and Business Models}
\label{subsec:Eco}

The transition from the IoT to the IoE requires not only technological advancements but also sustainable economic frameworks capable of supporting large-scale deployment across next-generation networks. Unlike conventional IoT paradigms, IoE ecosystems involve complex interactions among people, processes, data, and physical objects, leading to new challenges in incentive design, resource valuation, and stakeholder coordination. These challenges are further amplified in 6G environments characterized by ultra-dense connectivity, heterogeneous services, and strong network externalities, which necessitate sophisticated economic and business models to ensure long-term scalability and profitability \cite{gowroju2025survey6g,Khilukha2024,oecd2024digital}. A systematic economic analysis of the IoE ecosystem can be structured into four interdependent layers, as summarized in Table~\ref{tab:ioe_economic_models}, which compares the corresponding economic objectives, modeling approaches, and key advantages.\vspace{0.2cm}

\noindent\textbf{Information Collection: }  
This layer focuses on crowdsensing participation, data acquisition incentives, and task assignment optimization. Mobile crowdsensing systems fundamentally rely on active user participation, which raises key challenges related to motivation, data quality, and resource consumption \cite{6069707}. To address these issues, economic mechanisms such as contract theory, reverse auctions, and game-theoretic incentive schemes are commonly employed to motivate users and devices to contribute sensing data while compensating for energy expenditure, computational costs, and privacy risks \cite{MADDIKUNTA2022103464}. In particular, reverse auction–based and timeliness-aware incentive schemes have been shown to effectively optimize task allocation and participant selection in dynamic environments such as vehicular crowdsourcing in smart cities \cite{9328457,wickramasinghe2024auction}. Recent studies have further explored advanced incentive models, including diffusion-based and prospect-theoretic approaches tailored for 6G IoT edge services \cite{wen2024incentive6g}, while AI-enabled edge intelligence frameworks support efficient sensing data processing and distributed learning in heterogeneous environments \cite{wang2019inedge_ai}. These developments collectively enable scalable, efficient, and intelligent information collection across next-generation IoT and IoE systems.\vspace{.2cm}

\noindent\textbf{Information Transmission: }  
The transmission layer addresses spectrum sharing, relay selection, and communication resource allocation under economic constraints. Hierarchical game-theoretic models, including Stackelberg and coalition-based frameworks, enable efficient coordination between infrastructure providers and end devices to optimize power allocation, latency, and reliability while maintaining economic efficiency \cite{6279592}. In this context, joint computation offloading and resource allocation strategies have been extensively studied in fog/cloud-enabled systems, where fairness-aware and delay-sensitive mechanisms are designed to balance network performance and user demands \cite{8240666}. Emerging research on digital twin–enabled 6G networks further highlights the importance of trust-aware and incentive-compatible communication strategies, supporting efficient computation offloading, real-time network optimization, and secure resource sharing across distributed edge environments \cite{9711524,bounaira2025trustworthy}.\vspace{.2cm}

\noindent\textbf{Information Fusion: }  
This layer involves data aggregation, multimodal fusion, and edge intelligence processing, where economic incentives play a crucial role in ensuring data quality and truthful reporting. In mobile crowdsensing systems, incentive mechanisms such as reverse auction–based schemes have been widely adopted to promote truthful participation and efficient data collection from distributed devices \cite{9075087}. In parallel, trust and reputation management frameworks are essential to evaluate the reliability of contributed data and support trust-aware decision-making in IoT environments \cite{yan2014survey_trust_iot}. Furthermore, the integration of edge intelligence and deep learning techniques enables efficient multimodal data fusion and real-time processing across distributed edge nodes, enhancing system scalability and responsiveness \cite{8763885}. Recent trust-centric resource management approaches in 6G IoT systems demonstrate that combining economic incentives with trust-aware optimization significantly improves system robustness and resource utilization efficiency \cite{computers14010010}.\vspace{.2cm}

\noindent\textbf{Service Application: }  
At the application level, economic considerations govern pricing strategies, service differentiation, and market value creation. New business paradigms such as usage-based pricing, service bundling, and platform-oriented ecosystems enable IoE providers to maximize revenue while enhancing user satisfaction. Macroeconomic analyses indicate that digital ecosystem integration and data-driven service monetization are expected to become dominant value drivers in the coming decade, particularly with the emergence of AI-native and edge-enabled services \cite{oecd2024digital,mckinsey2025tech}. Today, the IoE industry is increasingly evolving toward coordinated multi-stakeholder ecosystems where network operators, cloud providers, edge infrastructure owners, and service providers collaborate to maximize overall value creation. Cooperative economic models and value-chain coordination mechanisms have been shown to outperform independent operational strategies in terms of both provider revenue and user utility, especially in large-scale 6G deployments characterized by strong cross-layer dependencies and network effects. Recent 6G flagship initiatives emphasize that such collaborative business ecosystems will be essential for realizing sustainable IoE services and unlocking new revenue streams across multiple industry verticals \cite{hexa2025value,gowroju2025survey6g}.

\subsection{Green and Energy-Aware Intelligent IoE}
\label{subsec:GEAI}
Sustainable IoE requires joint optimization of communication, computation, and learning under energy constraints. Energy harvesting surveys provide foundational models and open issues for long-lived IoE deployments \cite{ma2020eh_iot}, while 6G visions highlight energy efficiency as a primary design axis for future intelligent networks \cite{saad2020vision6g,fl6g_survey}.\vspace{.2cm}

\noindent\textbf{Wireless-powered IIoE as an energy-efficiency enabler: }
A practical direction for long-lived massive IoE/IIoE deployments is to combine \emph{wireless power transfer} with \emph{status-update scheduling} under a central controller and the High-Altitude Platform (HAP).
In a wireless-powered IIoE architecture, the HAP collects global channel information and coordinates both
charging and update transmissions, providing an upper-bound guideline when perfect CSI is assumed.
To simplify deployment, each device can follow a \emph{harvest-then-transmit} protocol and Time Division Multiple Access scheduling,
which avoids simultaneous energy and data transmissions and reduces synchronization/decoding requirements.
\cite{zheng2024_aee_dldqn}\vspace{0.2cm}

\noindent\textbf{Design insight (metric tuning): }
A practical control mechanism in these systems is the reward/weighting factor, which dictates the multi-objective trade-off between information freshness (normalized AoI) and resource conservation (e.g., wasted energy or bandwidth). Smaller weighting values prioritize low-latency updates and high frequency sensing, whereas larger values emphasize energy efficiency and battery longevity \cite{zheng2024_aee_dldqn}. This trade-off is often non-linear; for instance, optimizing transmit power to minimize AoI estimation errors requires a joint consideration of the device's remaining energy and the data's "age-benefit" to the global model \cite{liu2026age}. Furthermore, in scenarios where multiple sensors compete for limited uplink resources, this weighting factor serves as a prioritization index, where the system must balance the urgency of "stale" data against the high energy cost of transmission over fading channels to achieve a Pareto-optimal operating point \cite{hamidi2025ota}.\vspace{0.2cm}

\noindent\textbf{Energy trading and storage/EV scheduling: }
Beyond communication-side energy efficiency, IoE deployments can reduce overall grid waste through \emph{user-level} optimization that jointly schedules renewable generation, ES, and EV charging/discharging while interacting with an IoE energy-trading platform. A representative approach models a single household equipped with PV/wind, ES, and EV, and formulates a MILP that accounts for both the \emph{energy-trading market price} and the \emph{real-time pricing} from the utility, then solves it via a genetic algorithm to handle the model scale and constraint feasibility. Simulation results over one week indicate a large reduction of grid energy waste (e.g., 794.6 to 260.8 in their setting), showing that trading-aware scheduling can stabilize demand and improve energy sustainability at the system level \cite{lin2017_ioe_energy_trading}.\vspace{0.2cm}

\noindent\textbf{Hybrid learning--optimization for green IoE-6G: }
A promising direction is to combine (i) learning-based control (e.g., DRL/edge intelligence) with (ii) metaheuristic/evolutionary optimization to obtain robust, near-optimal decisions under complex constraints and non-stationary workloads. Recent hybrid evolutionary designs (e.g., EEHEA, combining leader-based optimization with adaptive differential evolution) illustrate the potential to reduce energy expenditure while improving latency, coverage, and localization performance, suggesting that \emph{hybrid learning--optimization} can be a practical pathway toward green and scalable IoE-6G operation \cite{singh2024eehea}.
\section{Conclusion}

This paper presents a comprehensive overview of the IoE as a unifying paradigm that extends the IoT by integrating people, data, processes, and things into a holistic cyber--physical--social ecosystem. We structure the discussion around architectural foundations, enabling technologies, and emerging research frontiers. We discuss how IoE systems evolve toward distributed, intelligent, and large-scale infrastructures supported by the device--edge--cloud continuum. Thenm we show that meeting scalable IoE requires the tight integration of multiple technological pillars, including AI-native communication, edge intelligence, federated learning, advanced wireless connectivity, and energy-aware system design. In particular, learning-driven orchestration, semantic communication, and distributed intelligence emerge as key mechanisms to support real-time, adaptive, and context-aware operation in heterogeneous and data-intensive environments.\vspace{.2cm}

Beyond technological enablers, this survey emphasizes that IoE introduces fundamental cross-layer challenges related to scalability, interoperability, security and privacy, and sustainability. Addressing these challenges requires a holistic design approach that jointly considers communication, computation, data management, and economic incentives across multi-stakeholder ecosystems.
Looking ahead, the convergence of IoE and 6G technologies, digital twins, and semantic-aware networking is expected to enable fully autonomous and self-X systems capable of operating at planetary scale. Future research should therefore focus on integrating trustworthy AI, communication-efficient distributed learning, and energy-aware orchestration mechanisms to achieve scalable, resilient, and sustainable IoE infrastructures.
Overall, IoE represents a foundational paradigm for next-generation cyber--physical systems, where intelligence, connectivity, and trust are intrinsically embedded within communication networks, paving the way toward truly autonomous and adaptive digital ecosystems.

\clearpage
\section*{List of Acronyms}
\begin{acronym}[OTA-FFL] 
\acro{5G}{fifth-generation mobile technologies}
\acro{6G}{sixth-generation mobile technologies}
\acro{AEE}{Long-Term Age--Energy Efficiency}
\acro{AI}{Artificial Intelligence}
\acro{AAI}{Agentic Artificial Intelligence}
\acro{AP}{Access Point}
\acro{AR}{Augmented Reality}
\acro{BBN}{Bayesian Belief Network}
\acro{CAPEX}{Capital Expenditure}
\acro{CARS}{Cloud-Assisted Remote Sensing}
\acro{CJT}{Coherent Joint Transmission}
\acro{cmWave}{centimeter wave}
\acro{CNF}{Cloud-Native Network Function}
\acro{CPU}{Central Processing Unit}
\acro{CSI}{Channel State Information}
\acro{D2D}{Device-to-Device}
\acro{DLT}{Distributed Ledger Technology}
\acro{ECORA}{Energy-Efficient Computation Offloading and Resource Allocation}
\acro{EEHEA}{Energy-Efficient Hybrid Evolutionary Algorithm}
\acro{EaaS}{Energy-as-a-Service}
\acro{ES}{Energy Station}
\acro{EV}{Electric Vehicle}
\acro{FF}{Far-Field}
\acro{FL}{Federated Learning}
\acro{GWD}{Gaussian Wasserstein Distance}
\acro{HAP}{High-Altitude Platform}
\acro{IDS}{Intrusion Detection System}
\acro{IIoE}{Industrial Internet of Everything}
\acro{IoBNT}{Internet of Bio-Nano Things}
\acro{IoE}{Internet of Everything}
\acro{IoEn}{Internet of Energy}
\acro{IoNT}{Internet of Nano Things}
\acro{IoSp}{Internet of Space}
\acro{IoT}{Internet of Things}
\acro{IoV}{Internet of Vehicles}
\acro{IoX}{Internet of X}
\acro{ISAC}{Integrated Sensing and Communication}
\acro{KD}{Knowledge Distillation}
\acro{KPI}{Key Performance Indicator}
\acro{LEO}{Low Earth Orbit}
\acro{LLMs}{Large Language Models}
\acro{MAE}{Mean Absolute Error}
\acro{MC}{Molecular Communication}
\acro{MEC}{Multi-access Edge Computing}
\acro{MDP}{Markov Decision Process}
\acro{MIoT}{Massive Internet of Things}
\acro{MILP}{Mixed-Integer Linear Programming}
\acro{MIMO}{Multiple-Input Multiple-Output}
\acro{ML}{Machine Learning}
\acro{MQL}{Multi-Quantile Loss}
\acro{NF}{Near-Field}
\acro{NFV}{Network Function Virtualization}
\acro{NOMA}{Non-Orthogonal Multiple Access}
\acro{NTN}{Non-Terrestrial Network}
\acro{P2P}{Peer-to-Peer}
\acro{PPP}{Poisson Point Process}
\acro{PPM}{Patient State Prediction Modeling}
\acro{PPO}{Proximal Policy Optimization}
\acro{PSbD}{Privacy-by-Design}
\acro{PUF}{Physical Unclonable Function}
\acro{QoE}{Quality of Experience}
\acro{QoS}{Quality of Service}
\acro{QS}{Quorum Sensing}
\acro{RAN}{Radio Access Network}
\acro{RF}{Radio-Frequency}
\acro{RFF}{Radio-Frequency Fingerprinting}
\acro{RGG}{Random Geometric Graph}
\acro{RIS}{Reconfigurable Intelligent Surfaces}
\acro{SEI}{Specific Emitter Identification}
\acro{SenaaS}{Sensing-as-a-Service}
\acro{SFC}{Service Function Chain}
\acro{SLA}{Service Level Agreement}
\acro{SPbD}{Security-by-Design}
\acro{TD3}{Twin-Delayed Deep Deterministic Policy Gradient}
\acro{THz}{Terahertz}
\acro{TRP}{Transmission--Reception Point}
\acro{UAD}{User Activity Detection}
\acro{UE}{User Equipment}
\acro{URLLC}{Ultra-Reliable Low-Latency Communication}
\acro{VLC}{Visible Light Communication}
\acro{VM}{Virtual Machine}
\acro{VMP}{Virtual Machine Placement}
\acro{VNF}{Virtual Network Function}
\acro{VSF}{Virtual Security Function}
\acro{WET}{Wireless Energy Transfer}
\acro{XAI}{Explainable Artificial Intelligence}
\acro{ZSM}{Zero-Touch Network and Service Management}
\end{acronym}

\bibliographystyle{IEEEtran}
\bibliography{refs}

\begin{IEEEbiography}[{\includegraphics[width=1.05in,clip,keepaspectratio]{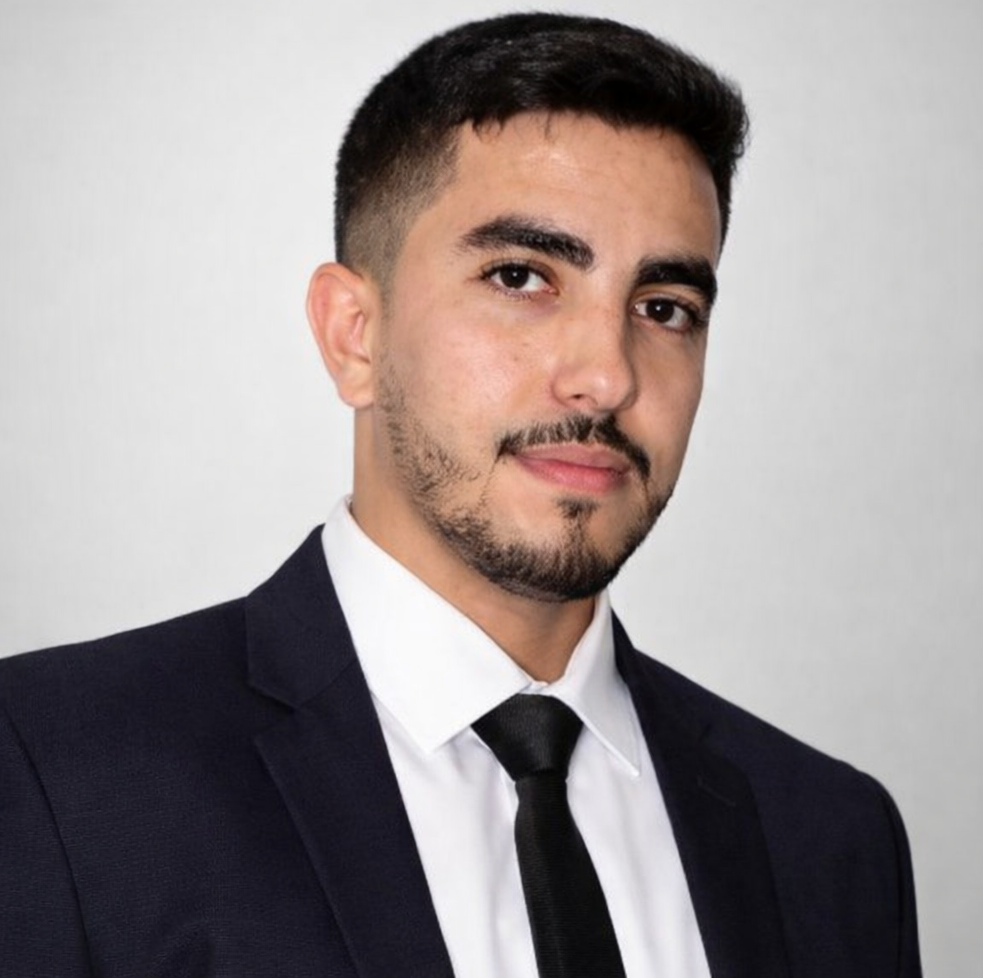}}]{Driss Choukri} (Student Member, IEEE) received the B.Sc. degree in Mathematics and Applications from the Faculty of Sciences Semlalia, Cadi Ayyad University, Marrakech, Morocco, in 2016, and the M.Sc. degree in Mathematics and Applications from the Faculty of Sciences and Techniques, Hassan I University, Settat, Morocco, in 2024. He is currently pursuing the Ph.D. degree in computer science with a focus on intelligent communication systems. Since 2017, he has been a high school mathematics teacher with the Ministry of National Education, Preschool and Sports of Morocco. His research interests include artificial intelligence for wireless communications, federated learning, over-the-air computation, the Internet of Things (IoT), the Internet of Everything (IoE), and energy-efficient design for 5G/6G networks.
\end{IEEEbiography}
\begin{IEEEbiography}[{\includegraphics[width=1.05in,keepaspectratio]{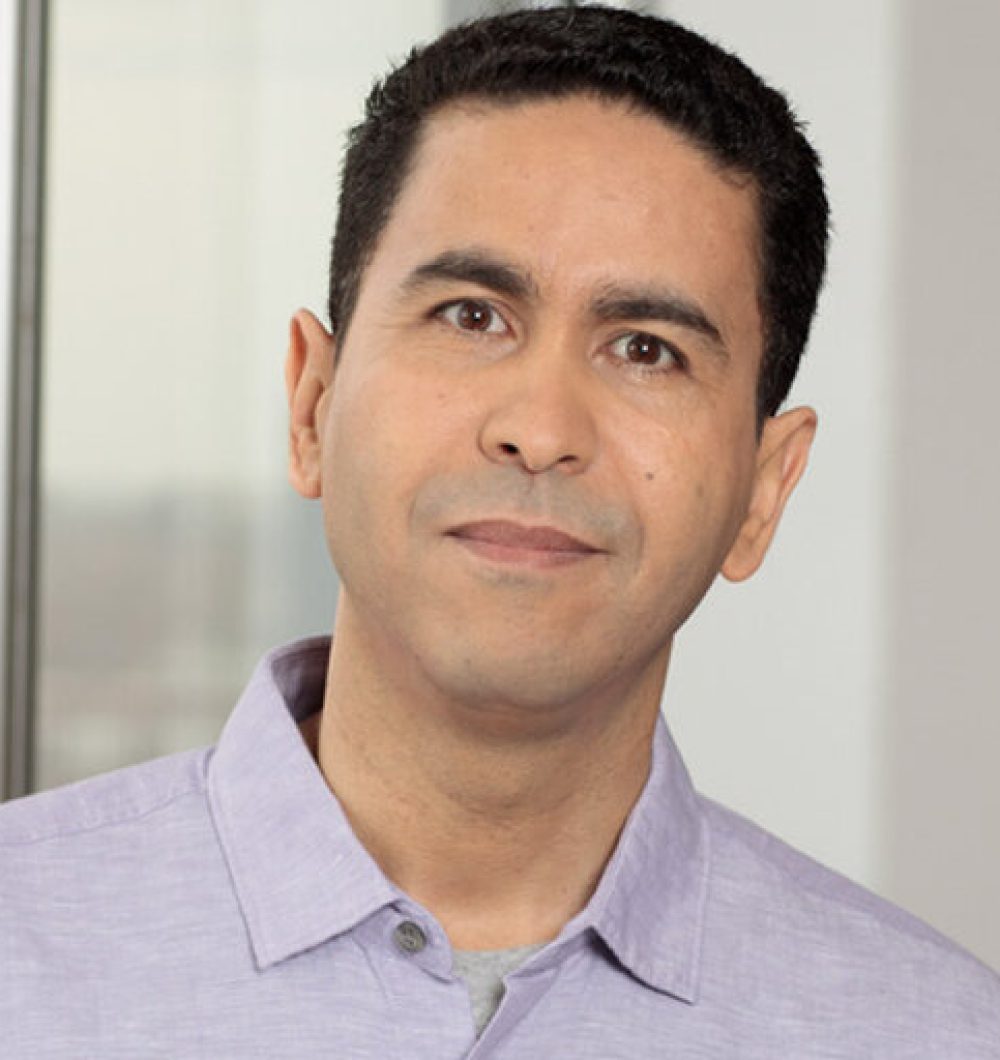}}]{Essaid Sabir } (Senior Member, IEEE) received the Ph.D. degree with honors in Networking and Computer Engineering from Avignon University, France, in 2010. From 2009 to 2024, he held various academic positions at Avignon University, Hassan II University of Casablanca, and Université du Québec à Montréal. Since 2024, he has been a Full Professor at TÉLUQ, Université du Québec. His research interests include 5G/6G networks, the Internet of Things (IoT) and Internet of Everything (IoE), ubiquitous networking, artificial intelligence and machine learning (AI/ML), and game theory. He has led and contributed to numerous national and international research projects and has received multiple awards in recognition of his work. He is actively engaged in the research community as the founder of the UNet Conference and co-founder of the WINCOM Conference. He also serves as an area, associate, and guest editor for several top-tier journals and regularly contributes to the organization of major international conferences.
\end{IEEEbiography}

\begin{IEEEbiography}[{\includegraphics[width=1.05in,clip,keepaspectratio]{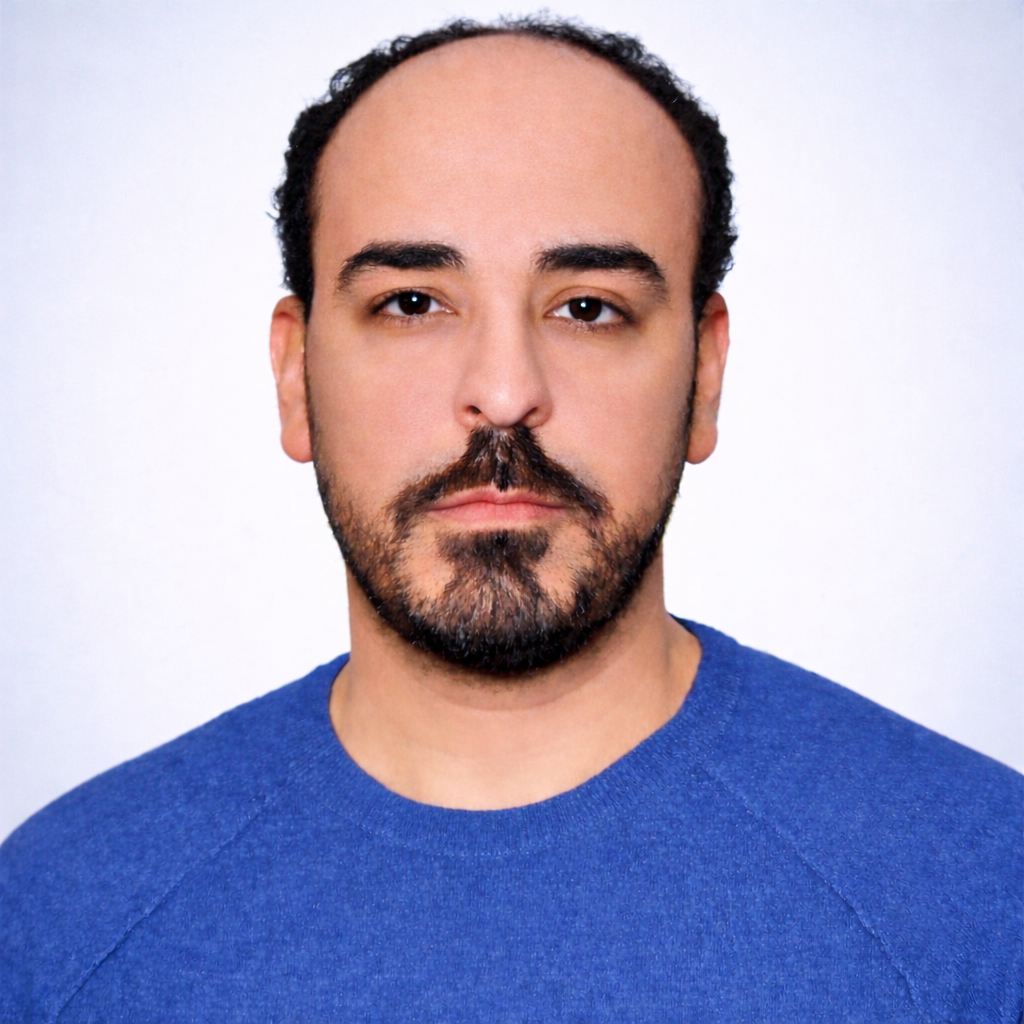}}]{Elmahdi Driouch } (Senior Member, IEEE) received the B.E. degree from the National School of Applied Sciences, Marrakech, Morocco, in 2006, and the M.Sc. and Ph.D. degrees in Computer Science from the Université du Québec à Montréal (UQAM), Canada, in 2009 and 2013, respectively. He held postdoctoral research positions at Concordia University from 2014 to 2015 and at UQAM from 2016 to 2017. From 2017 to 2019, he was an Assistant Professor with the Department of Computer Science at the Université de Moncton. He is currently an Associate Professor with the Department of Computer Science at UQAM. His research interests include wireless communications and networks, resource allocation, and algorithm design, with a particular focus on efficient and scalable solutions for next-generation communication systems. His work has been published in numerous peer-reviewed journals and international conferences, where it has contributed to both theoretical advancements and practical system design.
\end{IEEEbiography}

\begin{IEEEbiography}[{\includegraphics[width=1.05in,clip,keepaspectratio]{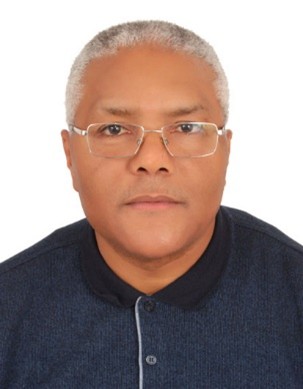}}]{Abdelkrim Haqiq } (Senior Member, IEEE) received a High Study Degree (Diplôme des Etudes Supérieures de troisième cycle) and a PhD (Doctorat d'Etat), both in the field of modeling and performance evaluation of computer networks, from Mohammed V University, Faculty of Sciences, Rabat, Morocco. Since September 1995, he has been working as a Professor at the department of Applied Mathematics and Computer at the Faculty of Sciences and Techniques, Settat, Morocco. He is a member of Machine Intelligence Research Labs (MIR Labs), Washington, USA and a member of the International Association of Engineers (IAENG).
He is an Expert Scientific Reviewer at the National Center for Scientific and Technical Research, Morocco.
He was the Director of Computer, Networks, Mobility and Modeling laboratory (IR2M). From 2010 to 2024, he served as a co-director of a NATO Multi-Year project entitled “Cyber Security Analysis and Assurance using Cloud-Based Security Measurement system”, having the code: SPS-984425.
His interests lie in the areas of modeling and performance evaluation of communication networks, mobile networks, cloud computing and security, emergent technologies, queueing theory, Markov decision processes theory, and game theory. He (co)authored more than 200 journal/conference papers.
He supervised 24 PhD thesis and co-supervised 3 others. Currently, he is supervising and co-supervising more 9 PhD thesis, some of them with the University of Castilla-La Mancha, Albacete, Spain, the TELUQ University and the Carleton University in Canada.
He is an associate editor of the International Journal of Computer Information Systems and Industrial Management Applications, an editorial board member of the International Journal of Intelligent Engineering Informatics, and of the International Journal of Blockchains and Cryptocurrencies, an international advisory board member of the International Journal of Smart Security Technologies, and of the International Journal of Applied Research on Smart Surveillance Technologies and Society. He is also an editorial review board of the International Journal of Fog Computing and of the International Journal of Digital Crime and Forensics.
He was a chair and a technical program committee chair/member of many international conferences and scientific events. He was also a Guest Editor and Co-Editor of special issues of numerous journals, books and international conference proceedings.


\end{IEEEbiography}

\end{document}